\documentclass[structabstract]{aa_arxiv}
\usepackage{txfonts}
\usepackage{graphicx}
\usepackage{natbib}
\usepackage{hyperref}
\bibpunct{(}{)}{;}{a}{}{,} 

\begin{document}
\title{Automatic stellar spectra parameterisation \\
in the IR \ion{Ca}{ii} triplet region}
\author{G.~Kordopatis \inst{\ref{inst1}}
  \and A.~Recio-Blanco \inst{\ref{inst1}}
  \and P.~de~Laverny \inst{\ref{inst1}}
  \and A.~Bijaoui \inst{\ref{inst1}}
  \and V.~Hill \inst{\ref{inst1}}
  \and G.~Gilmore \inst{\ref{inst2}}
  \and R.F.G.~Wyse \inst{\ref{inst3}}
  \and C.~Ordenovic \inst{\ref{inst1}}
}
\institute{Universit\'e de Nice Sophia Antipolis, CNRS, Observatoire de la C\^ote d'Azur, Cassiop\'ee UMR 6202, BP 4229, 06304 Nice, France  \email{georges.kordopatis@oca.eu} \label{inst1}
\and Institute of Astronomy, University of Cambridge, Madingley Road, Cambridge CB3 0HA, UK \label{inst2}
\and Johns Hopkins University, Baltimore, MD, USA  \label{inst3}
 }
\date{Received / Accepted}

 
  \abstract
  {Galactic archaeology aims to determine the evolution of the Galaxy
    from the chemical and kinematical properties of its individual
    stars. This requires the analysis of data from large spectroscopic
    surveys, with sample sizes in tens of thousands at present, with
    millions of stars being reached in the near future. Such large
    samples require automated analysis techniques and classification
    algorithms to obtain robust estimates of the stellar
    parameter  values. Several on-going and planned spectroscopic surveys have
    selected their wavelength region to contain the IR \ion{Ca}{ii}
    triplet ($\sim \lambda \lambda \ 8500 \ \AA$) and the work
    presented in this paper focuses on the automatic analysis of such
    spectra.}
  {We aim to develop and test an automatic method by which one can obtain
    estimates of values of the stellar atmospheric parameters
    (effective temperature, surface gravity, overall metallicity) from a stellar spectrum.  We also
    explore the degeneracies in parameter space, estimate the
    uncertainties in the derived parameter values and investigate the
    consequences of these limitations for achieving the goals of
    galactic archaeology. }
  {We investigated two algorithms, both of which compare the observed
    spectrum to a grid of synthetic spectra, but each uses a
    different mathematical approach for finding the optimum match and
    hence the best values of the stellar parameters.  Our
    investigation of these algorithms' robustness can be widely
    applied because it amplifies the main problems that the other
    methods can encounter.
    The first algorithm, MATISSE, derives the values of each stellar
    parameter through a local fit to the spectrum such that each pixel
    in wavelength space is treated separately.  The sensitivity of the
    flux at each wavelength to the value of a given stellar parameter
    is determined from the synthetic spectra. The observed spectrum is
    then projected using these sensitivity vectors to give an
    estimated value of the stellar parameters. This value depends on
    finding the true minimum in the fit and the algorithm must avoid
    being trapped in false local minima.  The second algorithm, DEGAS,
    uses a pattern-recognition approach and consequently has a more global
    vision of the parameter space. The best-fit synthetic spectrum is
    derived through a series of comparisons between the observed and
    synthetic spectra, summed over wavelength pixels, with additional
    refinements in the set of synthetic spectra after each stage,
     i.e. a decision tree.}
  {We identified physical degeneracies in different regions of the
    H--R diagram: hot dwarf and giant stars share the same
    spectral signatures.  Furthermore, it is very difficult to determine
    an accurate value for the surface gravity of cooler dwarfs. These
    effects are intensified when the lack of information increases,
    which happens for low-metallicity stars or spectra with low
    signal-to-noise ratios (SNRs). Our results demonstrate that the
    local projection method is preferred for spectra with high SNR,
    whereas the decision-tree method is preferred for spectra of lower
    SNR.  We therefore propose a hybrid approach, combining these
    methods, and demonstrate that sufficiently accurate results for
    the purposes of galactic archaeology studies are retrieved down to
    SNR$\sim$20 for typical parameter values of stars belonging to the
    local thin or thick disc, and for SNR down to $\sim$50 for the
    more metal-poor giant stars of the halo.}
  {If unappreciated, degeneracies in stellar parameters can introduce
    biases and systematic errors in derived quantities for target
    stars such as distances and full space motions. These can be
    minimised using the knowledge gained by thorough testing of the
    proposed stellar classification algorithm, which in turn lead to robust
    automated methods for the coming extensive spectroscopic surveys of
    stars in the Local Group. }


\maketitle

%

\section{Introduction}
Understanding the formation and evolution of the Milky Way Galaxy from
the properties of its long-lived constituent stars (also known as the
field of galactic archaeology) requires the collection of photometric
and/or spectroscopic data for statistically significant samples of
stars throughout the Galaxy.  Photometry has the advantage of faster
completion of deep wide-field/all-sky surveys but spectroscopy
provides more accurate, detailed information about the target stars.
For example, depending on the spectral resolution, one can more easily
determine overall metallicities ([M/H]), the enhancements of $\alpha$-elements
with respect to iron (with respect to the Sun, [$\alpha$/Fe]),
individual elemental abundances and the fundamental stellar parameters
effective temperature ($T_\mathrm{eff}$) and surface gravity
(log~$g$). The combination of these parameters allows one to derive line-of-sight
distances  through a comparison of a given star's
position on the H-R diagram and an appropriate set of theoretical
isochrones.  Full 6D phase-space coordinates can be determined if
proper motions are available, which additionally constrain  
models of Galaxy formation and evolution.

The wavelength range around the IR \ion{Ca}{ii} triplet is more and
more frequently used in studies of galactic archaeology. The IR
\ion{Ca}{ii} triplet ($\lambda \lambda 8498.02, 8542.09, 8662.14$~\AA) 
lines are strong for most stellar spectral types and
luminosity classes, as well as for very metal-poor stars \citep[see,
for example][]{Gaia_Range, Wilkinson_RVS}, providing ideal features
for a robust determination of the star's radial velocity. The usefulness
of this wavelength range goes beyond line-of-sight kinematics, because
there are numerous absorption lines from many species, including iron
and several $\alpha$-elements (\ion{Ca}{ii}, \ion{Si}{i}, \ion{Mg}{i})
and these remain visible even at relatively low spectral resolution
and low signal-to-noise ratio (SNR\footnote{Throughout the paper, we define the SNR as being 
the signal-to-noise ratio per pixel.}).  The pattern of the ratios of
these $\alpha$-elements to iron can be used to trace the star formation
time-scale in the parent system owing to distinctive roles played by
the supernovae of different type (and explosion timescale) in creating
and ejecting the different elements.  In addition, an estimate of
overall metallicity can be derived from empirical calibrations between
metallicity ([M/H]) and the equivalent widths of the \ion{Ca}{ii}
triplet
\citep{CaII_calibration_Battaglia,CaII_calibration_low_meta,Fulbright_2010}.
Furthermore, Paschen lines (for example $\lambda \lambda 8502.5,
8545.4, 8598.4,8665.0,  8750.5$~\AA) are visible for stars hotter than spectral
type G3. The \ion{Mg}{i} ($\lambda \lambda 8807$~\AA) line, which is a
useful indicator of surface gravity \citep[see][]{MgI_line, Battaglia11}, is also
visible, even in spectra of low SNR. Finally, lines from molecules
such as TiO and CN can be seen in spectra of the cooler stars.
 
The collection of very large samples of spectroscopic data to
undertake studies in galactic archaeology has become feasible in
 recent years. For example, multifiber instruments such as the
GIRAFFE/FLAMES spectrograph, mounted on Kueyen, the second-unit
telescope of the Very Large Telescope (VLT) of the European Southern
Observatory (ESO) allows spectra of  more than a hundred objects to be obtained
at a time with adequate SNR  (typically higher than $\sim$20)
 after only a few hours of exposure time.
The first galactic archaeology survey to use the IR \ion{Ca}{ii}
triplet region has been the RAdial Velocity Experiment survey
\citep[RAVE, $\lambda \lambda 8410-8795$~\AA,
see][]{presentation_RAVE}. This targets bright (limit of $I\sim13$)
stars and uses the 6dF multi-object spectrograph on the UKSchmidt
telescope with a resolution of R$= \lambda / \Delta\lambda \sim7500$;
the RAVE project has already collected more than four hundred thousand
spectra. In addition, several studies of the stellar populations of
the Milky Way and its satellite dwarf galaxies have been conducted
with the FLAMES multifiber spectrograph of ESO, using instrumental
setups centred on the IR \ion{Ca}{ii} triplet (specifically, the LR8
and HR21 setups).  The series of papers related to the ESO large
programmes DART \citep[][]{Tolstoy_DART, Battaglia_DART_2006,
  Battaglia_DART_2010} and that of Gilmore et al. 171.B-0520(A)
\citep[see][and references therein]{wyse_2006,Koch_2007,Koch_2008}
are good examples of the use of the IR \ion{Ca}{ii} triplet for
galactic archaeology. Furthermore, in the near future the ESA Gaia
mission will collect several tens of millions of spectra with its
Radial Velocity Spectrometer \citep[RVS,][]{Wilkinson_RVS}, at a
spectral resolution R$\sim 11500$ and two different samplings --
0.02453~nm and 0.07359~nm  (effective resolution of $\sim$7000) --
depending on the brightness of the targets.  \\

Despite this extensive use of the IR \ion{Ca}{ii} triplet region and
its previously mentioned advantages, the determination of the values
of stellar atmospheric parameters from low- to medium-resolution
spectra in this wavelength domain is far from trivial. In particular,
as we will see below, there exists a strong degeneracy between the
effective temperature and the surface gravity, in that varying either
produces similar changes in the normalised stellar flux of several
spectral features. This degeneracy can lead to the misclassification
of dwarf stars as giant stars, and {\it vice versa}, a problem that becomes more severe for
low-metallicity stars.  This misclassification obviously creates
errors in the derived stellar distances based on the estimated values
of the atmospheric parameters. The ability to mitigate this degeneracy
must be built-in to the automated methods of stellar parameter
determination and that is the only feasible way of dealing with the massive
datasets produced by the coming galactic archaeology surveys.  

The goal of this paper is to show which astrophysical information can
be retrieved from spectra observed around the calcium triplet.  We
investigate the performances of two methods of automated estimation of
the values of stellar parameters based on very different mathematical
approaches, which allows us to describe the main problems encountered in
this spectral region.  As a specific example of the application of
these two algorithms, we consider the case of spectra obtained with
the LR8 setup of FLAMES (8206-9400 \AA, R$\sim$6500,
sampling=0.2\AA). The spectra obtained using the other instruments
and/or setups mentioned above (RAVE - RVS - FLAMES/HR21) have either
broader wavelength ranges or higher spectral resolutions, and each of
these provides more information that can help to disentangle some of
the degeneracies. Consideration of the degeneracies resulting from
spectra obtained with the LR8 setup will therefore cover those likely
to result from the other setups.

The structure of this paper is as follows: in
Sect.~\ref{sec:automatic_algorithms} we introduce the two methods
whose performances we test, and in
Sect.~\ref{sec:training_testing_grid} we explain how the synthetic
spectra used as training and testing sets have been computed. In
Sect.~\ref{sec:performances} we present the performances that are
achieved by each of the two applied algorithms and present the final
adopted strategy for the derivation of the values of the stellar
atmospheric parameters.  Section~\ref{sec:final_pipeline} is dedicated
to the application of the method to real spectra, dealing with errors
in the derived parameter values that are introduced by uncertainties
in the radial velocity and in the continuum normalisation.  Finally,
in Sect.~\ref{sect:conclusions} we discuss the effects of the errors
in the parameter determinations on the science goals of galactic
archaeology.

\section{Automatic parameter estimation methods}
\label{sec:automatic_algorithms}
The parameter estimation problem consists in finding the stellar
atmospheric parameters (mainly effective temperature, surface gravity,
global metallicity and individual chemical abundances) that define a
synthetic spectrum that is an optimal fit to an observed
spectrum. This estimation cannot be made analytically, because of the
complexity of the theoretical spectra, which involve the very complex
physics included in stellar model atmosphere and spectral line
formation theory. As a consequence, the stellar spectra
parameterisation has to be performed by using synthetic spectra grids which 
span the parameter space that we are concerned with here.

We point out that we favour in our study parameter estimations
from synthetic spectra grids rather than grids of real observed stars,
mainly because a sufficiently accurate parameterisation of large samples of
different types of stars does not exist. Therefore, the isolation of
flux variations caused by variations of only one of the atmospheric
parameters cannot be easily performed, which is different from synthetic spectra.

The main problems encountered by the different automated estimation
methods are caused by the possible non-linearity of the model spectra and
the non-convexity of the distance function, which quantifies the
difference between a model and the data. On one hand, a change in the
atmospheric parameters can induce non-linear variations of the
spectral flux, which is more important for large-scale
parameter variations: two sets of parameters, distant in parameter
space, can result in very similar spectra.  On the other hand, the
flux changes caused by the variation of a given parameter, for instance,
the effective temperature, can be very similar to those induced by
another parameter, usually the surface gravity or the metallicity.
Therefore, secondary minima and multiple solutions to
the stellar spectra fitting problem may exist . Parameter degeneracy is usually
more severe when the available information about the parameters
decreases: e.g., with a more narrow spectral range, lower spectral resolution,
lack of spectral signatures, and so on. In addition, secondary minima
can also be artificially generated by noise disturbing the distance
function.

As presented in \citet{Bijaoui_automatic_classification_methods},
several methods exist in the literature for the automatic parameter
estimation from stellar spectra. The methods can be separated into {\it i)}~solving 
an optimisation problem (e.g. minimum distances, Nelder-Mead
algorithm, Gauss-Newton algorithm), {\it ii)}~optimising a projection on
given vectors (e.g. principal component analysis, MATISSE algorithm,
penalized $\chi^2$), or {\it iii)} as a classification problem
(e.g. artificial neural networks, support vector machines, decision
trees).

Optimisation and projection algorithms usually rely either on the
convexity of a distance function or the linearity of the
models. Nevertheless, a local parameter space treatment is sometimes considered 
 to minimise the failure of these conditions.  On the other
hand, classification algorithms tackle the parameterisation problem
from a completely different point of view: the pattern recognition
one. This different approach can have many advantages in severe
conditions of parameter degeneracy.

As the work of this paper shows, each method has its optimal
application field, and a combination of algorithms can be necessary
for achieving the best results.  We will illustrate
the difficulties that we encountered when performing the stellar
parameterisation of IR \ion{Ca}{ii} triplet region low-resolution
spectra.  To do this, we will apply two different parameterisation
methods: a projection algorithm, MATISSE \citep{Matisse_MNRAS} and a
classification method, DEGAS
\citep{Bijaoui_automatic_classification_methods}, based on an oblique
k-d decision tree. Those two algorithms, which are representative of two very
different mathematical approaches, exemplify the main problems that
the other methods can encounter when dealing with this kind of
spectra.

\subsection{The MATISSE method}
\label{sec:matisse}

\begin{figure*}[!t]
\centering
\includegraphics[width=16.6cm,height=10.cm]{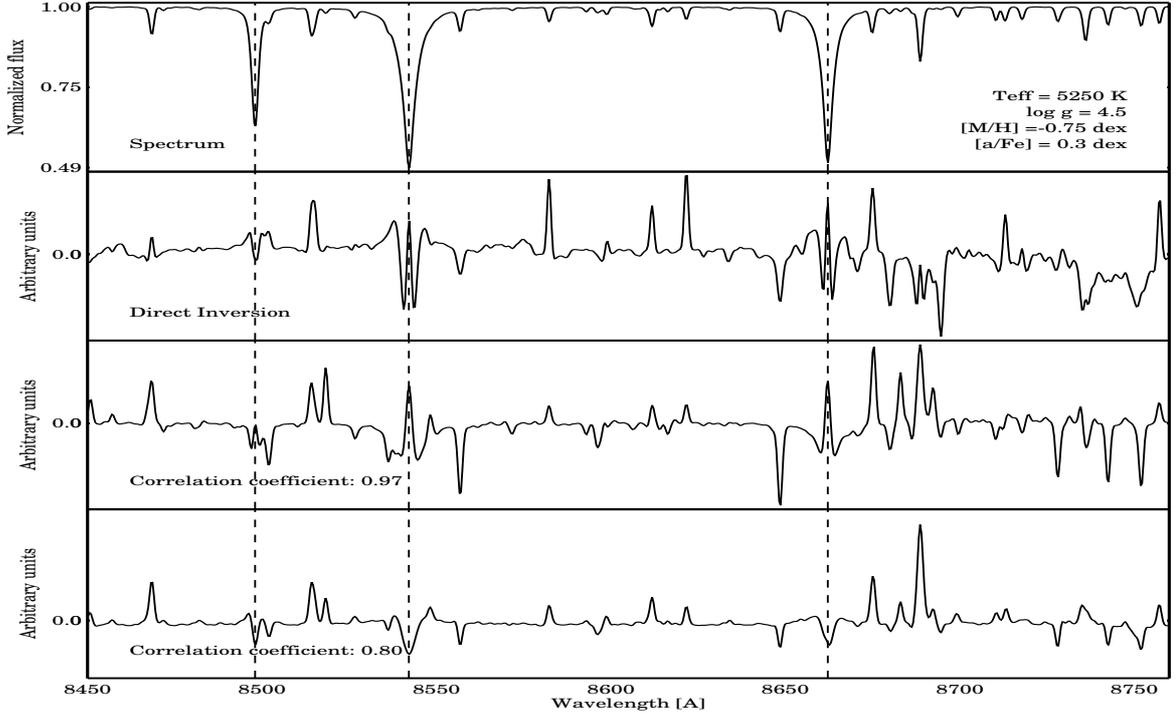}
  \caption{Illustration of the MATISSE basis functions. For every set of
    parameters $\theta_i$, corresponding to a spectrum in a grid of
    synthetic spectra, a set of $B_\theta(\lambda)$ was
    computed. Here, we can see the computed
    $B_{T_\mathrm{eff}}(\lambda)$ functions (and the corresponding
    synthetic spectrum) for a star with $T_\mathrm{eff}$=5250~K,
    $\log~g$=4.5, [M/H]=$-$0.75~dex, [$\alpha$/Fe]=0.3~dex. For clarity
    we represent only a part of the wavelength range in the
    plot. The strongest features in the spectrum (top panel),
    identified with dashed vertical lines, correspond to the
    \ion{Ca}{ii} triplet.  In the second row we show the
    $B_{T_\mathrm{eff}}(\lambda)$ computed with a direct inversion of
    the correlation matrix, whereas the last two panels show the
    $B_{T_\mathrm{eff}}(\lambda)$ functions computed with the
    Landweber iterative algorithm, imposing a coefficient of
    correlation between the input and the output parameters of 0.97
    and 0.80, respectively. As described in Sect.~\ref{sec:matisse},
    the smaller the coefficient, the less weight is given to the
    second-order variations of the spectral flux. The spectral
    analysis can therefore be optimised according to the quality of the
    spectrum, which is typically quantified in the signal-to-noise
    ratio.}
  \label{fig:Bfunctions}
\end{figure*}

The MATISSE algorithm (MATrix Inversion for Spectral SynthEsis) is a
local multi-linear regression method.  We briefly recall
now the basic equations of the method, but we encourage the reader
to review \citet{Matisse_MNRAS} and \citet{Bijaoui_Matisse} for a more 
comprehensive description. \\
MATISSE estimates a $\hat{\theta_i}$ stellar atmospheric parameter
($T_\mathrm{eff}$, $\log~g$, [M/H]) by projecting the observed spectrum
$O(\lambda)$ on a particular vector $B_\theta(\lambda)$ associated to
a theoretical $\theta_i$ parameter, as follows:

\begin{equation}
\hat{\theta_i}=\sum_\lambda B_{\theta_i}(\lambda) \cdot O(\lambda) .
\label{eq:general_matisse} 
\end{equation}

These vectors, called $B_\theta(\lambda)$ functions hereafter, are
computed during a learning phase from a library of synthetic spectra
 that cover the same wavelength range and have the same spectral
resolution and wavelength sampling as the observed spectra.  They
relate in a quantitative way the pixel-pixel flux variations in a
spectrum to a given variation of the $\theta_i$ parameter. 
If the $B_\theta(\lambda)$ are orthogonal, the effects from each
parameter affect the spectrum in a different way, and therefore the
atmospheric parameters are derived accurately.  When this is not the
case, possible degeneracies in parameter space can occur, which cause
a correlation of the parameter errors.  The $B_\theta(\lambda)$
functions are computed from an optimum multi-linear combination of
theoretical, synthetic spectra $S(\lambda)$, as follows:

\begin{equation}
B_{\theta_i}(\lambda)=\sum_j \alpha_{ij} \cdot S_j(\lambda),
\label{eq:bfunctions_matisse} 
\end{equation}
where the $\alpha_{ij}$ factor is the weight associated with each
synthetic spectrum $S_j(\lambda)$ to retrieve the
$\hat{\theta_i}$ parameter. To compute these weights during the
learning phase, Eq. \ref{eq:general_matisse} is applied to a subset of
synthetic spectra. Accordingly, we obtain by combining Eq. \ref{eq:general_matisse} and
\ref{eq:bfunctions_matisse}:

\begin{equation}
 \Theta_i = C \alpha_i ,
\label{eq:correlation_matrix_matisse} 
\end{equation}
where $C=[c_{jj'}]$ is the correlation matrix  between $S_j$ and $S_{j'}$, and $\Theta_i$ as well as $\alpha_i$ 
the vectors of the parameters $\theta_i$ and the weights $\alpha_{ij}$ for all considered
 spectra. The weights $\alpha_{ij}$ are then computed by inverting the
correlation matrix $C$.

We have as many $B_\theta(\lambda)$ functions as there are spectra in the
library of synthetic spectra. Each of them is computed using a small
parameter range, for which we assumed the parameter variations to
have a  linear effect on the spectral flux variations.  In
practice, to converge to a parameter sub-space, one can use
either generic $B^0_\theta(\lambda)$ functions, or impose \textit{a
  priori} the $B_\theta(\lambda)$, if a first estimation of the
parameters is available\footnote{The first estimates can come
  either from photometric measurements or from results of other
  algorithms.}.  When applying Eq.~\ref{eq:general_matisse} to
an observed spectrum, the method iterates for as long as the derived
parameters are not included in the considered parameter sub-space for
which the $B_\theta(\lambda)$ functions were computed.  Usually,
the final convergence is attained after a few iterations.

Here, the MATISSE method deals with normalised spectra and
consequently all the necessary information for deriving the atmospheric
parameters is provided by the spectral lines, i.e. their relative
strengths. This strength changes according to the spectral type, the
luminosity class and the metal content of the stars.  In noiseless
spectra and when the synthetic spectra perfectly match real stellar
spectra, all astrophysical information (i.e. spectral lines) can be
taken into account during the training phase of MATISSE. Nevertheless,
for noisier spectra or if there is a mismatch between the observed and
the synthetic stellar spectra, it is better to consider only the most
relevant features,  and give less weight to second-order variations.

The possibility of identifying a specific subset of the spectral
signatures that are to be used for an atmospheric parameter
derivation, which is a particular feature of MATISSE, allows us to
adopt an optimised strategy according to the SNR and the type of the
star, and therefore to optimise the analysis according to the quality of
each spectrum.  This optimisation is achieved when computing the
$B_\theta(\lambda)$ functions, and more precisely while inverting the
correlation matrix $C$ of Eq.~\ref{eq:correlation_matrix_matisse}. A
direct inversion would take into account all the {\it n}-order variations
caused by the parameters.  If one only considers the
first order variations of the spectral flux though, one has to approximate
the inverse of $C$.  The degree of approximation can be controlled by
using an iterative algorithm, such as that of Landweber
\citep{Landweber}. In that case, one can impose a correlation
coefficient between the input and the output parameters to
be equal to the desired degree of approximation. The higher the factor
of correlation, the lower the degree of approximation. A correlation
coefficient of one would be similar to a direct inversion.

In practice, the correlation matrix can  sometimes be ill-conditioned, and 
can therefore imply many near-zero eigenvalues. The Landweber algorithm adapts the 
inversion to the matrix conditioning, in the sense that the first eigenvectors 
are inversed during the first iterations, and the inversion of the smallest ones 
need more iterations to be accomplished. The iteration number is linked
 with the correlation coefficient cited above.

We took the grid of the synthetic spectra described in
Sect.~\ref{sec:grid}, stopped the iterations for the different values
of correlations of 0.75, 0.80, 0.90, 0.95 and 0.98, 
 and computed the
$B_\theta(\lambda)$ functions with the respective approximations of
$C^{-1}$.  This is illustrated in Fig.~\ref{fig:Bfunctions}, where we
plot the same $B_{T_\mathrm{eff}}(\lambda)$ function computed with
different correlation factors.  As we notice in this figure, 
all minor features of the spectrum are given a very low 
weight for a correlation factor of 0.80
(bottom panel of Fig.~\ref{fig:Bfunctions}).
These approximated $B_\theta(\lambda)$ functions are fully
justified as long as the spectra are of low quality,
where these minor features are anyway lost in the noise.  \\

Application of the various approximated $B_\theta(\lambda)$ functions
on noisy synthetic spectra allowed us to select the most suitable
combination, as a function of spectral type, metallicity, and SNR (see
Sect. \ref{subsec:performances_MATISSE}).

\subsection{DEGAS: an oblique k-d decision-tree method}

In the limit of the sampling precision, the parameter estimation is a
recognition problem.  The grid of synthetic spectra can be treated as a
known set of patterns among which we aim to identify the observed
spectra. In the learning phase, the recognition rules are established
using the grid of theoretical spectra.

Decision trees are commonly used for data mining
\citep{Quinlan_decision_trees}. At each tree node, a decision is taken
to split the data subset into two or more subsets. The leaf
level corresponds to the identified classes.  A k-d tree is a basic
space-partitioning structure in a k-dimensional space (in our case,
k=3, $T_\mathrm{eff}$, $\log~g$ and [M/H]). The decisions result from the
projection of the observations on a node vector. Classical, or
axis-parallel decision trees, check only one variable at each node.
In the particular case of oblique decision trees, the node vectors
(called {$\bf D_n$} hereafter) are obtained from a linear combination
of the structural features (i.e. the atmospheric parameters),
resulting in a simpler and more accurate tree
\citep{White_decision_trees}.

The DEGAS (DEcision tree alGorithm for AStrophysics) algorithm is an
oblique k-d decision tree, for which a preliminary version has already
been presented in \citet{Bijaoui_automatic_classification_methods}.
The recognition rules at each node are built during the learning
phase as follows:

\begin{enumerate}
\item The mean vector {\bf M} of the flux values per pixel is computed.

\item For each spectrum {$\bf S_j$} associated to the node, we calculate the scalar product
$c_j = {\bf S_j} \cdot {\bf M}$. Let \~c be the median value of $c_j$. 

\item The data are bisected into two subsets, $T_1$ and $T_2$, according to the following
criteria:\\
$S_j$ belongs to the subset $T_1$ if $c_j \leqslant $ \~c \\
$S_j$ belongs to the subset $T_2$ if $c_j$ $>$ \~c

\item The mean vectors {$\bf M_1$}  and {$\bf M_2$} of each subset are then computed, 
and the difference vector ${\bf D} = {\bf M_1} - {\bf M_2}$ is determined.

\item According to the Huygens theorem, the best separation between $T_1$ and
  $T_2$ corresponds to the case for which the dispersions around $M_1$
  and $M_2$ are minimal, which is when {\bf M} and {\bf D} are
  parallel.  Indeed, we examined the variation $V$ of the dispersions
  that resulted from spectra exchange between the two sets $T_1$ and
  $T_2$. Because the $V$ expression introduced the vector {\bf D}, we chose
  to get {\bf M} and {\bf D} parallel in order to maximise V. This
  dramatically improved the separation compared to the use of {\bf M}.
  Hence, in practice, if the angle between {\bf M} and {\bf D} is too
  wide (correlation coefficient smaller than 0.999), we iterate until
  convergence, re-separating the data by the hyperplane defined by
  {\bf D} (going back to step 2, replacing {\bf M} by {\bf D}).

\end{enumerate}

Once this procedure has converged for a particular node {\it n}, we determine
the final projection node vector {$\bf D_n$}, which will display
  the features that allow one to separate the data at that node, and the final median value $\tilde{c_n}$ of $c_j = {\bf S_j} \cdot {\bf D_n}$.

In this way, the recognition tree with $\mathrm{log_2}(N)$
levels is built, where $N$ is the number of spectra of the training grid. At
the lowest level nodes of the tree, the leaves, only one training
spectrum remains associated with each node.  During the application
phase, the target data {$\bf O_i$} passes through all levels of
the recognition tree, and a template is associated to it.

Nevertheless, noise can induce possible misclassifications. At each
node one has to take into account that the projection coefficient
$c_i={\bf O_i} \cdot {\bf D_n}$ is distributed according to a Gaussian
law, defined for the entire real axis. So, theoretically both branches
can be chosen, and the algorithm should fully explore the decision
tree. This would lead to an  inefficient method.  To allow the
assignation of a probability to each of the directions, we chose to
replace the Gaussian distribution by an Epanechnikov kernel, which
corresponds to a truncated parabola.  Let us consider
\begin{equation}
u_i=\frac{c_i-\tilde{c_n}}{\sigma_{c_i}},
\end{equation}
where $\sigma_{c_i}=\frac{1}{\mathrm{SNR}} \cdot \sqrt{\Sigma_\lambda
  D_n(\lambda)^2}$.  If $u_i \leqslant -k$ we decide that the correct
direction is 1. If $u_i \geqslant k$  direction 2 is chosen.  If
$-k < u_i < k$ both directions are considered.  Owing to the threshold,
generally only one direction is chosen, and in the end only few
leaves are selected.

After scanning all nodes, a subset of synthetic templates
is selected, and the distances of the templates from the observed spectrum are
computed.  Then, a weighted mean is evaluated on the parameters taking into
account these distances, setting
\begin{equation}
W_{i}^{n} =(1- |{\bf O_i} - {\bf S_n}|^2)^{p}.
\label{eq:dichodif_weight_distances} 
\end{equation}
The value $p$ of the polynomial exponent is fairly arbitrary. If 
the noise is important, the distances between the selected
templates and the observed spectrum are quite similar, and therefore a
strong exponent is needed to put more weight to the most similar
solution.  We tested several values ($p=16, 32, 64, 128$) on a
set of noisy synthetic spectra and decided to use $p=64$ throughout.

\section{Creating the training and testing spectroscopic set}
\label{sec:training_testing_grid}

As  noted earlier, most automatic spectral analysis methods,
e.g. artificial neural networks or maximum likelihood algorithms,
require a library of synthetic or observed
spectra with well-known parameters for their learning phase. These  reference spectra have to cover the whole
parameter space $T_\mathrm{eff}$, $\log~g$, [M/H] where the selected
targets are expected. No available observed libraries with those
characteristics that  covered our wavelength range with sufficiently accurate 
parameters were available in our case.  We therefore chose to compute a
synthetic library to fulfil our objectives.

\subsection{Grid of synthetic spectra for the learning phase.}
\label{sec:grid}

\begin{table}[!t]
\centering
\caption{Atmospheric parameters of the synthetic spectra of the learning grid. }
\begin{tabular}{lcc}
\hline
\hline
& Range & Step \\ \hline
$T_\mathrm{eff}$ & 3000 ; 8000 K     & 200 K between [3000;4000] \\ 
                &                   & 250 K  between [4000;8000] \\ \hline
log~$g$         & 0.0 ; 5.0         & 0.5 dex\\ \hline
[M/H]           & $-$5.0 ; +1.0 dex & 1.00 dex  between [$-$5;$-$3]\\ 
                &&0.50 dex  between [$-$3;$-$1]\\
                && 0.25 dex  between [$-$1;+1]\\\hline
\end{tabular}
\label{tab:param_range}
\end{table}

A library of synthetic spectra spanning the parameter space as
detailed in Table \ref{tab:param_range} was computed using MARCS
model atmospheres \citep{VALD_Suede} and the Turbospec code
\citep[][and further improvements by B. Plez]{turbospec}.  We assumed
a coupling between the overall metallicity and the $\alpha$-element
abundances\footnote{The chemical species considered as
  $\alpha$-elements are \ion{O}{}, \ion{Ne}{}, \ion{Mg}{}, \ion{Si}{},
  \ion{S}{}, \ion{Ar}{}, \ion{Ca}{} and \ion{Ti}{}.} according to the
commonly observed enhancements in metal-poor galactic stars. We considered
\begin{itemize}
\item $[\alpha$/Fe]=0.0 dex for 0.0 $\leq$ [M/H] $\leq$ +1.0 dex
\item $[\alpha$/Fe]=+0.1 dex for [M/H]=$-$0.25 dex
\item $[\alpha$/Fe]=+0.2 dex for [M/H]=$-$0.50 dex
\item $[\alpha$/Fe]=+0.3 dex for [M/H]=$-$0.75 dex
\item $[\alpha$/Fe]=+0.4 dex for [M/H] $\leq$ $-$1.0  dex.
\end{itemize}

Each spectrum was computed assuming hydrostatic and local
thermodynamic equilibrium (LTE), covering the wavelength range
8390-8860~\AA~ with a wavelength step of 0.02~\AA. For stars with
$3.5<$ $\log~g <5.0$~(cgs units), plane-parallel models were used, with
a microturbulence parameter $\xi$=1~\mbox{km s}$^{-1}$. For giant
stars ($\log~g$ $ < 3.0$), spherical symmetry, with $\xi$=2~\mbox{km
  s}$^{-1}$ was preferred. These models, for which the sphericity
effects may be considerable for low gravities
\citep[see][]{ref_microturbulence}, have been calculated for a mass of
1$M_{\sun}$.  The final library contains 2905 spectra of 23501
pixels. Let us note that a few models are missing in the synthetic
library owing to the approach to the Eddington flux limit or poor
convergence, as described in \citet{VALD_Suede}.

This paper deals with spectra that will be obtained with the LR8 setup
of FLAMES \citep{Presentation_FLAMES}. Therefore, this library had to be
adapted to the observational setup. The spectra were restricted to the
wavelengths 8400-8820~\AA, which contain all the predominant lines, and
we trimmed out spurious effects in the CCDs (border effects, presence of a
glow in the red part) and possible sky residuals. For this reason, the
range between 8775~\AA~ and 8801~\AA, which contains few iron lines
and possibly has important sky residuals, was also removed.
The spectral feature corresponding to \ion{Mg}{i} around 
$\sim$8807~\AA ~was  kept though. \\
The spectra were then convolved with a Gaussian kernel and re-binned to
match the sampling and resolution of the LR8 spectra.  In addition,
eight re-sampled pixels\footnote{This corresponds to less than 1\% of
  the total pixels in the spectrum.} corresponding to the cores of the
strong \ion{Ca}{ii} lines were also removed from the spectra (two
pixels for the first line, and three for the other two lines,
corresponding to 0.8 and 1.2~\AA, respectively), because a disagreement
is expected between the synthetic spectra and the true ones for these
pixels. Indeed, the cores of strong lines have a significant contribution
from the upper layers of the stellar atmospheres, where LTE, one of
the assumptions of the model atmospheres which were used, is not
applicable any more.  Removing them will accordingly avoid possible biases
to the final parameter extraction. We checked this with the
tests presented in Sect.~\ref{sec:performances}.  The final spectrum
contains in the end only 957 pixels with a sampling of 0.4~\AA.

\subsection{Line-list calibration}
\label{sec:linelist}
In order to compute realistic synthetic spectra, it is necessary to
know the atomic and the molecular parameters
of the existing lines as accurately as possible . Whereas central wavelengths and excitation
potentials are fairly well established for most of the atomic lines,
the probabilities of electronic transitions (illustrated by the
oscillator strengths $\log gf$) are more subject to uncertainties.
The combination of a line-list, a set of model atmospheres and a
spectral synthesis method has to be calibrated on standard stars of
different spectral types to check the line-list quality and
obtain reliable synthetic stellar spectra. \\
For that purpose, we used the set extracted from the VALD
database\footnote{\url{http://vald.astro.univie.ac.at/}} in May
2009 \citep{VALD_Kupka}  as an initial atomic line-list and combined it with MARCS model atmospheres to reproduce the observed high
resolution, high SNR spectra of the Sun \citep{Brault_Solar} and
Arcturus \citep{Sun_Hinkle, Prieto_S4N} as precisely as possible .

All radiative transfer computations were made assuming LTE and
hydrostatic equilibrium, and therefore no effort was made to fit the cores of strong spectral features better (e.g. \ion{Ca}{ii}
triplet or \ion{Mg}{i} line).  Instead we removed the pixels corresponding to the cores of the \ion{Ca}{ii}
triplet as described in the previous section.

The VALD atomic line-list set was first modified, adopting the
oscillator strength corrections by \citet{VALD_Suede} for some
lines.  The molecular line-list includes \element{ZrO}, \element{TiO},
\element{VO}, \element{CN}, \element{C_2}, \element{CH},
\element{SiH}, \element{CaH}, \element{FeH} and \element{MgH} lines
with their corresponding isotopic variations (kindly provided by
B.~Plez).  The adopted solar abundances were those of
\citet{solar_abundance}, except for \element{CNO}, in which case we
used \citet{Asplund_Sun}. In addition, we considered $\xi$=1~\mbox{km
  s}$^{-1}$ and $V\sin i$=2~\mbox{km s}$^{-1}$ .  We then calibrated
manually more than 250 lines to match the observed solar
spectrum, checking that the overall $\chi^2$ between the synthetic
template and the observed spectrum was decreasing with each
adjustment.

In addition, roughly 50 lines were calibrated on Arcturus under the
condition that this calibration did not increase the overall $\chi^2$
for the Sun. We also note that the calibration process did not
consider a correction of the molecular bands, even though
some incorrect minor features were noticed.

For Arcturus  we used the parameters and the
abundances of \citet{Smith_Arcturus}: $T_\mathrm{eff}$=4300~K,
$\log~g$=1.7, [M/H]=$-$0.6~dex, [$\alpha$/Fe]=+0.3~dex, $V\sin
i$=1.5~\mbox{km s}$^{-1}$, macroturbulence parameter
$\eta$=5.2~\mbox{km s}$^{-1}$ with a radial-tangential profile, and
$\xi$=1.7 ~\mbox{km s}$^{-1}$. We note though that the observed
spectrum of \citet{Sun_Hinkle} was already normalised, but had an
imperfect normalisation around the second line of the \ion{Ca}{ii}
triplet, noticeable as clearly asymmetric wing strengths.

Finally, because of the fairly poor constraints on its atmospheric
parameters, no line-list calibrations additional to those made
from the Sun and Arcturus were made on Procyon~A (spectral type
F5-IV). Although we checked if the previously made calibrations were
reproducing the Procyon observed spectrum correctly, which is available from
the $ S^4N$ library \citep{Prieto_S4N}. As expected, we found that the
overall fitting quality was improved compared to the computed
spectrum using the non-calibrated line-list of VALD.

\subsection{Testing set}
\label{sec:random_grid}
 
\begin{figure}
  \begin{center}
      \includegraphics[width=7cm,height=7cm]{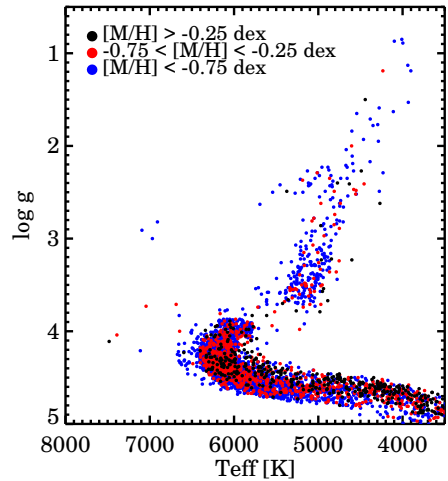} 
      \caption{Input parameters for the testing set. The spectra were computed based on a catalogue of pseudo-stars requested
        from the website of the Besan\c{c}on Galaxy model.}
\label{fig:random_grid}
  \end{center}
\end{figure}

In order to test the MATISSE and the DEGAS algorithms, we required a set of
synthetic spectra that did not sit on the learning grid nodes.
We used the specificities of the local $B_\theta(\lambda)$ functions
of MATISSE associated with a synthetic spectrum $S_0(\lambda)$ with 
$\theta_{{_0}k}$ parameters to compute a set of interpolated
spectra. Indeed, the variations from a grid's spectrum $S_0(\lambda)$
with $\theta_{{_0}k}$ parameters can be computed according to
the following expression:

\begin{equation}
  S_0(\lambda) - S(\lambda) =\sum_{k=1,K} (\theta_{{_0}k}-\theta_k) \sum_{k^{'}=1,K} (B^{-1})_{kk^{'}} B_{\theta_{k^{'}}}(\lambda),
\label{eq:spectra_interpolation1} 
\end{equation}
where $B_{kk^{'}}$ is the correlation matrix between the basis vectors, defined as 

\begin{equation}
B_{kk^{'}}= \sum_{\lambda} B_{\theta_{k}}(\lambda) B_{\theta_{k^{'}}}(\lambda).
\label{eq:spectra_interpolation2} 
\end{equation}
We used Eq.~\ref{eq:spectra_interpolation1} to compute a synthetic
spectrum of the Sun, and compared it to the observed one.  The spectra agreed well, validating in this way the
interpolation routine.

To test the methods efficiently, these interpolated spectra have to
represent realistic cases, i.e. they have to follow a plausible H--R diagram.  For that
purpose, we made  use of the Besan\c{c}on model of the
Milky Way \citep{Modele_besancon}, which returns atmospheric
parameters of simulated stars towards a given line-of-sight.  We
queried from the website of the Besan\c{c}on
model\footnote{\url{http://model.obs-besancon.fr/}} simulated mock
catalogues of stars towards the galactic bulge, the north galactic
pole and towards intermediate latitudes ($l=245^\circ, b=45^\circ$)
to model the range of stellar parameters
encountered in different galactic stellar populations.  $10^4$ of
these stars were selected randomly, to create our testing
sample (see Fig.~\ref{fig:random_grid}).  For each of these
pseudo-stars, two different metallicity values were considered 
to obtain a catalogue spanning the whole metallicity range: the
one given by the model of Besan\c{c}on, and another lowered by
--0.75~dex.  The $2~10^4$ spectra of these pseudo-stars, were then
interpolated from the learning grid using
Eq.~\ref{eq:spectra_interpolation1}, at the sampling and resolution of
the FLAMES LR8 spectra.

Finally,  to test the robustness of the methods, four
different values of white Gaussian noise (SNR $\sim$10, 20, 50, 100~pixel$^{-1}$)
were used to degrade these spectra, raising the final number of
testing spectra to $8~10^4$.

\section{Individual performances of the two methods}
\label{sec:performances}
 We considered that no photometric information is available for the stars whose spectra we analysed. We therefore aimed to explore the performances of each method without additional data. We recall that often this is not true. In particular, this will not be the case for the Gaia space mission, because the two spectrophotometers BP/RP together with astrometric information will provide a handful of information, constraining that way the allowed ranges for $T_{\rm eff}$ and  $\log~g$.  

\subsection{Performance of MATISSE}
\label{subsec:performances_MATISSE}

\begin{figure}
  \begin{center}
    \begin{tabular}{ll}
      \includegraphics[width=4.3cm,height=4.3cm]{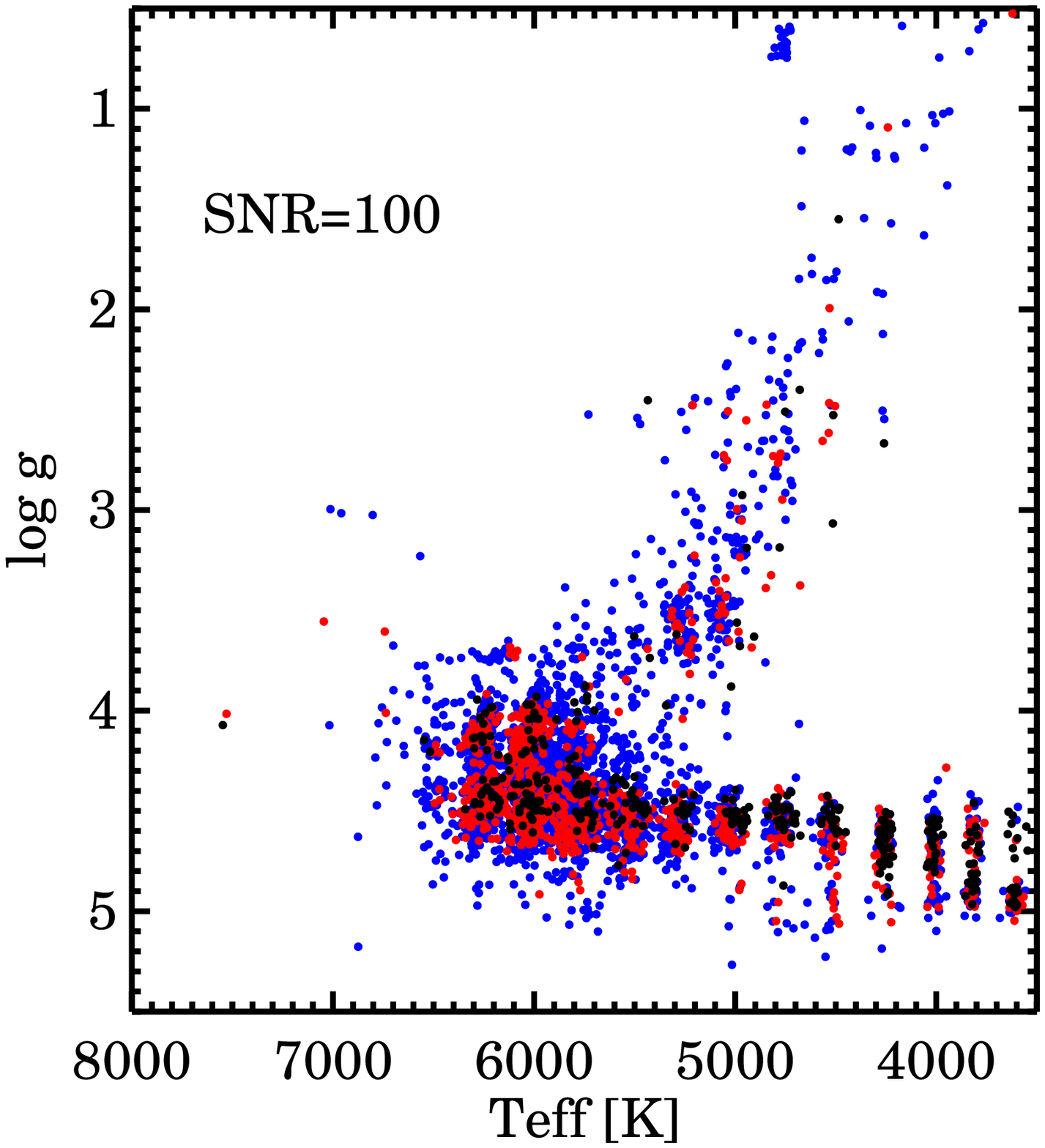} & \includegraphics[width=4.3cm,height=4.3cm]{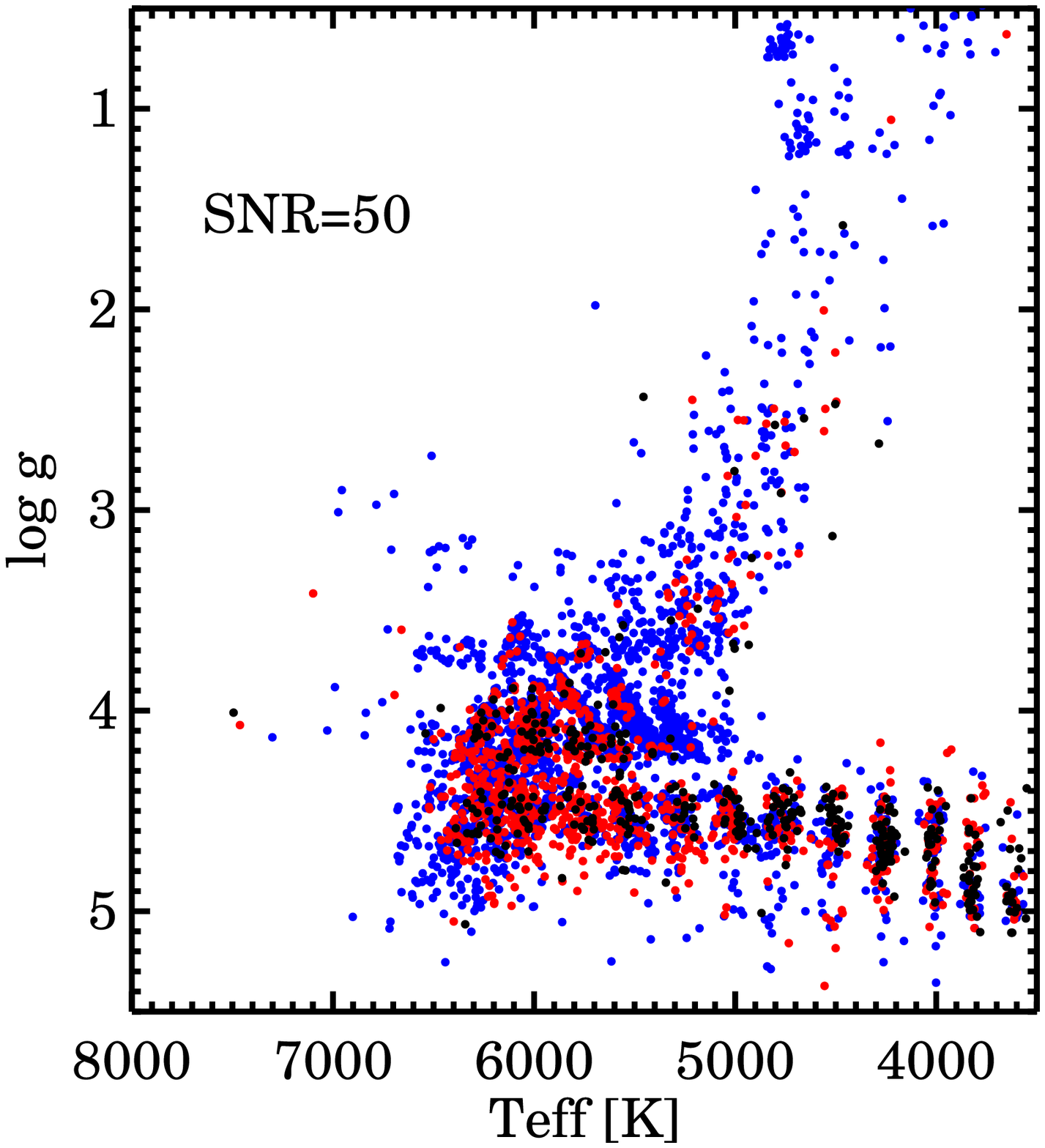}  \\
      \includegraphics[width=4.3cm,height=4.3cm]{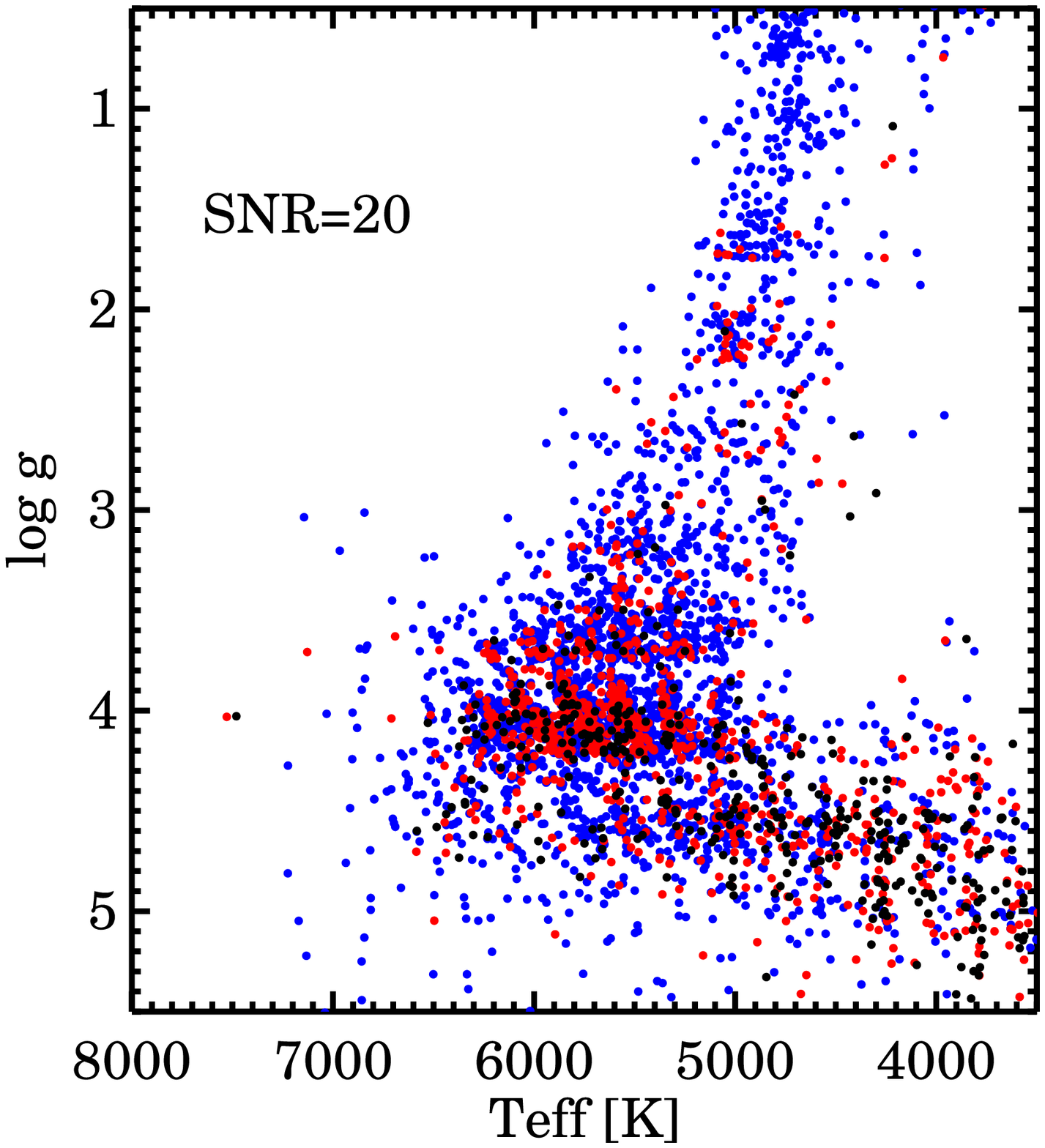} & \includegraphics[width=4.3cm,height=4.3cm]{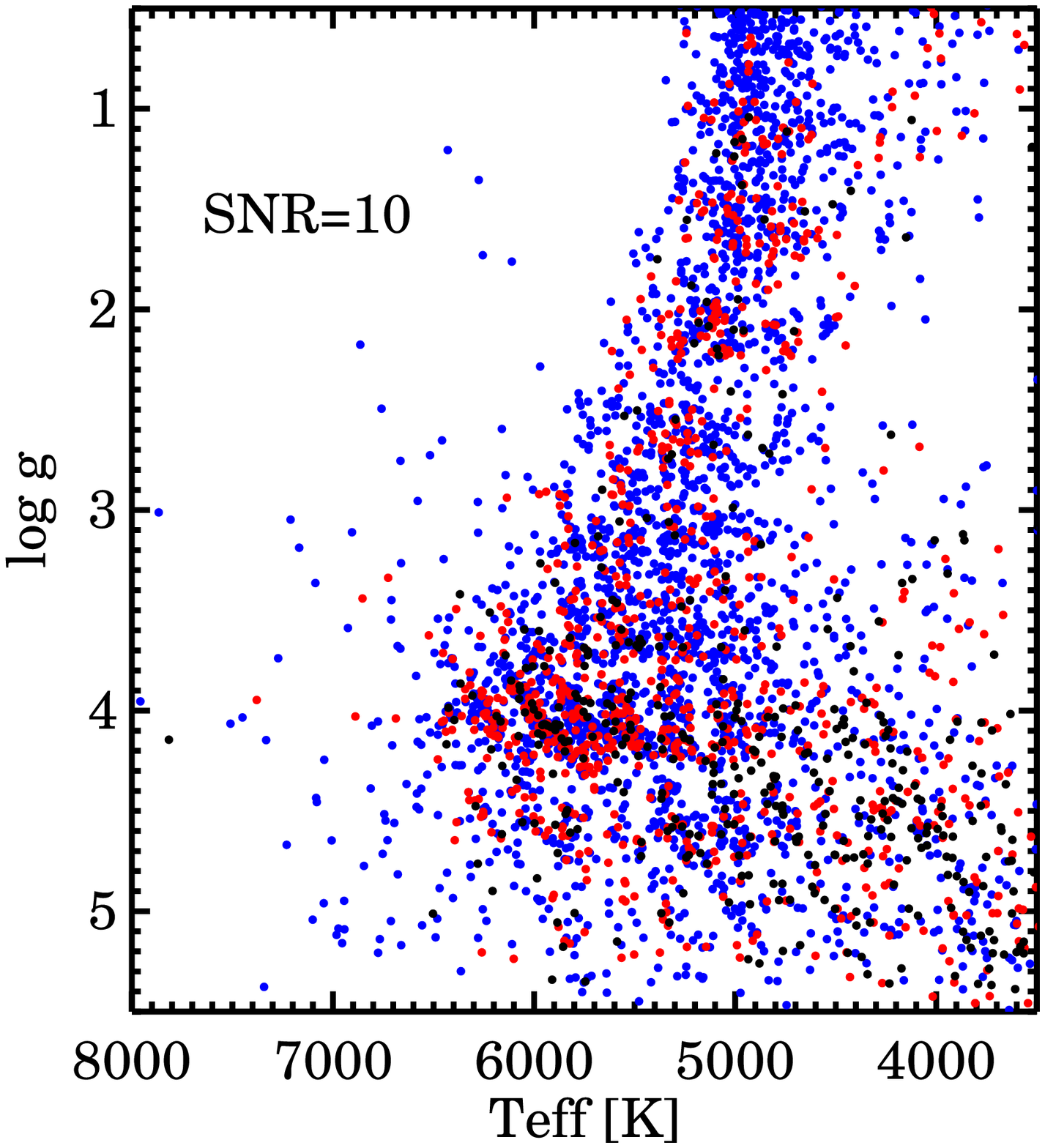}\\
    \end{tabular}
    \caption{Dependence of the H--R diagram for spectra analysed with the MATISSE algorithm
      as a function of signal-to-noise ratio. The colour code, quantifying
      metallicity, is the same as in Fig.~\ref{fig:random_grid}. As
      expected, stars with lower metallicities and hence 
      less spectral information are those with the greatest scatter.
      }
\label{fig:HR_Matisse}
  \end{center}
\end{figure}

\begin{figure*}
  \begin{center}
    \begin{tabular}{ccc}
      \includegraphics[width=4cm,height=3.5cm]{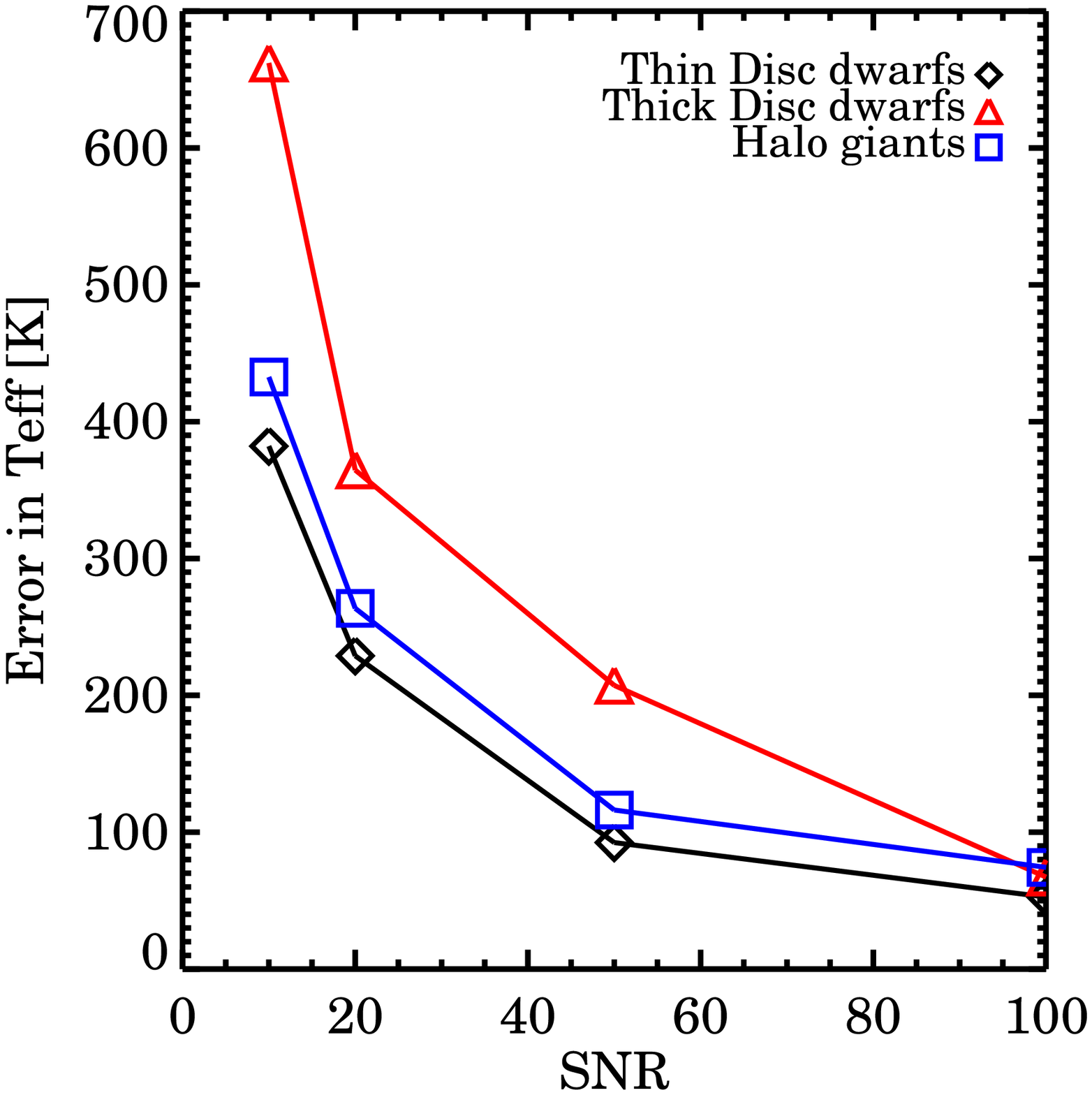} & \includegraphics[width=4cm,height=3.5cm]{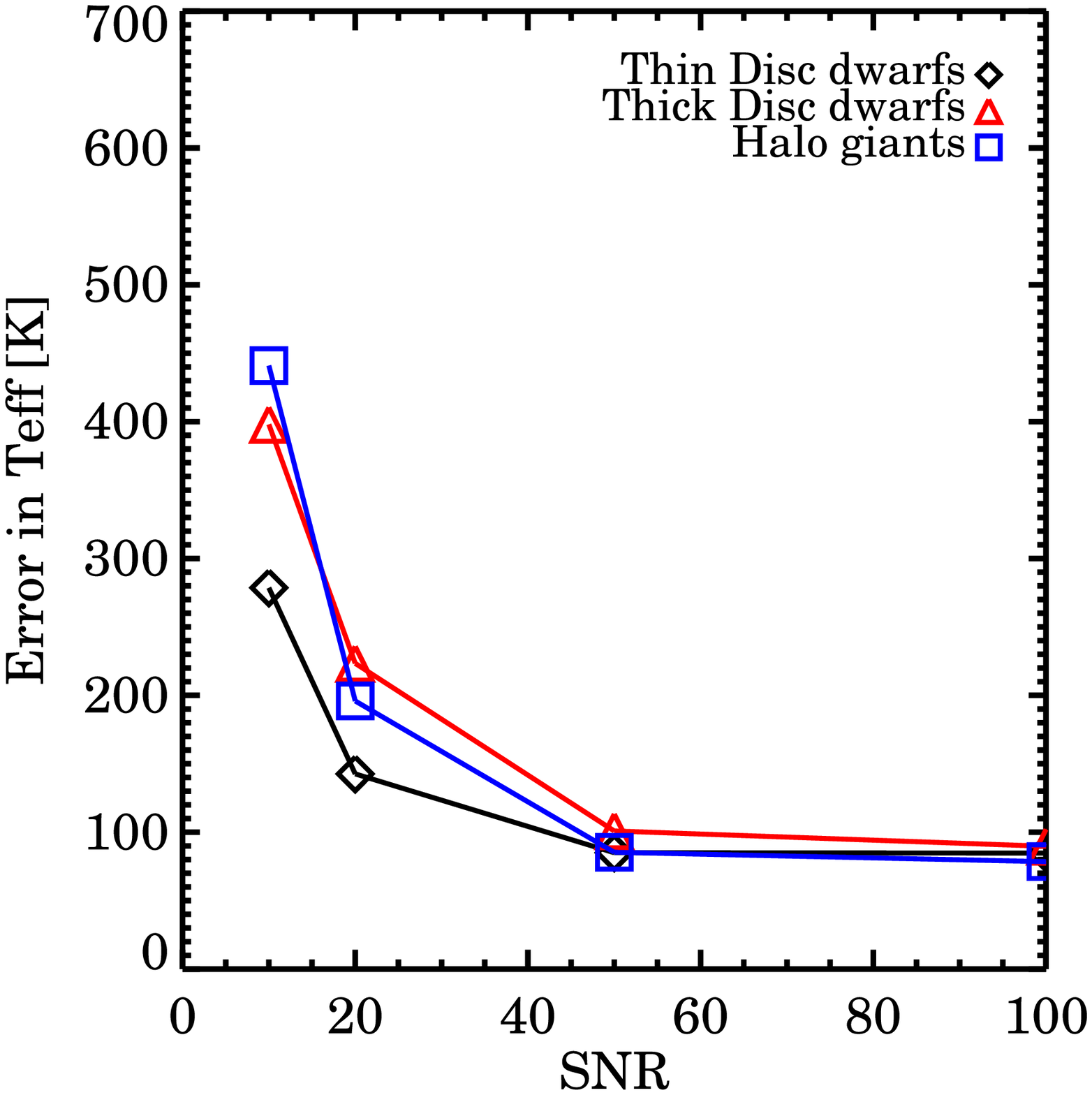} 
      \includegraphics[width=4cm,height=3.5cm]{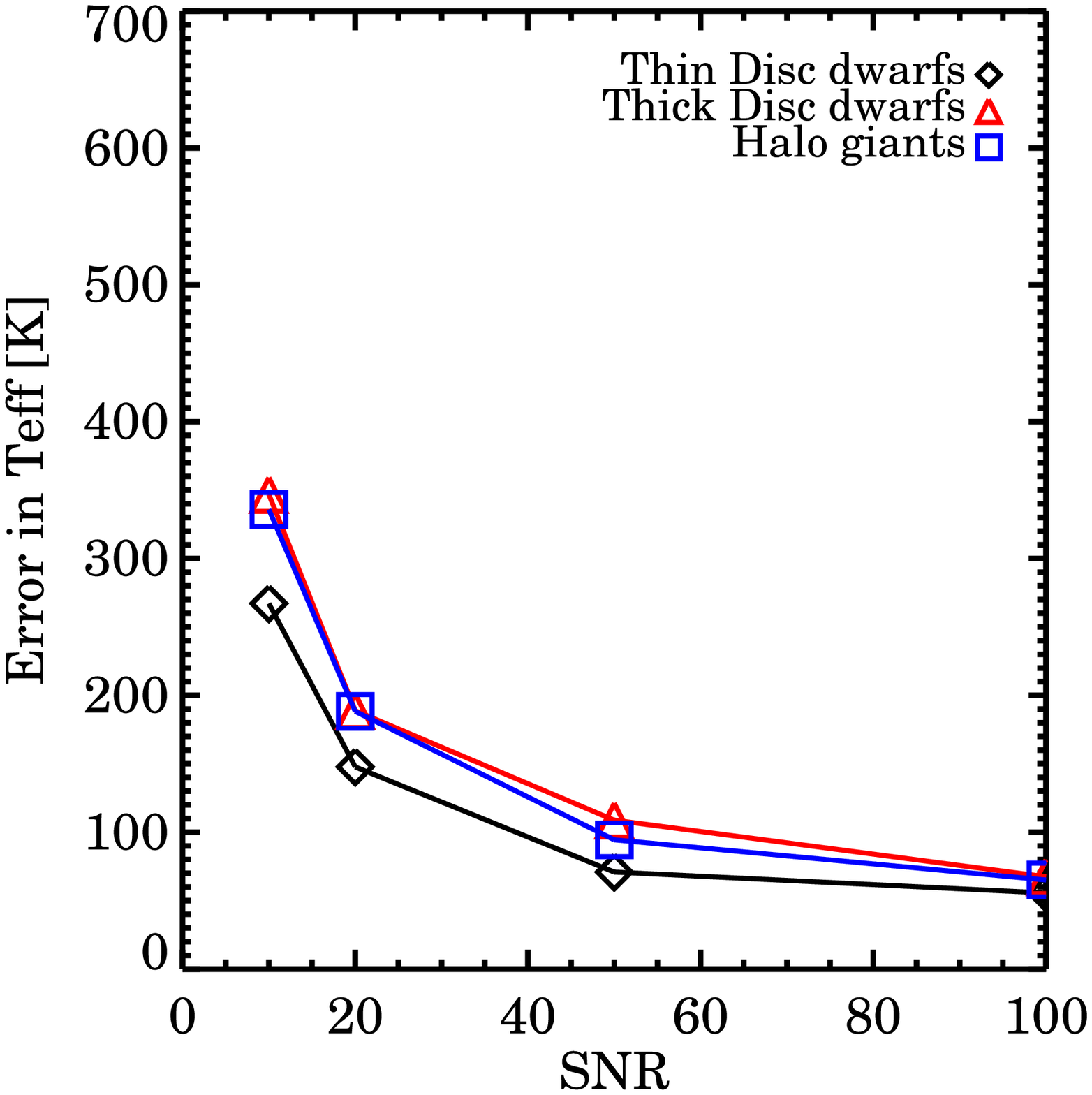} \\
\includegraphics[width=4cm,height=3.5cm]{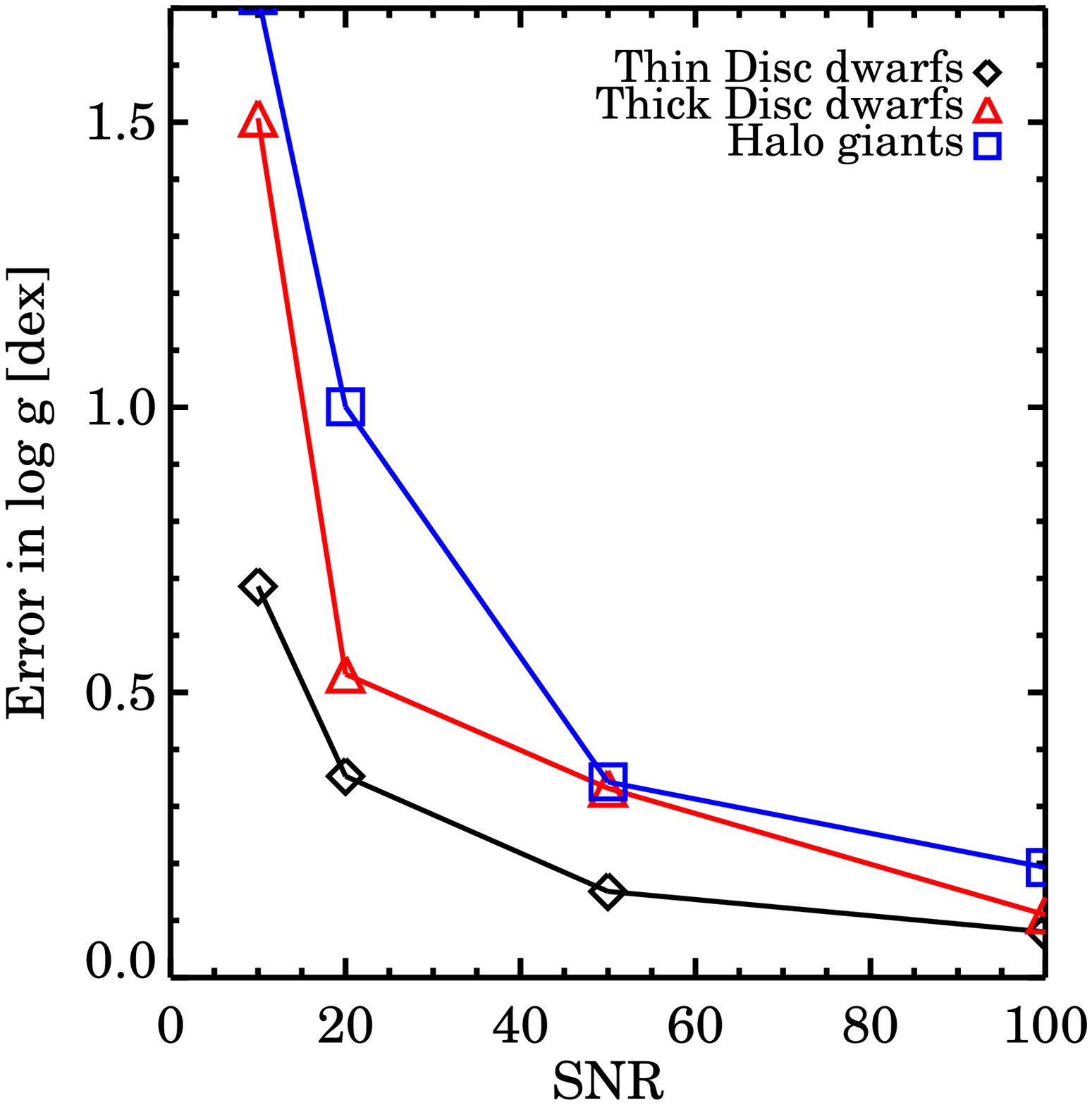} & \includegraphics[width=4cm,height=3.5cm]{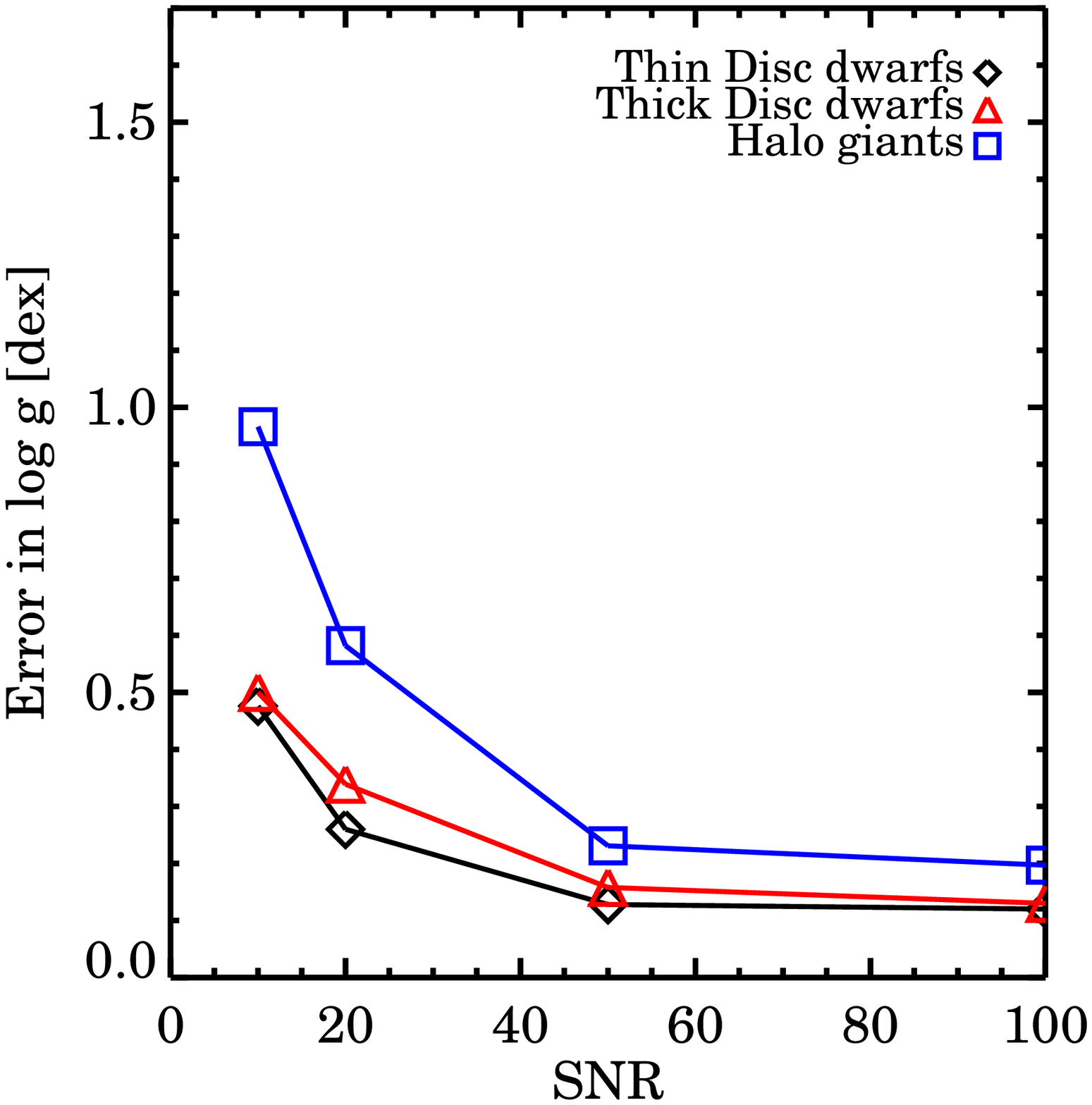} 
      \includegraphics[width=4cm,height=3.5cm]{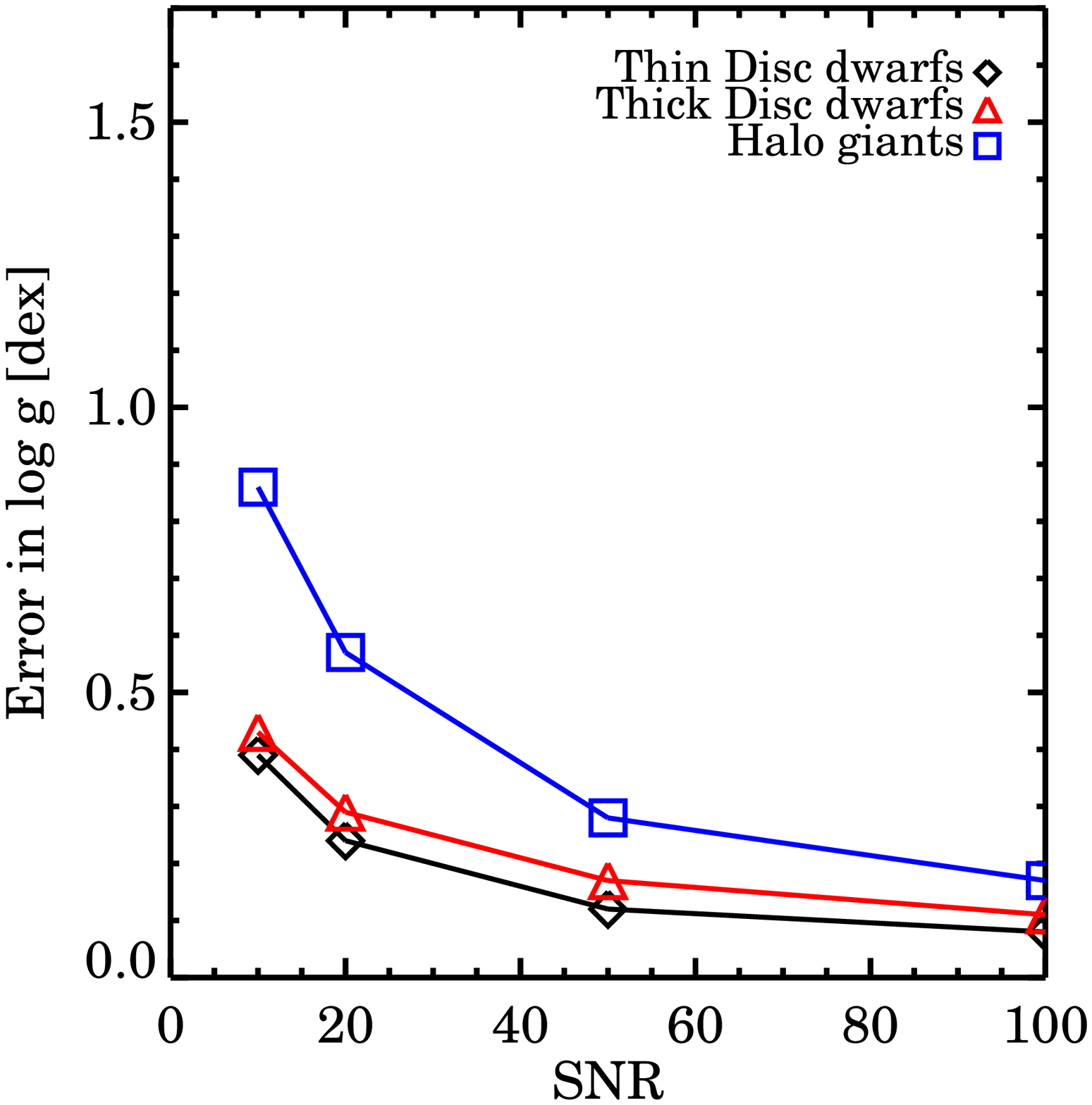} \\
\includegraphics[width=4cm,height=3.5cm]{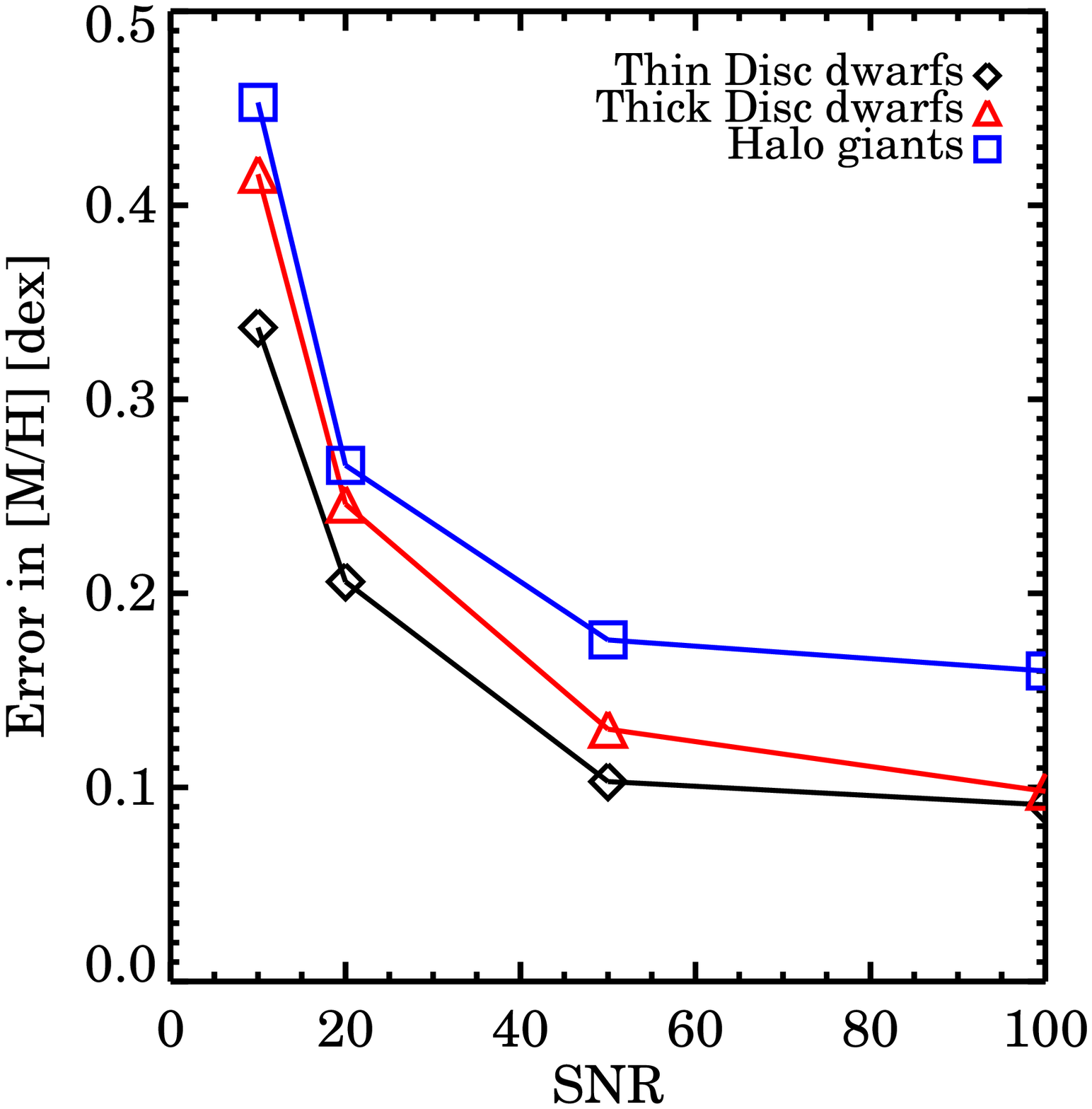} & \includegraphics[width=4cm,height=3.5cm]{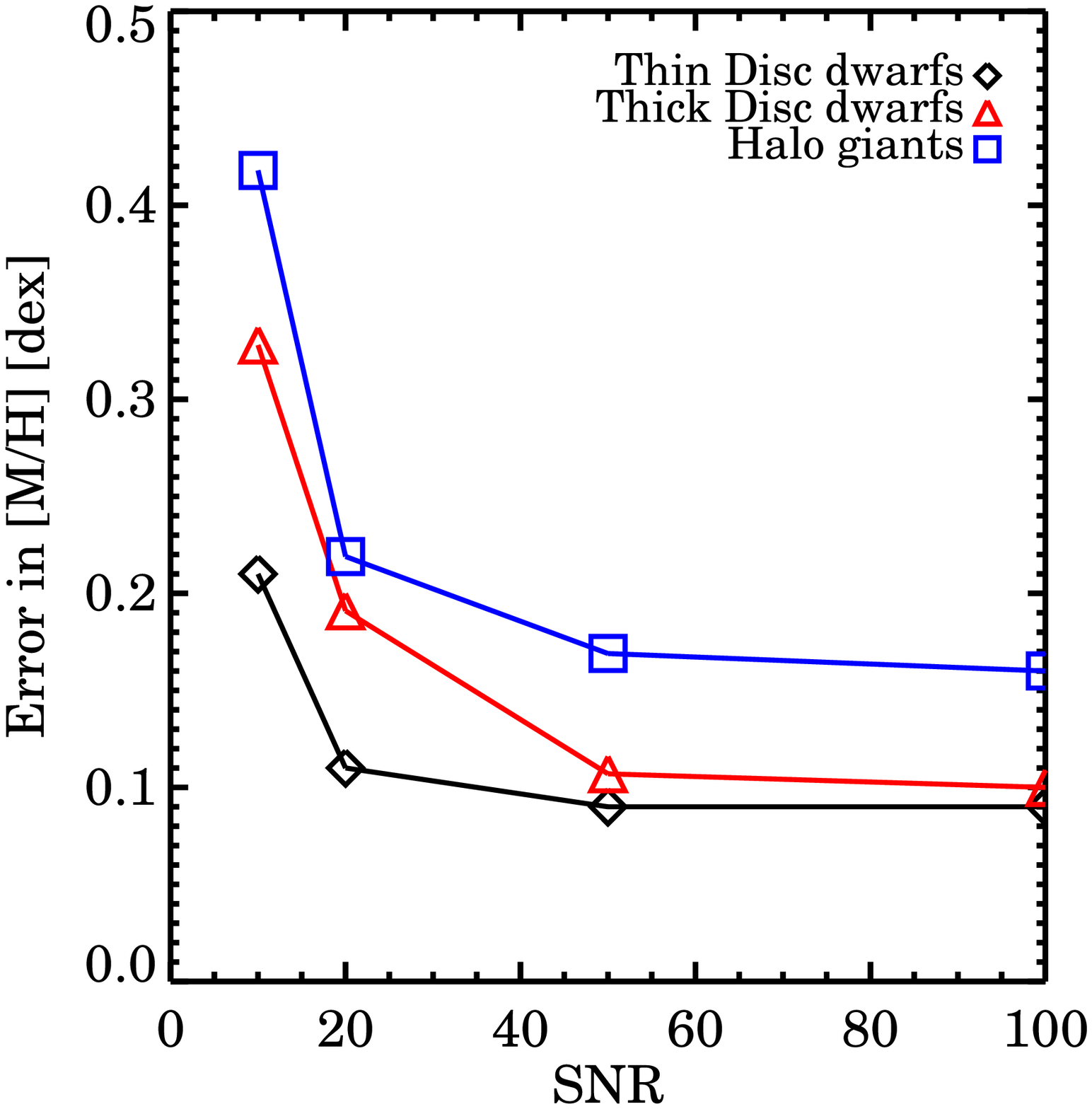} 
      \includegraphics[width=4cm,height=3.5cm]{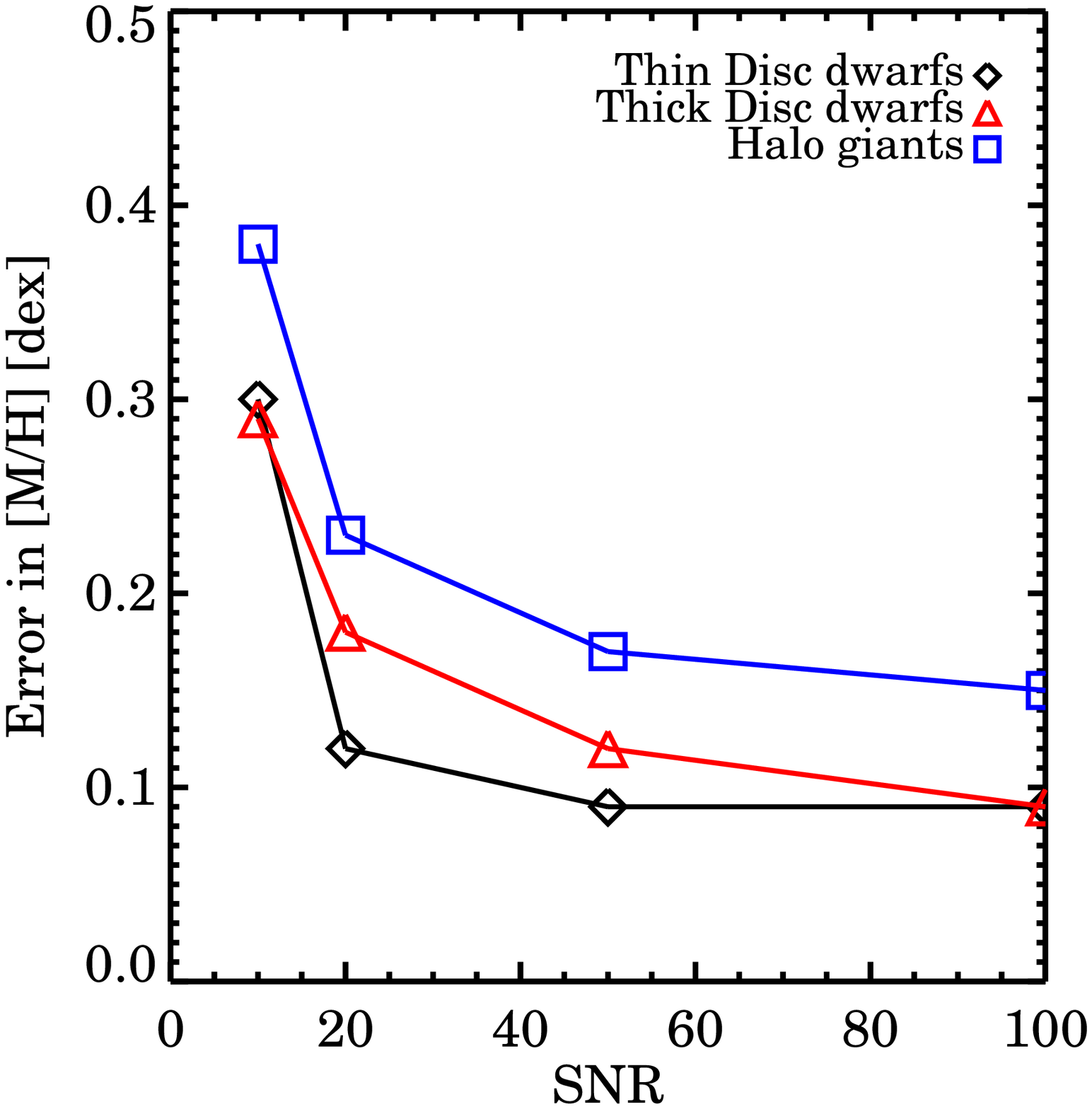} \\
    \end{tabular}

    \caption{Relative errors of the stellar parameters derived by
      MATISSE (left column), DEGAS (middle column), and the final
      adopted pipeline (right column), at 70\% of the error
      distribution as a function of the signal-to-noise ratio. Intermediate temperature
      (5000~$<T_\mathrm{eff}<$6000~K) thin disc
      ($-0.25<$[M/H]$<0.5$~dex) and thick disc
      ($-1.5<$[M/H]$<-0.25$~dex) dwarf stars are represented as black
      diamonds and red triangles symbols, respectively. Halo giant
      stars ($T_\mathrm{eff}<$~6000~K, $\log~g~<3.5$,
      --2.5$<$[M/H]$<$--1.25~dex) are blue squares.}
\label{fig:Q70_Matisse}
  \end{center}
\end{figure*}

We ran MATISSE with  the first solution found with the
$B^0_\theta(\lambda)$ functions as input (i.e. without any \textit{a priori}
information)  on the $8~10^4$ synthetic spectra of the
testing set described in Sect.~\ref{sec:random_grid}.
Figure~\ref{fig:HR_Matisse} and
Table~\ref{tab:internal_errors_matisse} show, respectively, the
evolution of the shape of the H--R diagram and the relative errors at
70~\% of the error distribution, as a function of the SNR for
different types of stars.  In the left column of
Fig.~\ref{fig:Q70_Matisse} we represent the evolution of the
typical errors according to the SNR for thin disc
($-0.25<$[M/H]$<0.5$~dex) and thick disc ($-1.5<$[M/H]$<-0.25$~dex)
dwarf stars and halo giants ($T_\mathrm{eff}<$~6000~K, $\log~g~<$3.5,
--2.5$<$[M/H]$<$--1.25~dex).  These values (also represented in the last
lines of Table~\ref{tab:internal_errors_matisse}) were obtained
for the above mentioned testing set and the best $B_\theta(\lambda)$
functions combination (see Sect.~\ref{sec:matisse}).  Two main
conclusions can be derived:
\begin{itemize}
\item Good results are obtained down to SNR$\sim$20 for spectra where
  a lot of spectral information is available, typically, metal-rich
  dwarfs. The errors increase with decreasing metallicity (see Table
  \ref{tab:internal_errors_matisse}). For GV type stars (dwarfs,
  $T_\mathrm{eff}> 5000-6000$~K), with SNR$\sim$50, the internal
  errors on $T_\mathrm{eff}$ rise from 90~K for metal-rich ([M/H]$>$
  --0.5~dex) to 290~K for metal-poor stars (-1$<$[M/H]$<$-2 dex).
  Errors on $\log~g$  are increased from 0.15~dex to 0.50~dex, and for metallicity
  from 0.10~dex to 0.21~dex.

\item Two degeneracy regimes occur. The first, between the hot
  sub-dwarfs ($T_\mathrm{eff}>$6000 K) and the giant branch, is caused by to
  the common spectral signatures shared by $T_\mathrm{eff}$ and
  $\log~g$.  The second concerns the cool dwarfs and originates in the
  poor constraints on $\log~g$ available with this wavelength range
  and resolution.
\end{itemize}

These two degeneracies are mainly caused by the behaviour of the
\ion{Ca}{ii} lines (the strongest spectral signatures) for different
spectral types.  Indeed, considering that pressure dependence can be
translated into approximate gravity dependences,  the wings of the \ion{Ca}{ii} 
lines grow proportionally to $g^{1/3}$ for cool main
sequence stars  \citep[see][]{Gray_book}, but strongly depend on the
$T_\mathrm{eff}$.  For dwarfs with 3000~$ < T_\mathrm{eff} \lesssim
$~5000~K, where no Paschen lines are developed, the only relevant
spectral signatures concerning $\log~g$  come from the weaker
metallic lines which, depending on the SNR and the global metallicity
of the star, can be lost in the noise.

\begin{figure}
  \begin{center}
      \includegraphics[width=5.5cm,height=5.5cm]{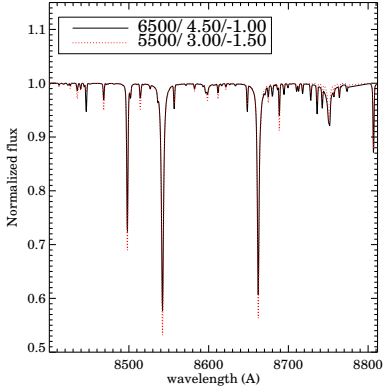} 
      \caption{Noise-free synthetic spectra showing the degeneracy in
        parameter space between a hot dwarf star (black, solid
        line) and a cool giant (red dotted line). The spectra differ
        significantly only at the cores of the strong lines and at
        Pa12~($\sim$~8750~\AA), but their wings differ weakly. }
\label{fig:Degeneracy}
  \end{center}
\end{figure}

On the other hand, parameter degeneracies (and also the secondary
minima) that affect hot dwarfs involve very distant locations in 
parameter space. Indeed, if we compare the spectra of a star with
$T_\mathrm{eff}$=6500~K; $\log~g$=4.5; [M/H]=--1.0~dex and another
that is significantly cooler and more giant, with $T_\mathrm{eff}$=5500~K;
$\log~g$=3.0; [M/H]=--1.5~dex (see Fig.~\ref{fig:Degeneracy}), we can see
that they are almost identical. Only the cores of the strong lines
(calcium, magnesium) are different, and their wings differ
weakly. In addition, the few Paschen lines that both spectra exhibit
are almost identical (only Pa12, around 8750\AA, is significantly
different).  This additionally supports our removal of the cores of the
\ion{Ca}{ii} lines in our automated analysis.  Indeed, our tests have
revealed that avoiding the \ion{Ca}{ii} line cores in the reference
grid helped to decrease the number of secondary minima.  The
disagreement between the synthetic templates and the true spectra, in
addition to the effect of noise on these line-core pixels, can induce
important biases, because spectra with similar cores will have a lower
overall $\chi^2$.  The price to pay to avoid these biases is a bigger
dispersion in the final parameter estimates.  This effect becomes
even more critical when the metallicity decreases, owing to the lack of
spectral information.

The absence of distinct spectral signatures can also be seen from the
angles between the $B_\theta(\lambda)$ functions at different regions
of the H--R diagram. We recall that if the $B_\theta(\lambda)$
functions are perpendicular, the parameters can be perfectly derived,
because they use different spectral signatures.  For example, a star
with $T_\mathrm{eff}$=6500~K; $\log~g$=4.5; [M/H]=--1.0~dex has an angle
between $T_\mathrm{eff}$ and $\log~g$ of $33.4^{\circ}$, $66.5^{\circ}$
between [M/H] and $T_\mathrm{eff}$, and finally $89.2^{\circ}$ between
$\log~g$ and [M/H].  We can therefore expect a degeneracy between the
$T_\mathrm{eff}$ and $\log~g$ determination, and a milder one between
$T_\mathrm{eff}$ and metallicity, as we noticed with our testing set.

On  the other hand, for the cool dwarfs, inaccurate $\log~g$ derivation is
only caused by the lack of spectral sensitivity, and not because 
the spectral signatures are shared between the parameters. Indeed, for
a star with $T_\mathrm{eff}$=4500~K; $\log~g$=4.5; [M/H]=--1.0~dex the
angle between $T_\mathrm{eff}$ and $\log~g$ is $89.7^{\circ}$, between
[M/H] and $T_\mathrm{eff}$ is $82.9^{\circ}$, and finally between
$\log~g$ and [M/H] is $109.9^{\circ}$.

In terms of $\chi^2$ between the MATISSE parameter solution and the
true reference values, the secondary minima for cool
dwarfs lie close to the parameter space formed by
$T_\mathrm{eff}$;~$\log~g$;~[M/H] and consequently MATISSE still derives
accurate parameter estimates, even at low SNR. For hot
dwarfs the low-SNR estimates are farther from the true parameter values,
resulting in higher overall errors.  Adding noise to the signal will
accentuate this behaviour, increase the number of secondary minima,
and hence decrease the accuracy of MATISSE.

In addition, Fig.~\ref{fig:HR_Matisse} shows that some
stars are estimated to have $\log~g >$5. This effect is more important
for low-metallicity stars, but also affects metal-rich stars at low
SNR. This effect, which is typical of the projection algorithms, appears
because MATISSE may perform extrapolations in parameter
space to obtain the atmospheric parameters. Indeed, the
$B_\theta(\lambda)$ functions on which the spectra are projected, 
still attach too much weight to certain spectral signatures
even when they are optimised to a spectral type and an SNR value, which enhances the sensitivity to
noise.  Of course, results lying outside the grid boundaries have to
be considered as unreliable and need to be treated separately (see
Sect.~\ref{subsec:adopted_strategy} for more details on how these
values are finally treated).

\begin{figure}
  \begin{center}
    \begin{tabular}{cc}
      \includegraphics[width=4cm,height=3.9cm]{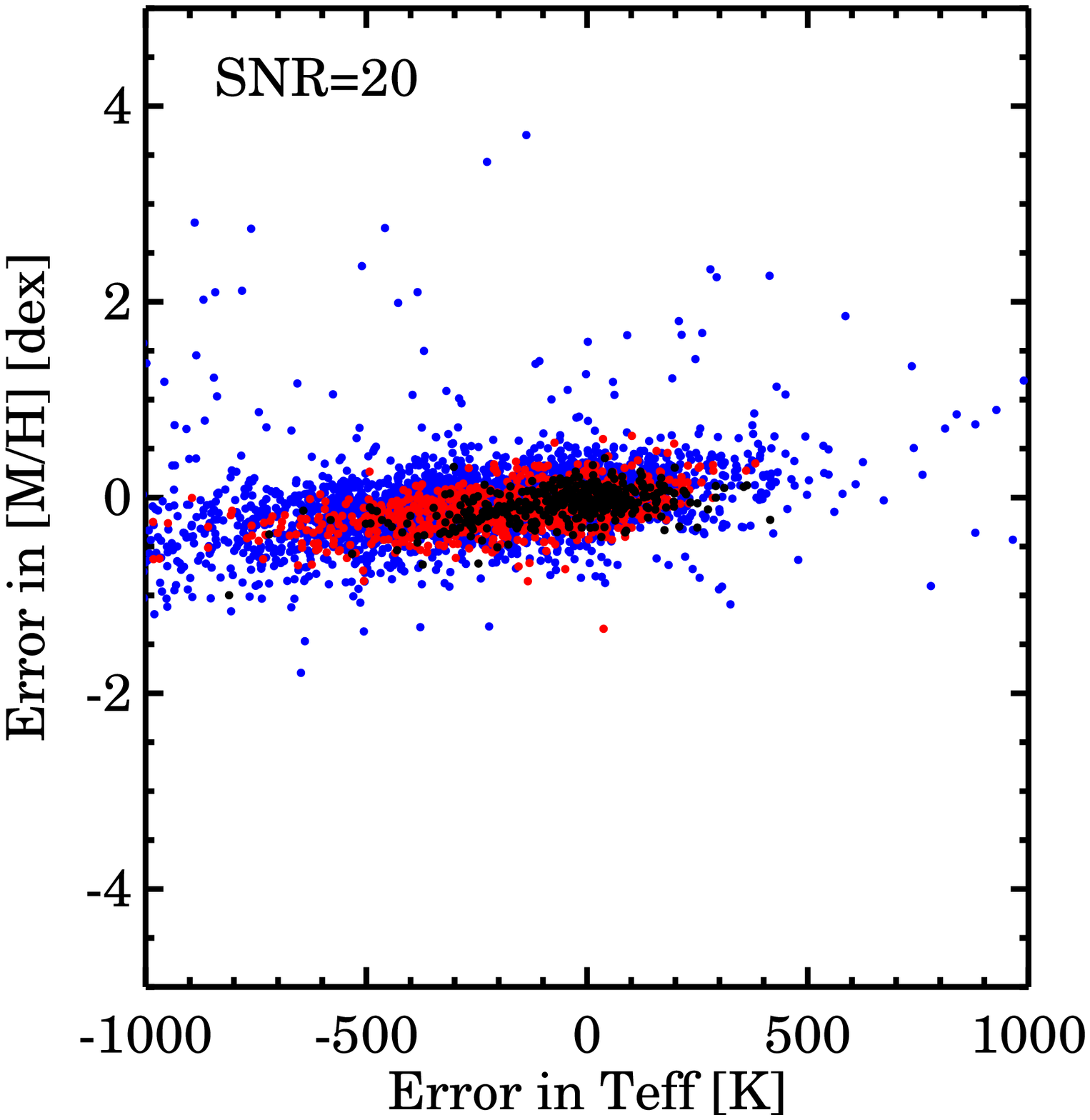} &  \includegraphics[width=4cm,height=3.9cm]{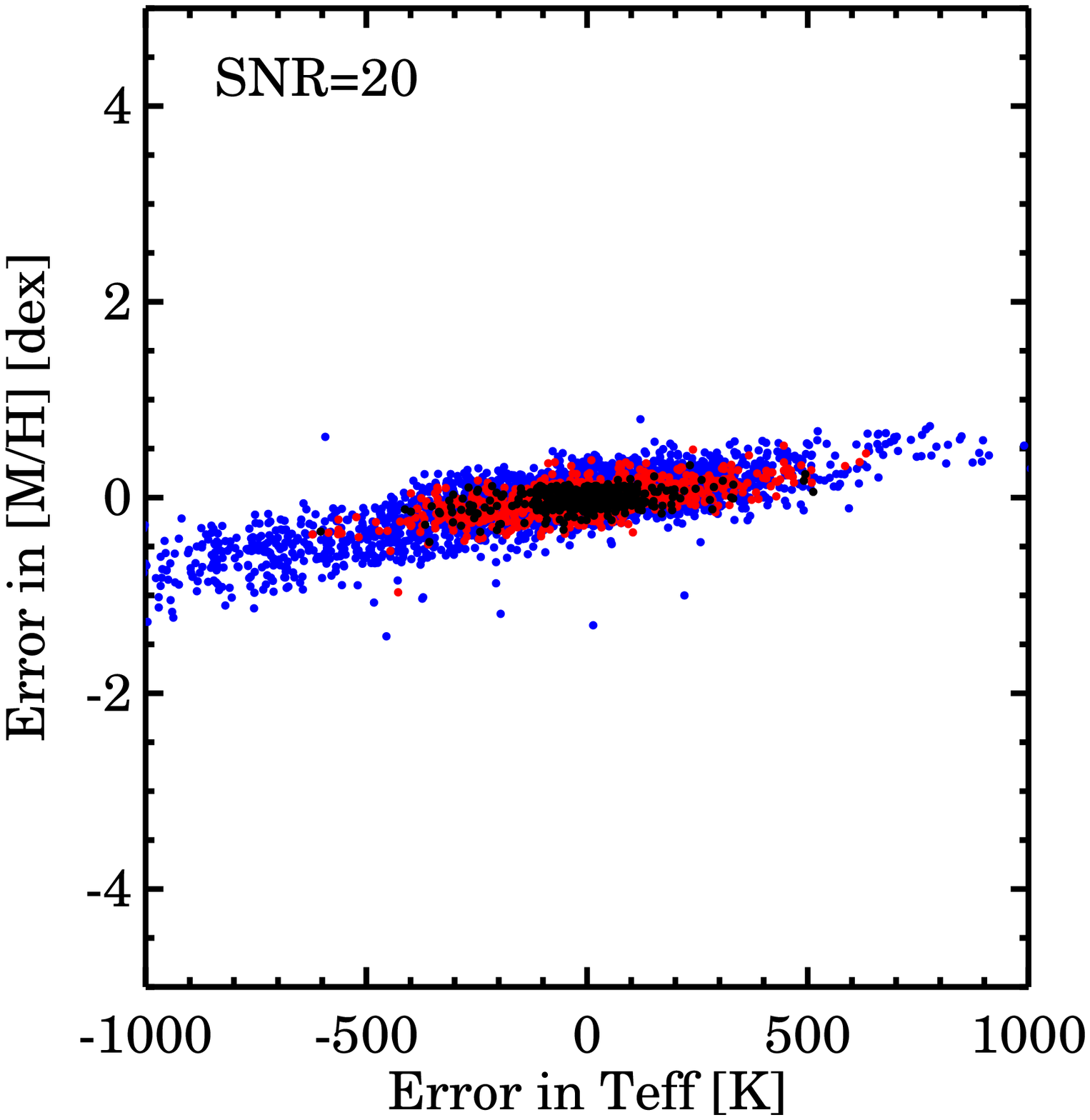}\\ 
      \includegraphics[width=4cm,height=3.9cm]{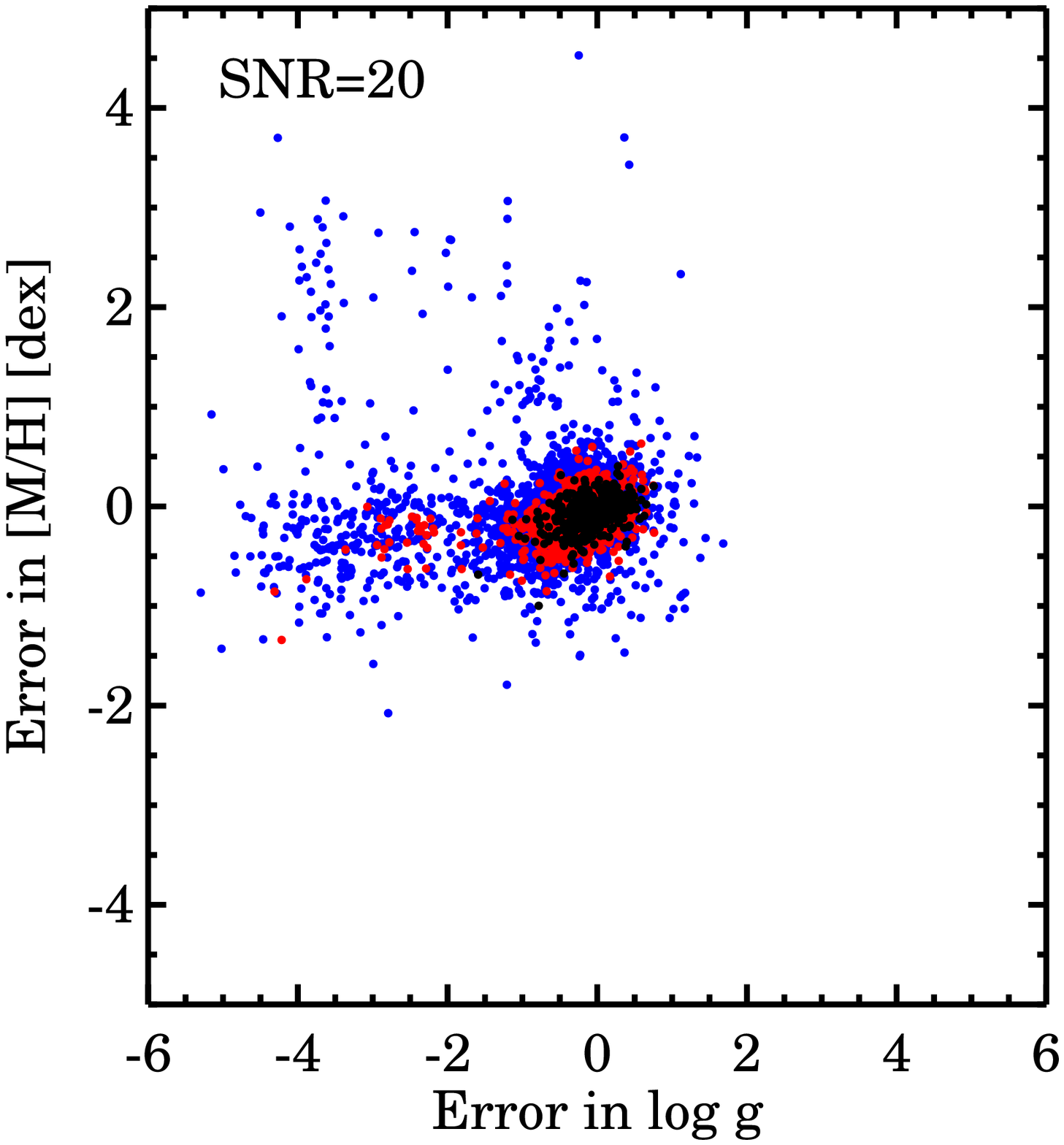} & \includegraphics[width=4cm,height=3.9cm]{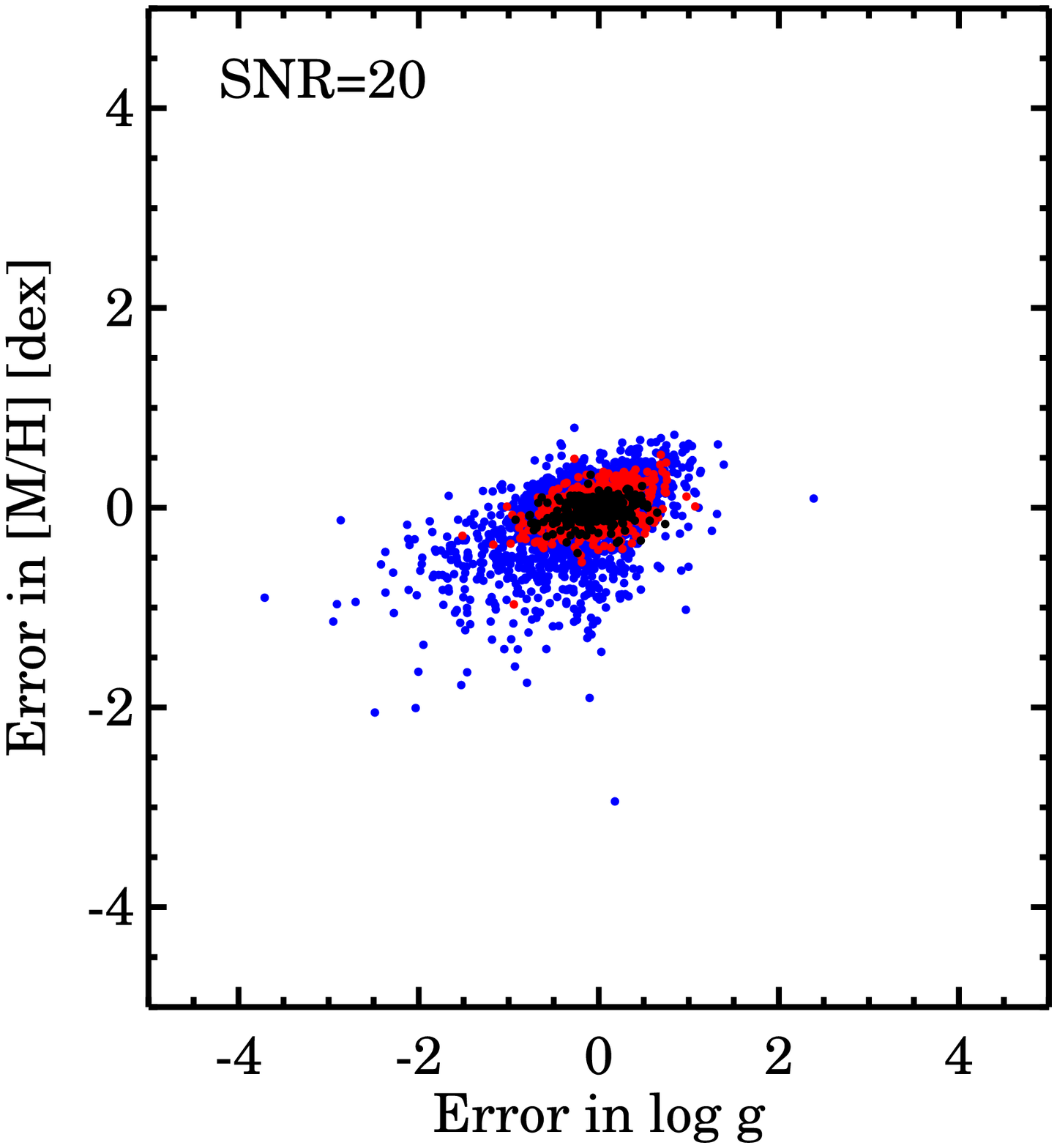} \\
      \includegraphics[width=4cm,height=3.9cm]{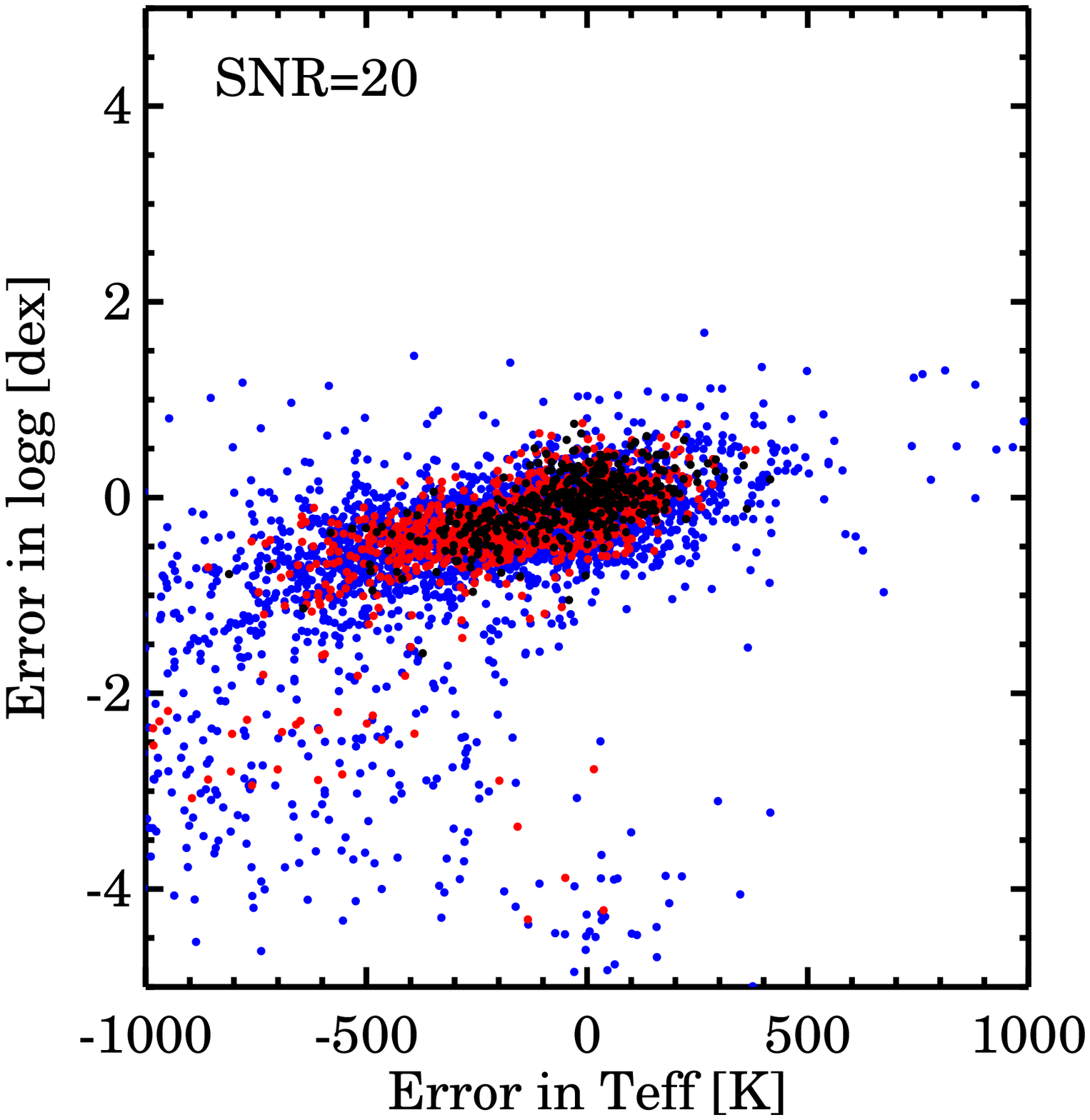} & \includegraphics[width=4cm,height=3.9cm]{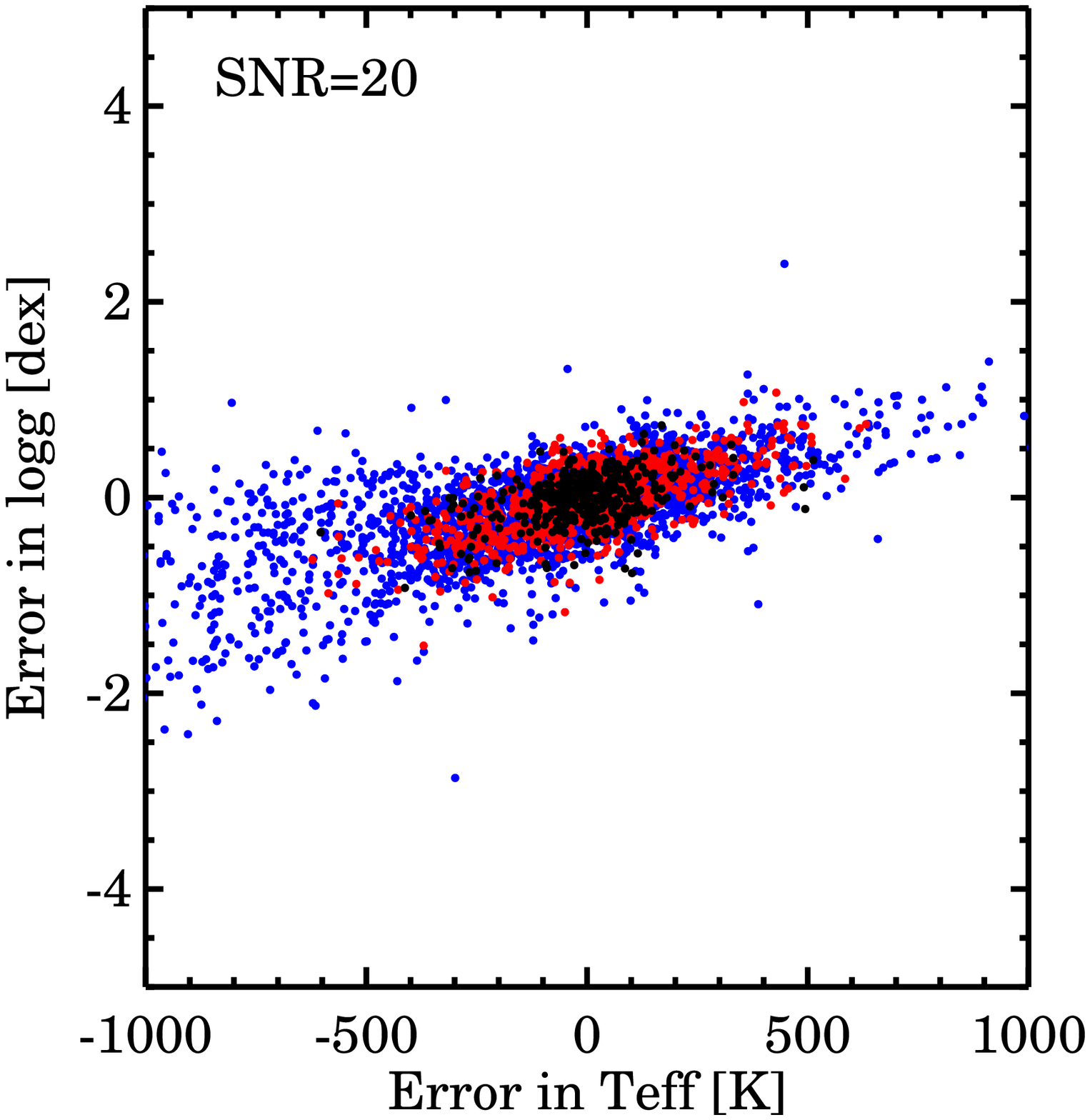} \\ 
    \end{tabular}
    \caption{Error correlations for MATISSE (left column) and DEGAS
      (right column) for spectra with SNR$\sim$20.  The colour code,
      which corresponds to the metallicity, is the same as in
      Fig.~\ref{fig:random_grid}. The general shape of these figures
      shows the degeneracy between the atmospheric parameters.  For
      higher SNRs the correlation slopes are similar, but less
      extended, because the errors are smaller. The specific properties of
      each analysis method lead to the somewhat different extension and dispersion
      of the correlations. }
\label{fig:correl_errors}
  \end{center}
\end{figure}

The correlation of the parameter errors is non-negligible, as seen in
Fig.~\ref{fig:correl_errors}, which illustrates again the effects of
the non convexity and the degeneracies of the distance
function. Indeed, we found the well-known $T_\mathrm{eff}$-[M/H]
correlation as well as the correlation between
$\log~g-T_\mathrm{eff}$. The correlation between $\log~g$ and [M/H], 
however, is almost insignificant for metal-rich and intermediate-metallicity stars, down to SNR$\sim$20.

\begin{table*}[h]
  \begin{center}
\caption{Relative errors of MATISSE at 70\% of the error distribution. }
\begin{tabular}{l||cccc||cccc||cccc}
\hline \hline
 & \multicolumn{4}{c||}{$T_\mathrm{eff}$ (K)} &  \multicolumn{4}{|c||}{$\log~g$ (dex)} &  \multicolumn{4}{|c}{[M/H] (dex)}\\ \hline
SNR   (pixel$^{-1}$)& 100 & 50 & 20 & 10 & 100 & 50 & 20 & 10 & 100 & 50 & 20 & 10 \\ \hline
     KII-IV, [M/H]$>$-0.5~dex &     50 &     48 &    103 &    157 &   0.07 &   0.14 &   0.30 &   0.52 &   0.09 &   0.10 &   0.12 &   0.21 \\
 KII-IV, -1$<$[M/H]$<$-0.5~dex &     56 &     67 &    119 &    234 &   0.13 &   0.18 &   0.42 &   0.79 &   0.08 &   0.09 &   0.18 &   0.30 \\
  KII-IV, -2$<$[M/H]$<$-1~dex &     72 &     85 &    175 &    328 &   0.16 &   0.26 &   0.78 &   1.32 &   0.15 &   0.17 &   0.26 &   0.43 \\
       KII-IV, [M/H]$<$-2~dex &     56 &     70 &    200 &    386 &   0.24 &   0.56 &   1.17 &   1.72 &   0.12 &   0.13 &   0.21 &   0.60 \\
     GII-IV, [M/H]$>$-0.5~dex &     63 &     73 &    133 &    272 &   0.10 &   0.12 &   0.29 &   0.81 &   0.08 &   0.08 &   0.18 &   0.29 \\  
GII-IV,  -1$<$[M/H]$<$-0.5~dex &     67 &     84 &    163 &    307 &   0.13 &   0.19 &   0.54 &   1.12 &   0.07 &   0.11 &   0.18 &   0.39 \\
  GII-IV, -2$<$[M/H]$<$-1~dex &     66 &    120 &    281 &    407 &   0.18 &   0.30 &   0.92 &   1.52 &   0.17 &   0.18 &   0.25 &   0.43 \\
      GII-IV, [M/H]$<$-2~dex  &    131 &    297 &    528 &    729 &   0.33 &   0.63 &   1.85 &   2.51 &   0.18 &   0.34 &   0.56 &   1.33 \\
            FII-IV, all [M/H] &     80 &    110 &    127 &    150 &   0.11 &   0.15 &   0.17 &   0.25 &   0.07 &   0.10 &   0.17 &   0.23 \\
\hline
         KV, [M/H]$>$-0.5~dex &     56 &     61 &     90 &    140 &   0.09 &   0.13 &   0.28 &   0.54 &   0.08 &   0.09 &   0.15 &   0.27 \\
 KV, -1$<$[M/H]$<$-0.5~dex    &     71 &     76 &    102 &    181 &   0.12 &   0.17 &   0.32 &   0.82 &   0.09 &   0.09 &   0.17 &   0.32 \\
      KV, -2$<$[M/H]$<$-1~dex &     88 &     99 &    138 &    260 &   0.17 &   0.25 &   0.52 &   1.78 &   0.15 &   0.18 &   0.24 &   0.40 \\
          KV, [M/H]$<$-2~dex  &     93 &    114 &    210 &    554 &   0.32 &   0.36 &   0.47 &   1.50 &   0.16 &   0.16 &   0.27 &   0.58 \\
               GV, $>$-0.5~dex&     55 &     90 &    254 &    439 &   0.08 &   0.15 &   0.42 &   0.86 &   0.09 &   0.10 &   0.23 &   0.38 \\
    GV, -1$<$[M/H]$<$-0.5~dex &     62 &    195 &    329 &    587 &   0.11 &   0.33 &   0.54 &   1.62 &   0.08 &   0.13 &   0.23 &   0.39 \\
       GV, -2$<$[M/H]$<$-1~dex&    110 &    296 &    502 &    798 &   0.17 &   0.50 &   0.80 &   2.52 &   0.17 &   0.21 &   0.33 &   0.51 \\
          GV, [M/H]$<$-2~dex  &    472 &    561 &    861 &   1136 &   0.71 &   0.95 &   2.64 &   3.56 &   0.26 &   0.34 &   0.69 &   1.28 \\
               FV, $>$-0.5~dex&     57 &    106 &    330 &    552 &   0.10 &   0.18 &   0.41 &   0.84 &   0.10 &   0.12 &   0.26 &   0.41 \\
    FV, -1$<$[M/H]$<$-0.5~dex &     71 &    165 &    405 &    730 &   0.12 &   0.27 &   0.50 &   1.14 &   0.10 &   0.15 &   0.31 &   0.46 \\
       FV, -2$<$[M/H]$<$-1~dex&    137 &    340 &    575 &   1075 &   0.18 &   0.43 &   0.75 &   1.88 &   0.17 &   0.23 &   0.42 &   0.68 \\
          FV, [M/H]$<$-2~dex  &   1249 &   1314 &   1389 &   1529 &   1.28 &   1.63 &   2.86 &   3.43 &   1.92 &   2.13 &   2.23 &   2.04 \\
\hline
             Thin disc dwarfs &     52 &     92 &    228 &    382 &   0.08 &   0.15 &   0.35 &   0.69 &   0.09 &   0.10 &   0.21 &   0.34 \\
            Thick disc dwarfs &     67 &    207 &    364 &    662 &   0.11 &   0.33 &   0.53 &   1.51 &   0.10 &   0.13 &   0.25 &   0.42 \\
                  Halo giants &     74 &    116 &    263 &    432 &   0.19 &   0.34 &   1.00 &   1.72 &   0.16 &   0.18 &   0.27 &   0.45 \\
\hline

\end{tabular}
\label{tab:internal_errors_matisse}
\tablefoot{ Because of the poor statistics on FII-IV stars, all the metallicities were considered to infer the published value. The three last lines represent the accuracies of the method for typical old thin disc dwarfs ($\log~g>$3.9, --0.5$<$[M/H]$<-0.25$~dex), thick disc dwarfs (($\log~g>$3.9, --1.5$<$[M/H]$<-0.5$~dex) and halo giants ($T_\mathrm{eff}<$6000~K, $\log~g<$3.5, --2.5$<$[M/H]$<-1.25$~dex).}  
\end{center}
\end{table*}

\subsection{Performances of the DEGAS algorithm}

\begin{figure}
  \begin{center}
    \begin{tabular}{ll}
      \includegraphics[width=4.3cm,height=4.3cm]{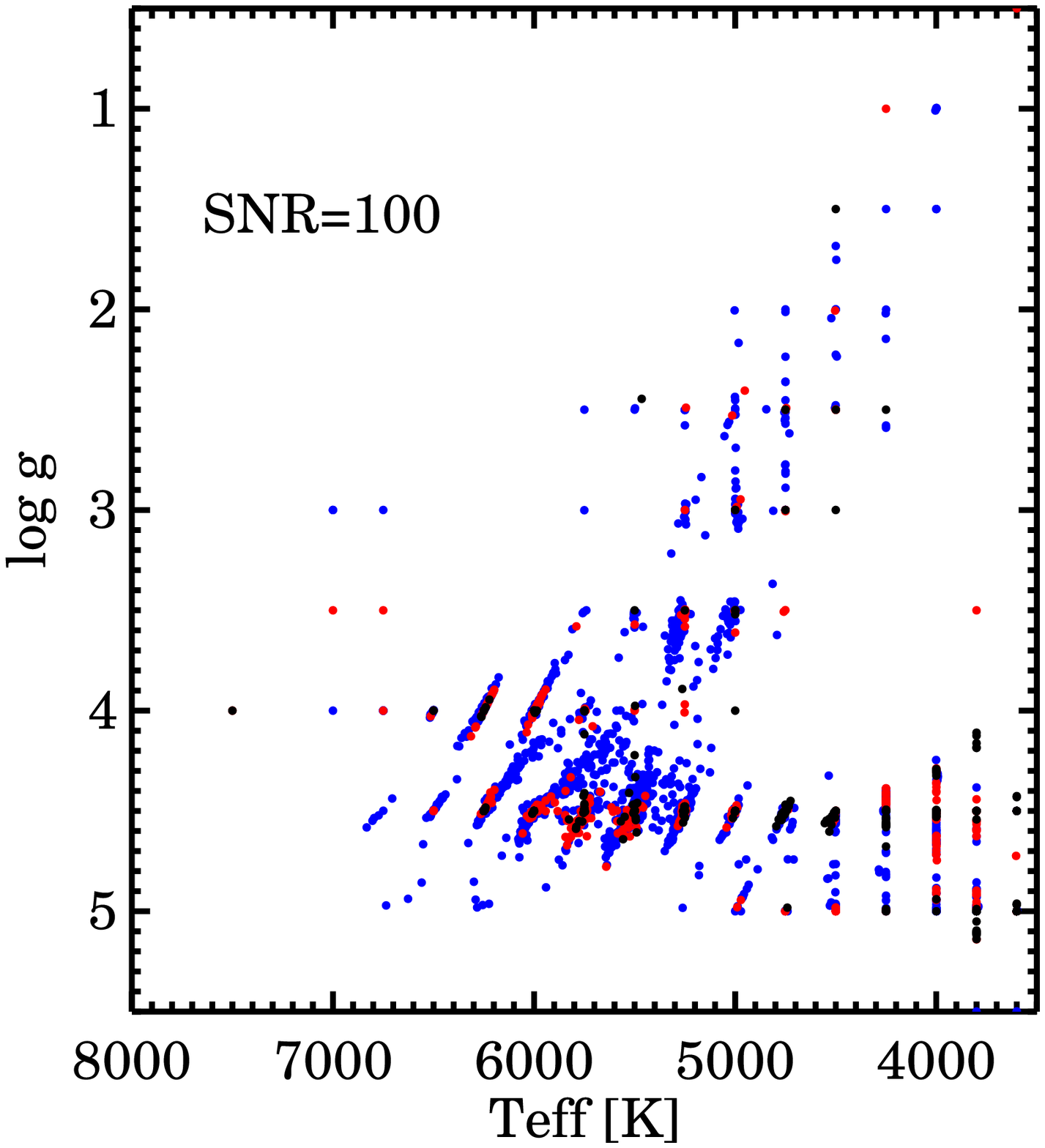} & \includegraphics[width=4.3cm,height=4.3cm]{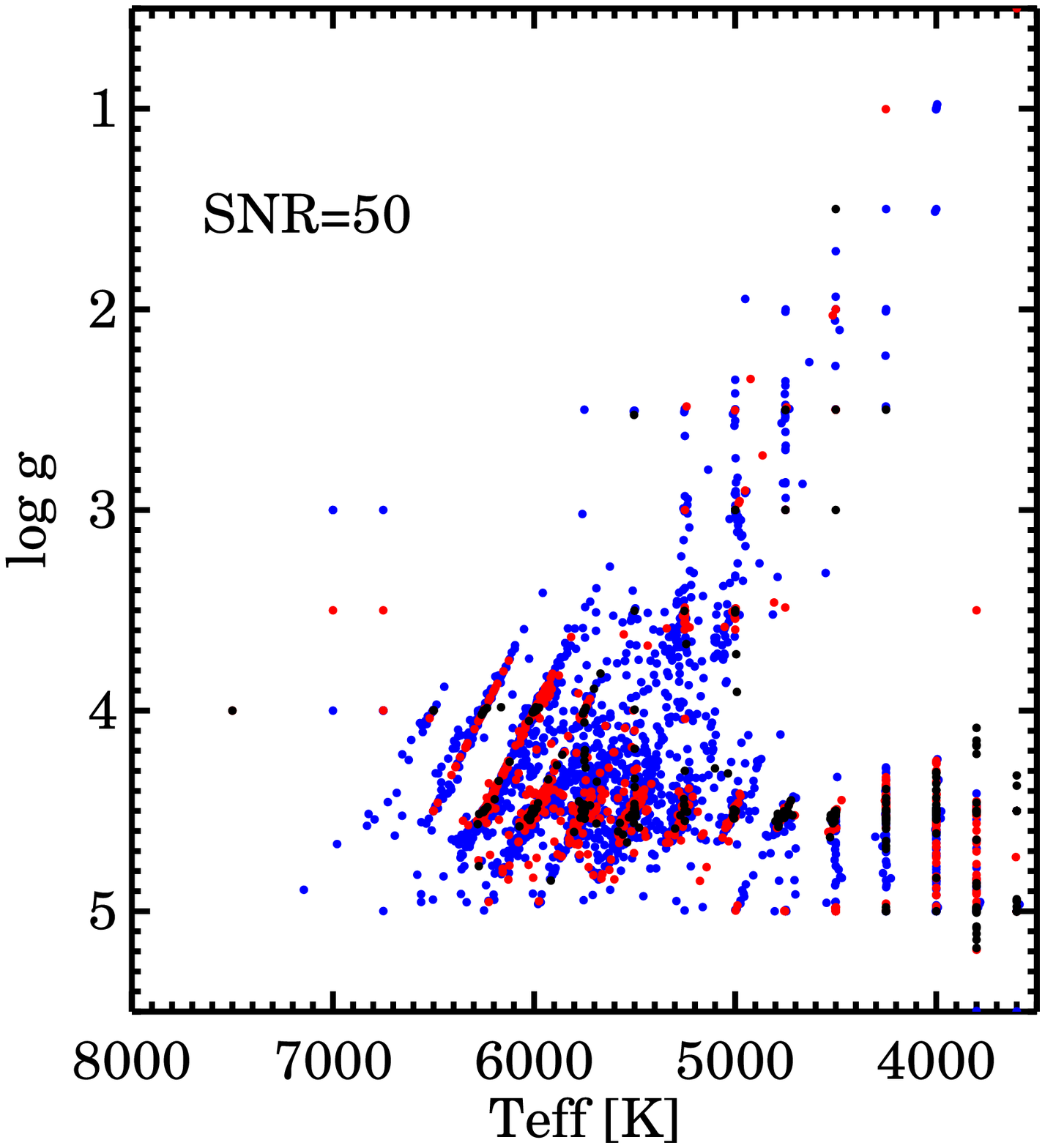}  \\
      \includegraphics[width=4.3cm,height=4.3cm]{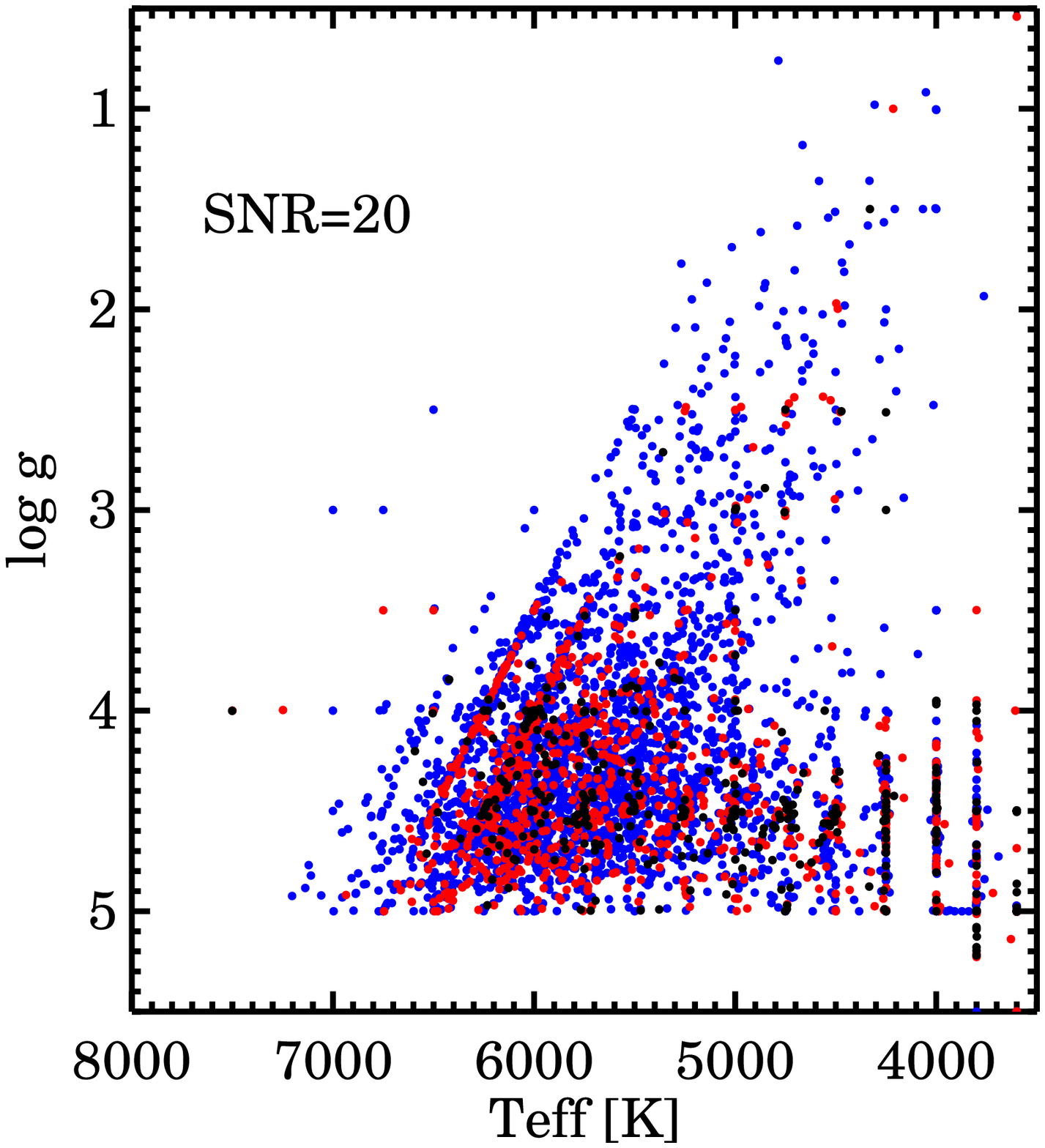} & \includegraphics[width=4.3cm,height=4.3cm]{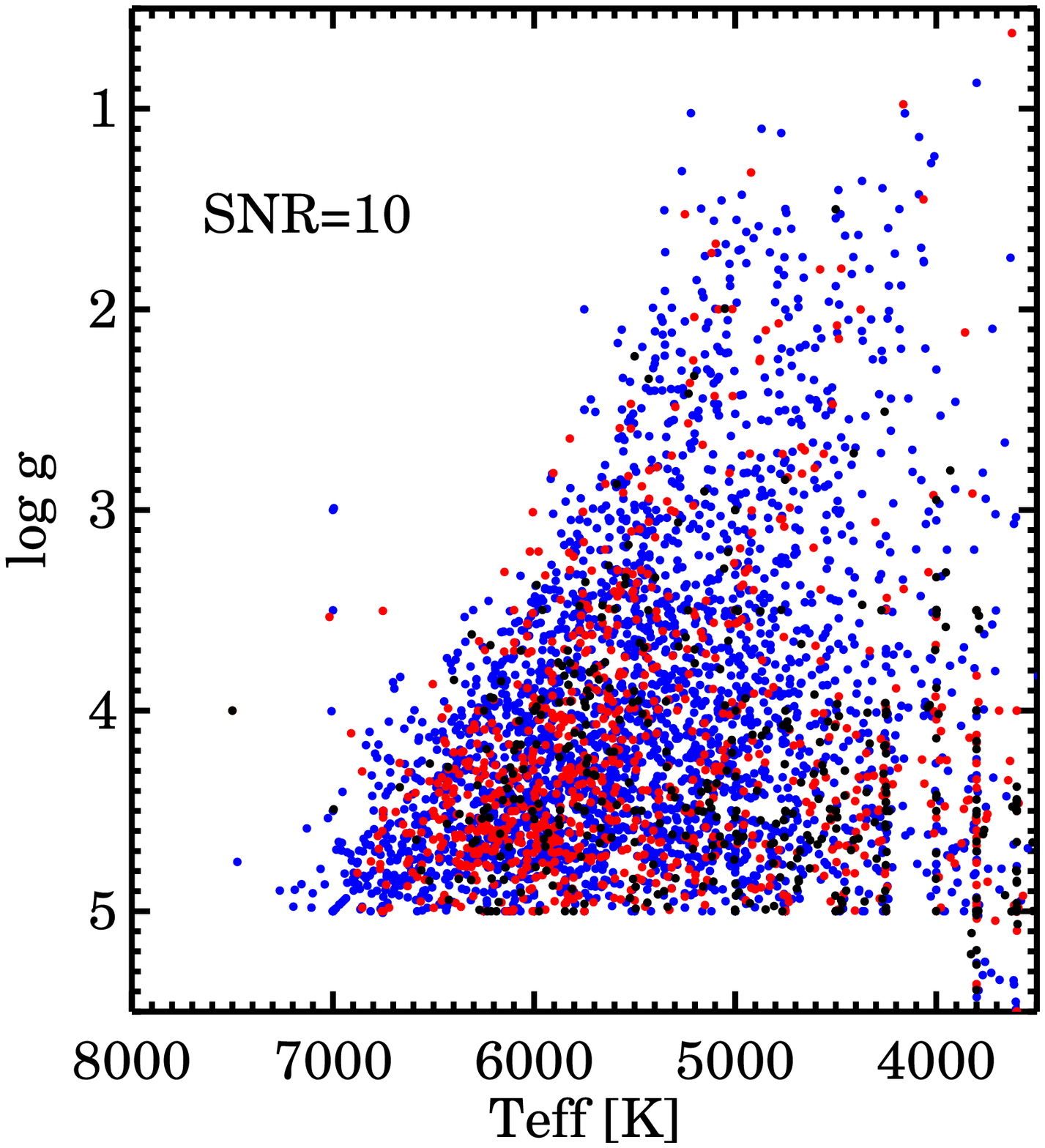}\\
    \end{tabular}
    \caption{Dependance of the H--R diagram structure recovered with the DEGAS
      algorithm as a function of the spectrum's signal-to-noise ratio. The colour code is the same as in
      Fig.~\ref{fig:random_grid}. }
\label{fig:HR_Dicho}
  \end{center}
\end{figure}

Unlike MATISSE and its local $B_\theta(\lambda)$ functions, the DEGAS
method allows a more global view of the parameter space, and
therefore, it is less affected by local minima traps. Accordingly
degradation of the parameter accuracies is slow with decreasing SNR
(see Table~\ref{tab:q70_dicho} and the middle column of
Fig.~\ref{fig:Q70_Matisse}). The critical SNR value for which DEGAS
gives better results than MATISSE is around SNR$\sim$35.  Above that
value, DEGAS is incapable of inferring an appropriate interpolated
parameter combination between the grid points
(Fig.~\ref{fig:HR_Dicho}) because Eq.~\ref{eq:dichodif_weight_distances} 
assigns too much weight to the
resulting leaves of the decision tree.  However, the closest grid
point (absolute minimum) is usually found.  Of course this effect
results in relatively higher errors compared to those of MATISSE.
This feature cannot be improved without affecting the results at lower
SNR.

Nevertheless, even at higher SNR, the pattern recognition approach
deals better with spectra having low metallities
($\mathrm{[M/H]}<-1$~dex).  For instance, at SNR$\sim$50, a G-type
giant with $\mathrm{[M/H]}<-2$~dex is expected to have errors of
182~K, 0.43~dex, 0.22~dex for $T_\mathrm{eff}$, $\log~g$ and [M/H],
respectively. These accuracies are 35\% better than those obtained
with MATISSE.  Furthermore, the same effect can be observed for 
KV  metal-poor type stars, resulting in a (partial) resolution of the
thickening of the cool part of the main sequence seen with MATISSE
(Fig.~\ref{fig:HR_Matisse}).

DEGAS gives quite good results down to SNR$\sim$20 for stars with
[M/H]$>-1$~dex. Typical errors for giants are 121~K, 0.29~dex,
0.13~dex for $T_\mathrm{eff}$, $\log~g$ and [M/H], respectively.  
These values are $\sim$168~K, 0.30~dex, 0.14~dex for dwarfs, 
and hotter dwarfs have lower accuracies compared to cooler ones.

Let us stress though that an additional reason why this method gives
smaller errors compared to MATISSE is that the derived parameter
values cannot lie outside the grid boundaries. This  is also
illustrated when comparing the dispersion of the error correlation for
the two algorithms in Fig.~\ref{fig:correl_errors}.  However, the
general shape of the correlations has not changed, as expected, because
the physical degeneracies are impossible to separate for any of the
methods.

\begin{table*}[h]
  \begin{center}
    \caption{Relative errors of DEGAS at 70\% of the error
      distribution.}
\begin{tabular}{l||cccc||cccc||cccc}
\hline \hline
 & \multicolumn{4}{c||}{$T_\mathrm{eff}$ (K)} &  \multicolumn{4}{|c||}{$\log~g$ (dex)} &  \multicolumn{4}{|c}{[M/H] (dex)}\\ \hline
SNR   (pixel$^{-1}$)& 100 & 50 & 20 & 10 & 100 & 50 & 20 & 10 & 100 & 50 & 20 & 10 \\ \hline
    KII-IV, [M/H]$>$-0.5 dex  &     91 &     87 &    107 &    213 &   0.16 &   0.17 &   0.22 &   0.53 &   0.08 &   0.08 &   0.09 &   0.22 \\
KII-IV, -1$<$[M/H]$<$-0.5 dex &     76 &     75 &    106 &    233 &   0.19 &   0.20 &   0.30 &   0.61 &   0.08 &   0.08 &   0.14 &   0.24 \\
 KII-IV, -2$<$[M/H]$<$-1 dex  &     81 &     81 &    125 &    315 &   0.19 &   0.20 &   0.49 &   0.93 &   0.17 &   0.18 &   0.21 &   0.35 \\
      KII-IV, [M/H]$<$-2 dex  &     77 &     78 &    263 &    399 &   0.20 &   0.24 &   0.54 &   0.98 &   0.10 &   0.16 &   0.20 &   0.43 \\
    GII-IV, [M/H]$>$-0.5 dex  &     91 &     91 &    100 &    274 &   0.16 &   0.18 &   0.24 &   0.60 &   0.08 &   0.08 &   0.11 &   0.24 \\
GII-IV, -1$<$[M/H]$<$-0.5 dex &     78 &     88 &    171 &    301 &   0.19 &   0.20 &   0.38 &   0.70 &   0.08 &   0.09 &   0.16 &   0.29 \\
 GII-IV, -2$<$[M/H]$<$-1 dex  &     75 &     84 &    249 &    468 &   0.19 &   0.23 &   0.61 &   1.04 &   0.17 &   0.17 &   0.24 &   0.46 \\
     GII-IV, [M/H]$<$-2 dex   &    107 &    182 &    469 &    695 &   0.27 &   0.43 &   1.03 &   0.85 &   0.18 &   0.22 &   0.52 &   0.66 \\
            FII-IV all [M/H]  &     75 &     75 &     75 &    138 &   0.12 &   0.12 &   0.13 &   0.14 &   0.12 &   0.12 &   0.20 &   0.26 \\
\hline
        KV, [M/H]$>$-0.5 dex  &     77 &     77 &     88 &    141 &   0.18 &   0.18 &   0.23 &   0.43 &   0.09 &   0.09 &   0.10 &   0.17 \\
  KV, -1$<$[M/H]$<$-0.5 dex   &     81 &     80 &     98 &    160 &   0.18 &   0.18 &   0.26 &   0.46 &   0.09 &   0.09 &   0.11 &   0.22 \\
     KV, -2$<$[M/H]$<$-1 dex  &     84 &     83 &    123 &    257 &   0.18 &   0.19 &   0.30 &   0.55 &   0.17 &   0.17 &   0.20 &   0.30 \\
          KV, [M/H]$<$-2 dex  &     91 &     95 &    204 &    429 &   0.13 &   0.21 &   0.27 &   0.78 &   0.11 &   0.11 &   0.23 &   0.33 \\
        GV, [M/H]$>$-0.5 dex  &     83 &     84 &    168 &    310 &   0.11 &   0.13 &   0.30 &   0.49 &   0.09 &   0.09 &   0.12 &   0.24 \\
   GV, -1$<$[M/H]$<$-0.5 dex  &     83 &     97 &    216 &    371 &   0.11 &   0.14 &   0.33 &   0.48 &   0.09 &   0.09 &   0.16 &   0.30 \\
     GV, -2$<$[M/H]$<$-1 dex  &     99 &    121 &    320 &    656 &   0.13 &   0.18 &   0.40 &   0.67 &   0.19 &   0.20 &   0.29 &   0.52 \\
          GV, [M/H]$<$-2 dex  &    264 &    365 &    612 &    943 &   0.26 &   0.34 &   0.73 &   1.09 &   0.18 &   0.26 &   0.46 &   0.78 \\
        FV, [M/H]$>$-0.5 dex  &     90 &     93 &    205 &    332 &   0.18 &   0.18 &   0.31 &   0.47 &   0.09 &   0.09 &   0.15 &   0.27 \\
   FV, -1$<$[M/H]$<$-0.5 dex  &     93 &    101 &    232 &    411 &   0.18 &   0.19 &   0.36 &   0.52 &   0.08 &   0.09 &   0.17 &   0.32 \\
     FV, -2$<$[M/H]$<$-1 dex  &     95 &    108 &    401 &    817 &   0.18 &   0.20 &   0.47 &   0.80 &   0.19 &   0.20 &   0.33 &   0.65 \\
          FV, [M/H]$<$-2 dex  &    156 &    264 &    622 &   1003 &   0.21 &   0.30 &   0.67 &   0.89 &   0.18 &   0.22 &   0.56 &   1.06 \\
\hline
             Thin disc dwarfs &     84 &     85 &    142 &    278 &   0.12 &   0.13 &   0.26 &   0.48 &   0.09 &   0.09 &   0.11 &   0.21 \\
            Thick disc dwarfs &     89 &    100 &    223 &    398 &   0.13 &   0.16 &   0.34 &   0.50 &   0.10 &   0.11 &   0.19 &   0.33 \\
                  Halo giants &     78 &     85 &    195 &    441 &   0.20 &   0.23 &   0.58 &   0.97 &   0.16 &   0.17 &   0.22 &   0.42 \\
\hline
\end{tabular}
\tablefoot{see Table~\ref{tab:internal_errors_matisse} for a description of the line labels.}
\label{tab:q70_dicho}
  \end{center}
\end{table*}

\begin{table*}[h]
  \begin{center}
\caption{Relative errors of the final adopted pipeline, at 70\% of the error distribution.}
\begin{tabular}{l||cccc||cccc||cccc}
\hline \hline
 & \multicolumn{4}{c||}{$T_\mathrm{eff}$ (K)} &  \multicolumn{4}{|c||}{$\log~g$ (dex)} &  \multicolumn{4}{|c}{[M/H] (dex)}\\ \hline
SNR  (pixel$^{-1}$)& 100 & 50 & 20 & 10 & 100 & 50 & 20 & 10 & 100 & 50 & 20 & 10 \\ \hline
    KII-IV, [M/H]$>$-0.5 dex  &     51 &     49 &    102 &    142 &   0.10 &   0.14 &   0.21 &   0.41 &   0.09 &   0.09 &   0.10 &   0.15 \\
KII-IV, -1$<$[M/H]$<$-0.5 dex &     55 &     65 &    109 &    225 &   0.12 &   0.17 &   0.31 &   0.60 &   0.08 &   0.09 &   0.12 &   0.20 \\
 KII-IV, -2$<$[M/H]$<$-1 dex  &     68 &     77 &    132 &    294 &   0.15 &   0.23 &   0.47 &   0.85 &   0.14 &   0.17 &   0.22 &   0.34 \\
      KII-IV, [M/H]$<$-2 dex  &     51 &     92 &    247 &    246 &   0.21 &   0.47 &   0.61 &   1.01 &   0.14 &   0.16 &   0.20 &   0.15 \\
    GII-IV, [M/H]$>$-0.5 dex  &     70 &     69 &    158 &    253 &   0.10 &   0.15 &   0.37 &   0.67 &   0.08 &   0.09 &   0.14 &   0.21 \\
GII-IV, -1$<$[M/H]$<$-0.5 dex &     65 &     74 &    164 &    309 &   0.12 &   0.19 &   0.44 &   0.65 &   0.07 &   0.11 &   0.11 &   0.25 \\
 GII-IV, -2$<$[M/H]$<$-1 dex  &     63 &     94 &    234 &    357 &   0.17 &   0.25 &   0.59 &   0.84 &   0.17 &   0.18 &   0.24 &   0.40 \\
      GII-IV, [M/H]$<$-2 dex  &     90 &    214 &    392 &    386 &   0.25 &   0.48 &   0.83 &   0.72 &   0.17 &   0.27 &   0.38 &   0.43 \\
            FII-IV all [M/H]  &     69 &    106 &     71 &     92 &   0.14 &   0.15 &   0.12 &   0.10 &   0.07 &   0.08 &   0.15 &   0.15 \\
\hline
        KV, [M/H]$>$-0.5 dex  &     59 &     64 &     87 &    119 &   0.09 &   0.12 &   0.19 &   0.23 &   0.08 &   0.08 &   0.09 &   0.14 \\
   KV, -1$<$[M/H]$<$-0.5 dex  &     73 &     79 &     95 &    158 &   0.12 &   0.16 &   0.20 &   0.26 &   0.08 &   0.09 &   0.10 &   0.18 \\
     KV, -2$<$[M/H]$<$-1 dex  &     84 &     87 &    120 &    253 &   0.14 &   0.17 &   0.21 &   0.34 &   0.15 &   0.16 &   0.19 &   0.31 \\
          KV, [M/H]$<$-2 dex  &     92 &     86 &    177 &    333 &   0.17 &   0.18 &   0.14 &   0.85 &   0.10 &   0.13 &   0.19 &   0.39 \\
        GV, [M/H]$>$-0.5 dex  &     57 &     76 &    160 &    275 &   0.08 &   0.12 &   0.27 &   0.40 &   0.09 &   0.10 &   0.14 &   0.29 \\
   GV, -1$<$[M/H]$<$-0.5 dex  &     64 &    104 &    190 &    295 &   0.10 &   0.16 &   0.28 &   0.43 &   0.08 &   0.11 &   0.14 &   0.22 \\
     GV, -2$<$[M/H]$<$-1 dex  &     89 &    136 &    317 &    614 &   0.14 &   0.18 &   0.39 &   0.70 &   0.16 &   0.19 &   0.30 &   0.52 \\
          GV, [M/H]$<$-2 dex  &    169 &    328 &    654 &    756 &   0.22 &   0.46 &   0.67 &   0.95 &   0.18 &   0.26 &   0.50 &   0.61 \\
        FV, [M/H]$>$-0.5 dex  &     54 &     83 &    204 &    361 &   0.11 &   0.16 &   0.27 &   0.43 &   0.10 &   0.12 &   0.19 &   0.34 \\
   FV, -1$<$[M/H]$<$-0.5 dex  &     70 &    116 &    197 &    404 &   0.12 &   0.20 &   0.27 &   0.43 &   0.09 &   0.12 &   0.15 &   0.29 \\
     FV, -2$<$[M/H]$<$-1 dex  &    100 &    174 &    529 &    945 &   0.16 &   0.27 &   0.48 &   0.85 &   0.16 &   0.20 &   0.40 &   0.79 \\
          FV, [M/H]$<$-2 dex  &    185 &    383 &    741 &    981 &   0.27 &   0.46 &   0.73 &   0.90 &   0.27 &   0.41 &   0.63 &   1.03 \\
\hline
             Thin disc dwarfs  &     55 &     70 &    147 &    267 &   0.08 &   0.12 &   0.24 &   0.39 &   0.09 &   0.09 &   0.12 &   0.30 \\
            Thick disc dwarfs  &     67 &    108 &    188 &    346 &   0.11 &   0.17 &   0.29 &   0.43 &   0.09 &   0.12 &   0.18 &   0.29 \\
                  Halo giants  &     65 &     94 &    188 &    335 &   0.17 &   0.28 &   0.57 &   0.86 &   0.15 &   0.17 &   0.23 &   0.38 \\
\hline
\end{tabular}
\tablefoot{see Table~\ref{tab:internal_errors_matisse} for a description of the line labels.}
\label{tab:final_quantiles}
  \end{center}
\end{table*}

\subsection{Errors caused by the radial velocity estimation}
The automated stellar parameterisation relies on a good radial velocity
correction of the observed spectra to the rest frame where the
synthetic templates are calculated. It is therefore necessary to test
the robustness of each method in the case where the spectra are not
perfectly at the rest frame.
 
\begin{figure}
  \begin{center}
    \begin{tabular}{ll}
      \includegraphics[width=4.2cm,height=4cm]{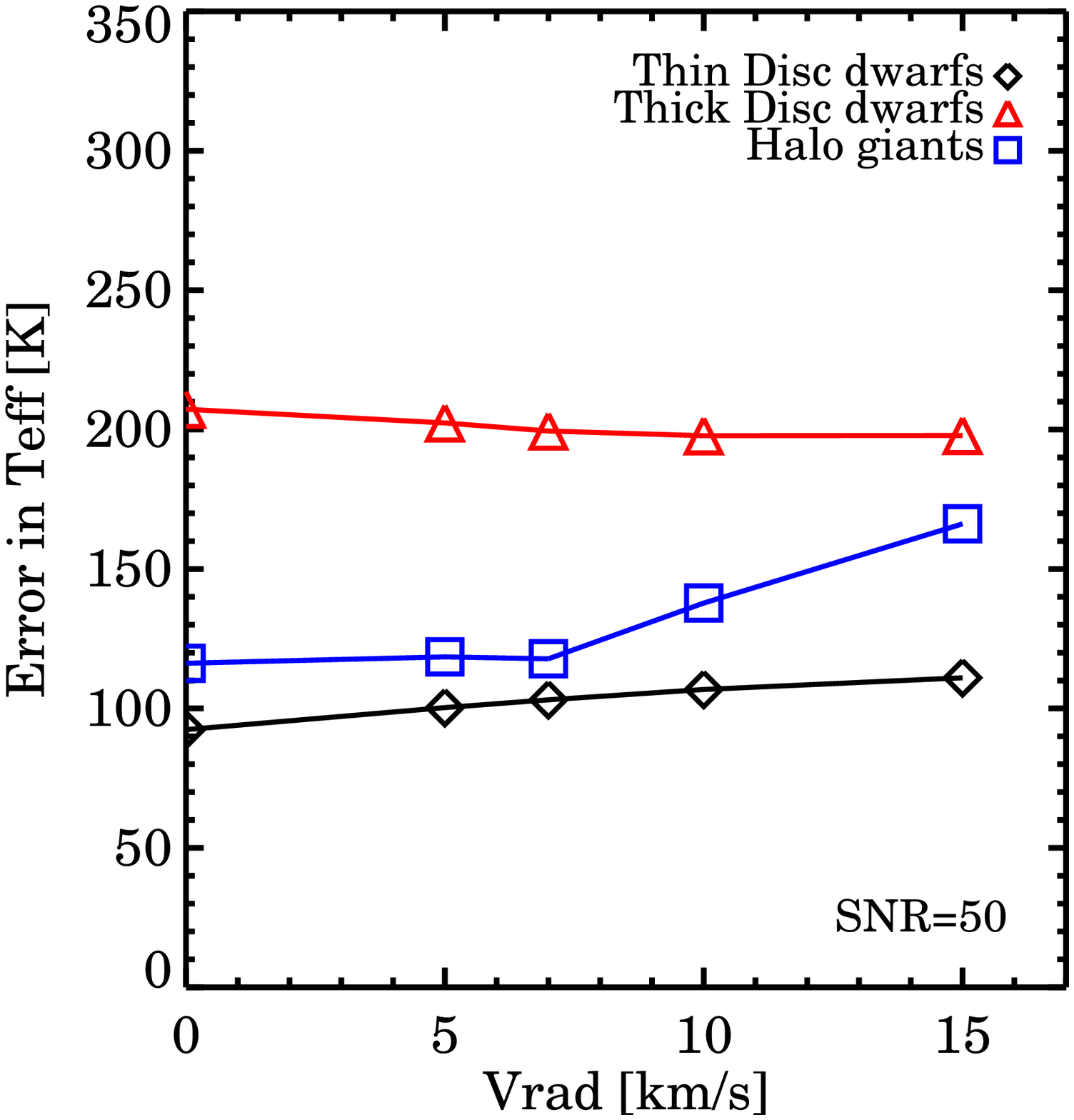} & \includegraphics[width=4.2cm,height=4cm]{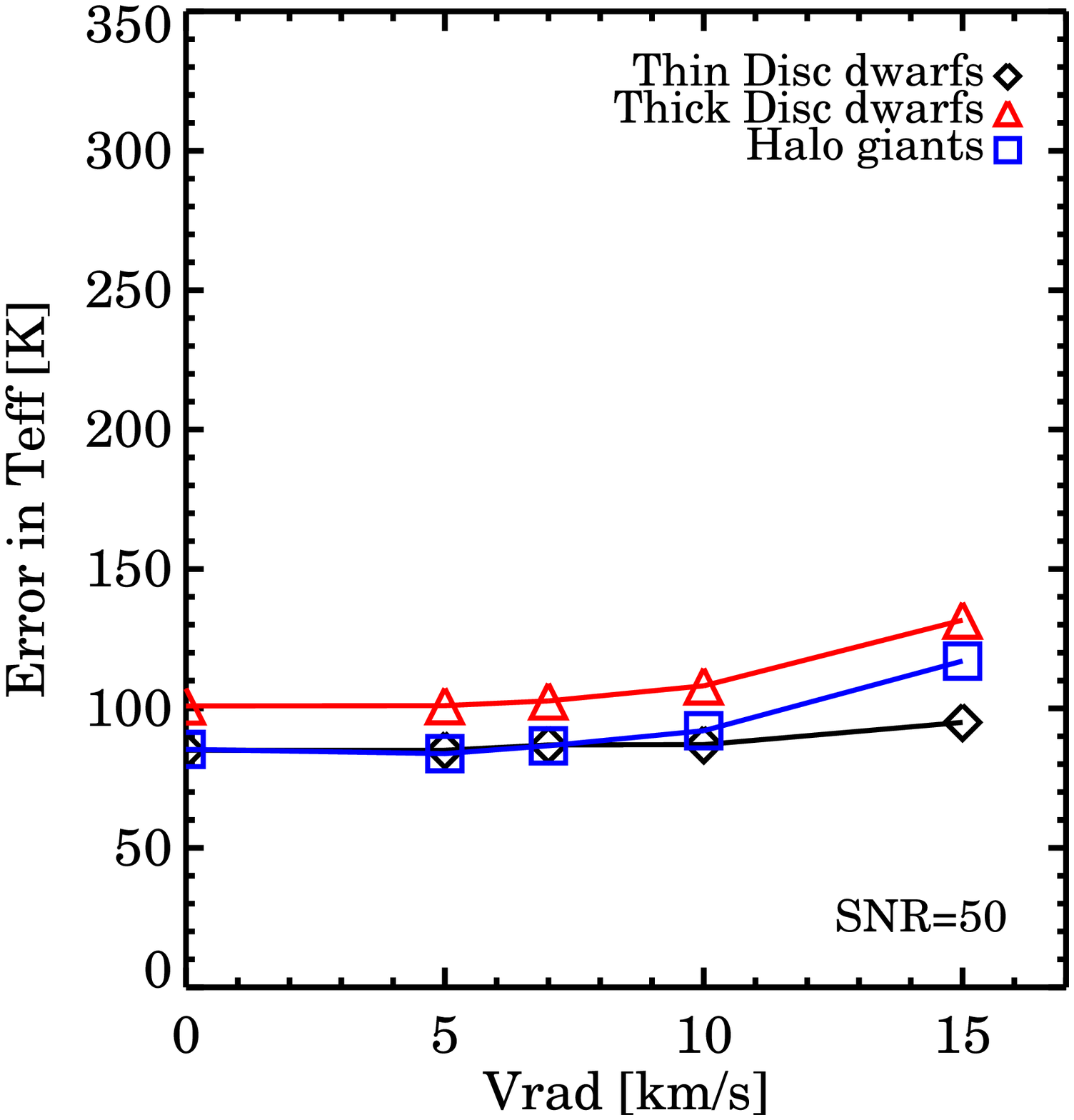}  \\
      \includegraphics[width=4.2cm,height=4cm]{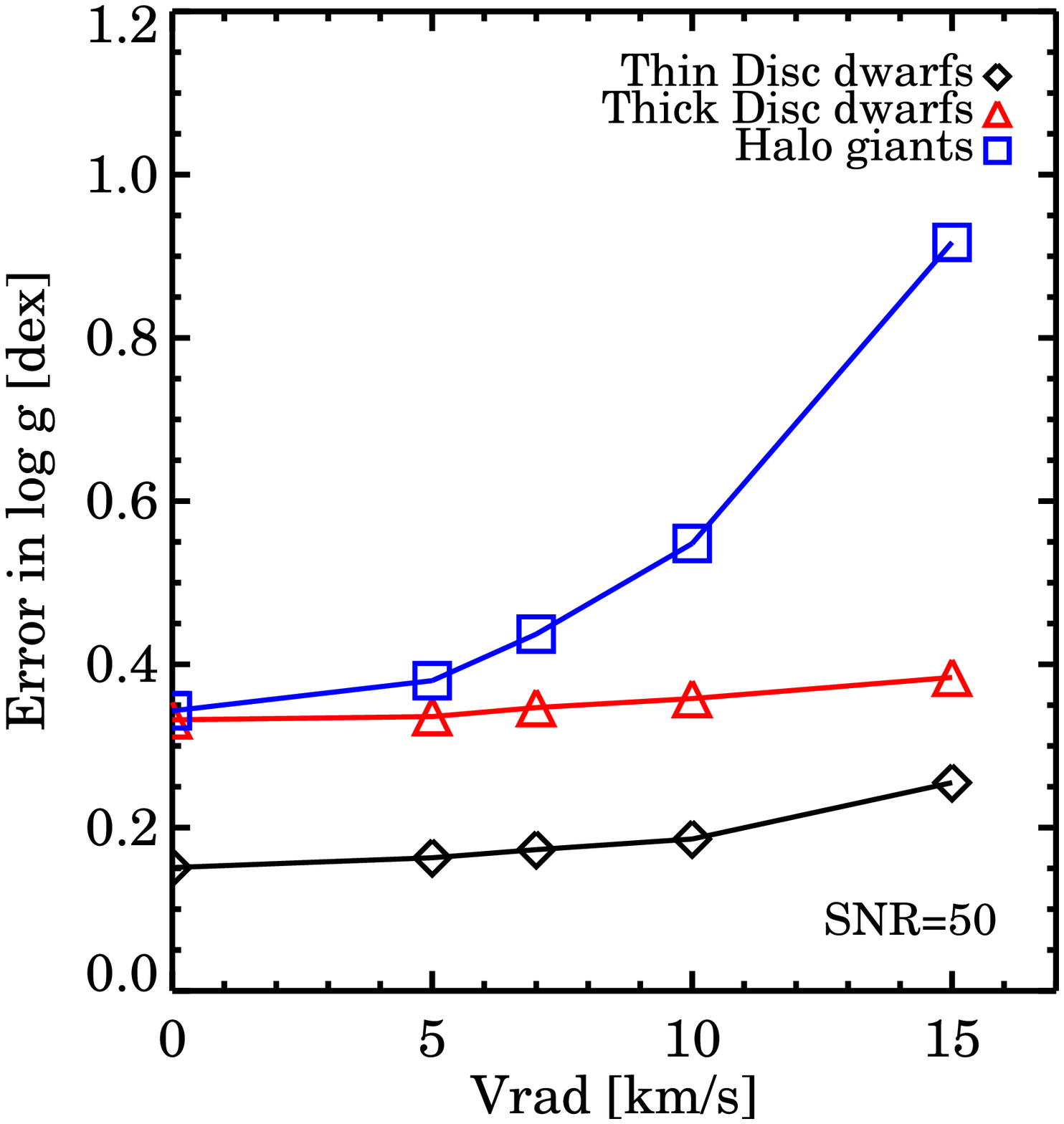} & \includegraphics[width=4.2cm,height=4cm]{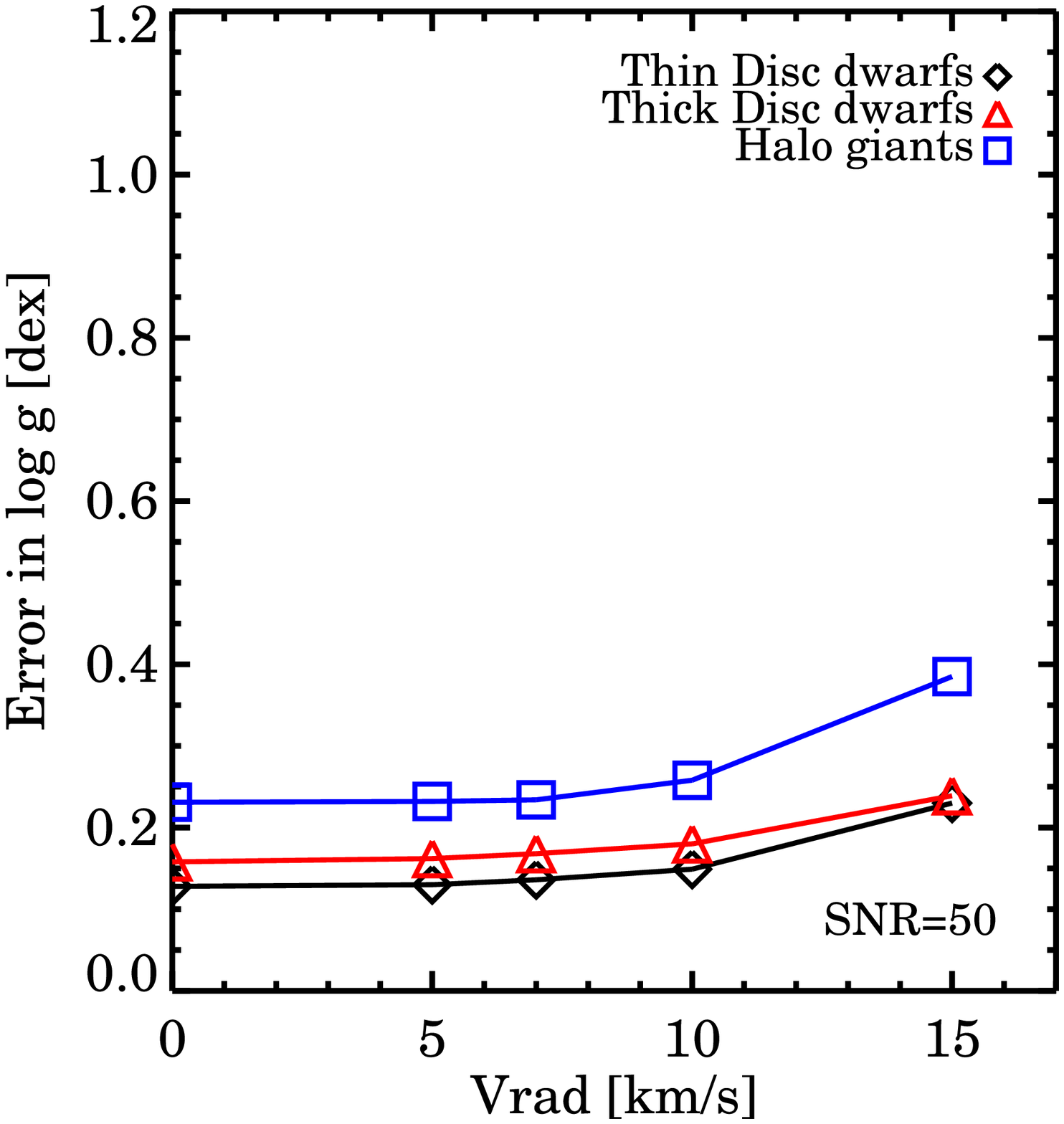}\\
      \includegraphics[width=4.2cm,height=4cm]{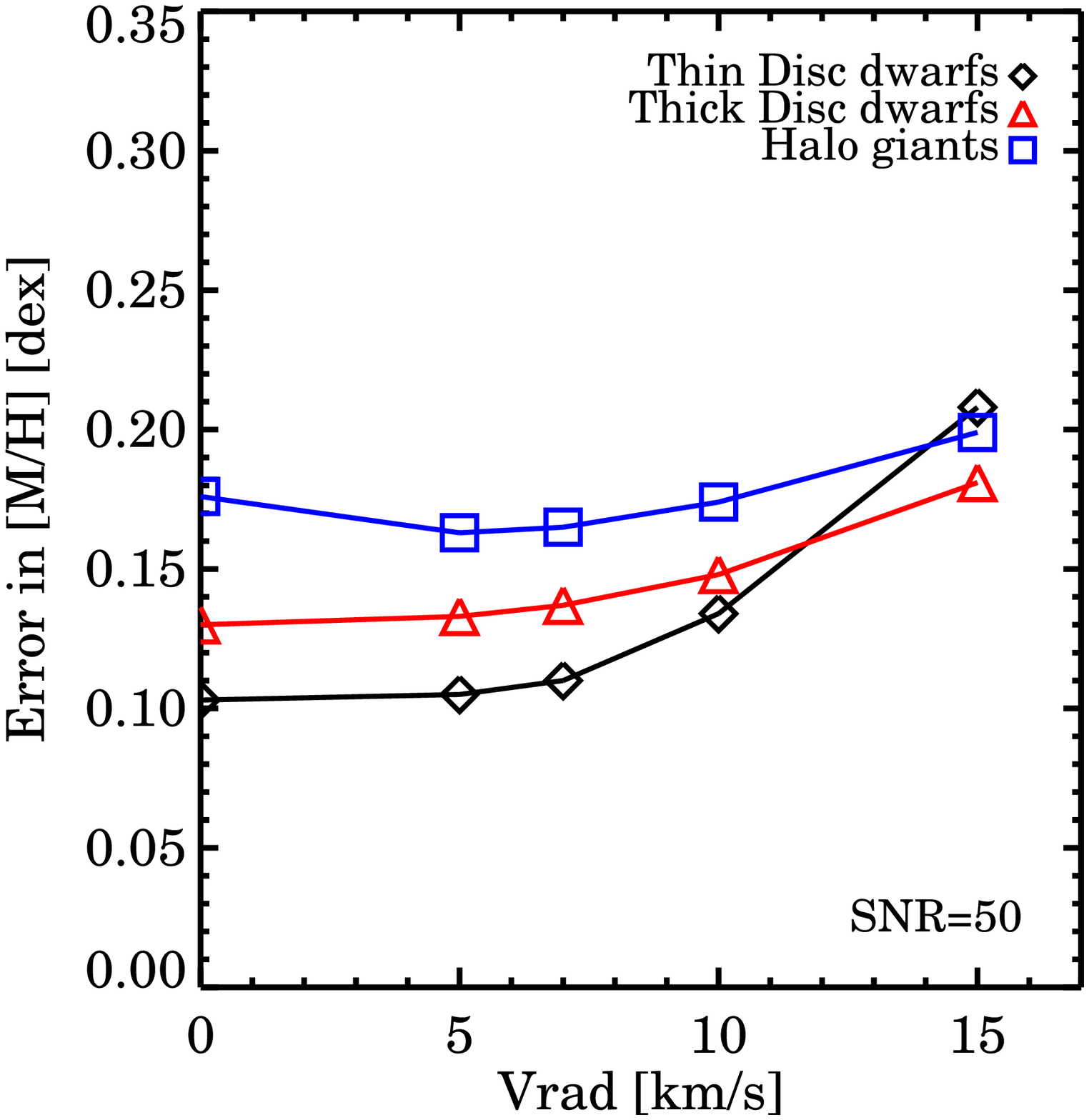} & \includegraphics[width=4.2cm,height=4cm]{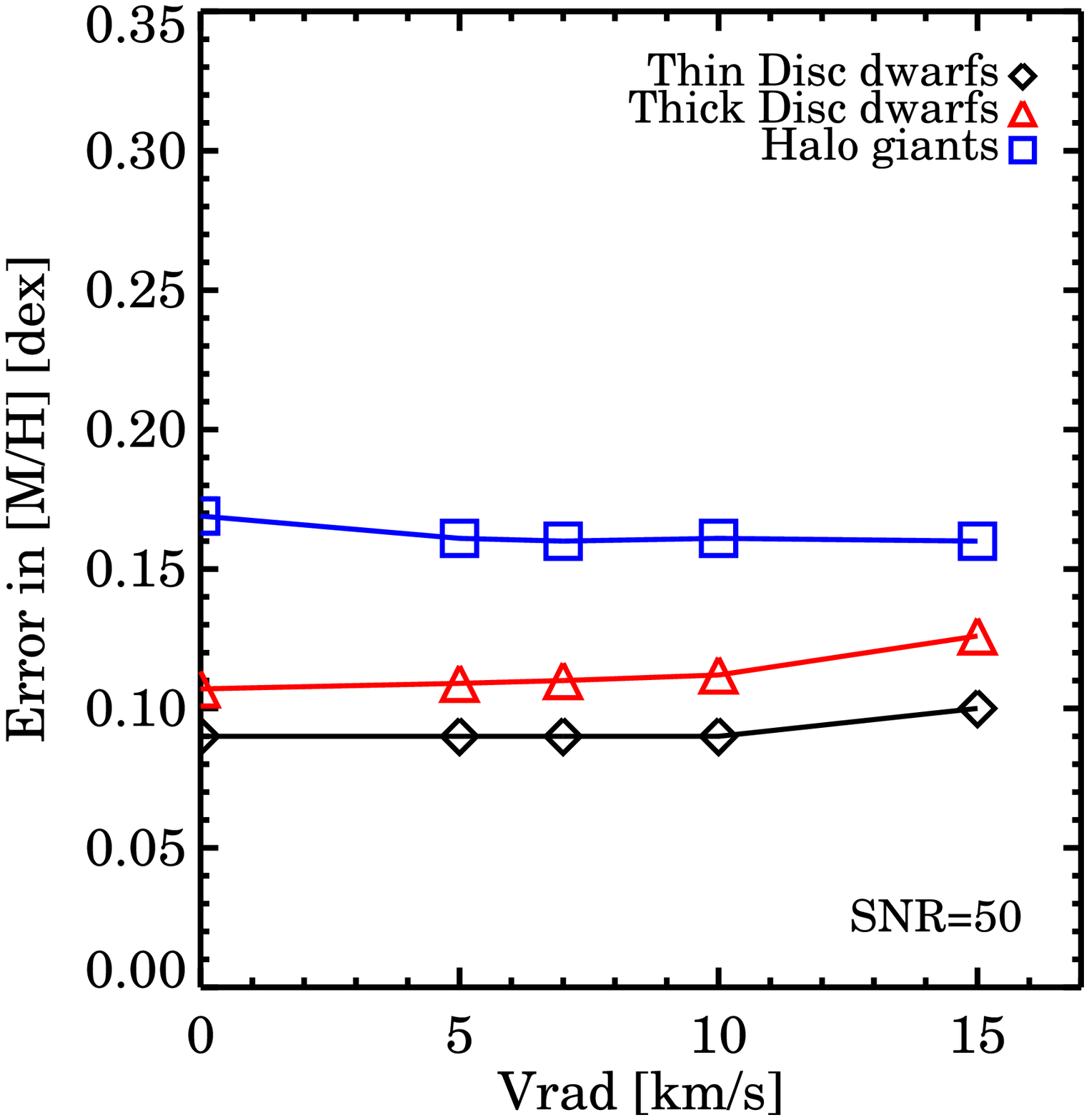}\\
    \end{tabular}
    \caption{Sensitivity to radial velocity errors for MATISSE
      (left) and DEGAS (right). The values at
      $V_\mathrm{rad}=0$~\mbox{km s}$^{-1}$ correspond to the values
      found in Fig.~\ref{fig:Q70_Matisse}. Additional errors caused by an incorrect
       $V_\mathrm{rad}$ correction are almost null as long as the
      Doppler shift remains lower than 7~\mbox{km s}$^{-1}$. For lower
      SNR the robustness of the algorithms is higher, which has an almost
      negligible effect even for an extreme velocity error of
      $V_\mathrm{rad}$=15~\mbox{km s}$^{-1}$.  }
\label{fig:Vrad_errors}
  \end{center}
\end{figure}

To this purpose we used the random set of spectra presented in
Sect. \ref{sec:random_grid}, with SNR$\sim$ 100, 50, 20 and 10~pixel$^{-1}$, and
introduced four different values of Doppler shifts ( $V_\mathrm{rad}=$5, 7, 10 and
15~\mbox{km s}$^{-1}$, corresponding to 1/3, 1/2, 2/3 and 1 pixel
shift).  We then ran the two algorithms with these spectra and checked
the resulting error distribution for each parameter.  In
Fig.~\ref{fig:Vrad_errors} we show the relative errors obtained for
the different radial velocity shifts for an SNR=50 and for both
methods. The results show that the spectra need to be corrected to better
than half a pixel (corresponding to a $V_\mathrm{rad}$ accuracy of
less than $\sim$ 7~\mbox{km s}$^{-1}$) to obtain estimates that are still
similar to those obtained for spectra at the rest frame.  Most of
the times, this condition is fulfilled, because the errors commonly
obtained on $V_\mathrm{rad}$ for FGK stars at SNR$\sim$20 in the LR8
setup are less than 5~\mbox{km s}$^{-1}$ (see the companion paper
Kordopatis et al. 2011b).  However, the atmospheric parameter errors,
induced by a non-perfect $V_\mathrm{rad}$ correction, tend to increase
more rapidly when the metallicity decreases. This is because 
 the few spectral signatures present in the signal have a
relatively small equivalent width (the spectral lines are spread on
only a few pixels), and even at high SNR their misplacement have a
serious effect on the parameter estimations.

Furthermore, good estimates are obtained for both  methods for
shifts up to 2/3 of a pixel at any SNR for intermediate- to high-metallicity 
stars. We point out though that DEGAS is in general more
stable than is MATISSE when dealing with spectra that are not in the rest
frame, especially for metal-poor stars.
  
At lower SNR ($\leqslant 20$) both algorithms are stable to
Doppler shifts and show a similar behaviour.  Indeed, for low-quality
spectra the errors induced by noise dominate those arising from a poor
radial velocity correction.

\subsection{Summary: adopted strategy}
\label{subsec:adopted_strategy}
We decided to set up a procedure that
would combine the benefits of MATISSE (very good results in a local
environment, easy interpretation) and DEGAS (global view and better
tackling of the secondary minima problem).  The MATISSE algorithm
gives better results if there are no significant secondary minima in
the distance function, because of the local application of its
learning $B_\theta(\lambda)$ functions.  In the particular case of 
low-resolution IR \ion{Ca}{ii} spectra, this condition is achieved for
data with an SNR value higher than $\sim 35$ and global metallicity
[M/H]~$>-2.0$~dex.  Nevertheless, the use of $B^0_\theta(\lambda)$ is
not the optimal approach to obtain the best possible results.  Indeed,
the number of local minima tends to increase with noise and the lack
of spectral signatures. Given the properties of the parameter space
around the \ion{Ca}{ii} region, the generic $B^0_\theta(\lambda)$
functions will not provide the absolute minimum and hence will
accentuate the degeneracies.  Therefore, to fully exploit the
capabilities of MATISSE, we decided to use the
results of DEGAS as an initial input. This operation reduces the effect caused by the
secondary minima, and can therefore provide more accurate results (up to
50\% accuracy improvement for $T_\mathrm{eff}$ and $\log~g$).

For lower SNR (less than 35) and very metal-poor stars, the non-convexity of 
the distance function is too severe to infer good results
with a projection method such as MATISSE. The use of DEGAS is therefore 
preferred. Indeed, DEGAS can derive  
errors lower by 75\% than those of MATISSE 
for a low-SNR metal-poor spectrum.

Finally, as seen in Fig.~\ref{fig:HR_Matisse} and
Fig.~\ref{fig:HR_Dicho}, because of  the non-convexity of the distance
function, both methods can return results in regions of the H--R
diagram where no stars can exist.  Removing 
these parameter combinations from the 
possible solution space will therefore reduce the
degeneracies and increase the method accuracy.  Based on the
$Y^2$ isochrones \citep{YY_isochrones} of stars spanning ages from
0.25~Gyr to 15~Gyr and metallicities from $-3$~dex to +0.8~dex, we
decided to exclude the $B_\theta(\lambda)$ functions and the grid
points used for DEGAS, where no possible isochrones were nearby. In
practice, to avoid too important astrophysical priors in
the derived parameters, we removed only the templates with $\log~g$=5
and $T_\mathrm{eff} > 6250$~K, those with $T_\mathrm{eff} \leq 4250$~K
and $4\leq \log~g \leq 3$~dex, as well as all  stars with
[M/H]~$\leq -3$~dex, $T_\mathrm{eff} \leq 4000$~K and
$\log~g~\leq$~4~dex.  We checked in
Sect.~\ref{subsec:normalisation_effects} that this selection did not
introduce biases in the final results.


\section{Final adopted pipeline}
\label{sec:final_pipeline}
Based on the results and the discussion in the above
sections, we adopted a final pipeline.  This procedure 
took into account the following points:
\begin{itemize}
\item
The applied parameterisation method has to be chosen according to its
optimal application conditions defined in Sect.~\ref{subsec:adopted_strategy}.

\item
Spectra normalisation (when the data are not flux-calibrated)
is coupled to the atmospheric parameter determination.

\item
The parameterisation algorithms are optimised for a given SNR value,
which implies a robust SNR measurement. 
\end{itemize}

However, the SNR determination itself depends on the spectra
normalisation and the atmospheric parameter estimation.  The final
adopted pipeline therefore iterates on the parameter estimation until
convergence of the normalisation solution and the SNR measurement, as
described in the following steps:

\begin{enumerate}
\item A rough sigma-clipping, polynomial fit normalisation is
  performed and an initial value of the SNR is assumed that is constant for all the
  spectra (we adopted an intermediate value of SNR $\sim$50).  
  This will give us a first rough estimation of the
  parameters using DEGAS to compute a synthetic spectrum
  with the parameters of the first guess. Using this template, the
  SNR is estimated for the first time, as described in
  Sect.~\ref{subsec:snr_measurement}.

\item
\label{pipeline:boucle}
DEGAS is re-applied with the newly found SNR values. The new
atmospheric parameters are then used to compute a synthetic template
and re-normalise the input spectrum. This step is repeated several
times to obtain a normalisation  as perfect as possible (see
Sect.~\ref{subsec:normalisation_effects}).  Convergence is achieved
when the continuum shape is unchanged compared to the previous
iteration.

\item The SNR is re-computed for the final normalised spectra. If its
  value is higher than 35 and the [M/H] value determined by DEGAS is
  higher then --2.0~dex, MATISSE is run with the set of
  $B_\theta(\lambda)$ functions corresponding to the spectral type and
  the SNR value. The  results of DEGAS from the previous step are kept
  for low-SNR or very metal-poor stars.

\item For other stars, the final atmospheric parameter estimations are
  obtained when the convergence of MATISSE is achieved. If the 
  MATISSE results are beyond the parameter grid boundaries
  shown in Table~\ref{tab:param_range}, the results of DEGAS are
  taken.  The final SNR is derived only at the end of this step, with
  a template corresponding to the adopted stellar atmospheric values.
\end{enumerate}

\subsection{Measurement of the signal-to-noise ratio}
\label{subsec:snr_measurement}
The SNR is estimated following the procedure described in
\citet{RAVE_second_data_release}. First, the atmospheric parameter
guesses obtained by a given method are used to obtain a synthetic
spectrum with the same atmospheric parameters.  The difference between
the template and the observed spectrum is then computed, keeping only
the pixels where the difference changes sign from the previous or the
next adjacent pixel and has a relative flux close to 1.  This
selection criterion allows us to keep only the continuum pixels, where
the difference is caused by pure noise, and avoids the selection of
pixels that are affected by systematic effects (for example due to a bad
template selection).  This difference is then divided by the
theoretical spectrum and the SNR estimate is obtained by taking the
inverse
of its standard deviation.  \\
As we previously pointed out, the measurement of the SNR strongly depends
 on the normalisation and the synthetic template selection. In
addition, the selection of the synthetic template depends on the
accuracy of the results of the used method (selected
$B_\theta(\lambda)$ functions for MATISSE or the considered leaves for
DEGAS), which also depend on the SNR.  The redundancy of the problem
therefore requires several iterations before a convergence of
the values can be achieved. 
For that reason, the SNR needs to be estimated several times during our procedure to obtain a good normalisation (see
Sect.~\ref{subsec:normalisation_effects}).

\subsection{Normalisation effects}
\label{subsec:normalisation_effects}

\begin{figure}
  \begin{center}
    \begin{tabular}{ll}
      \includegraphics[width=4.3cm,height=4.3cm]{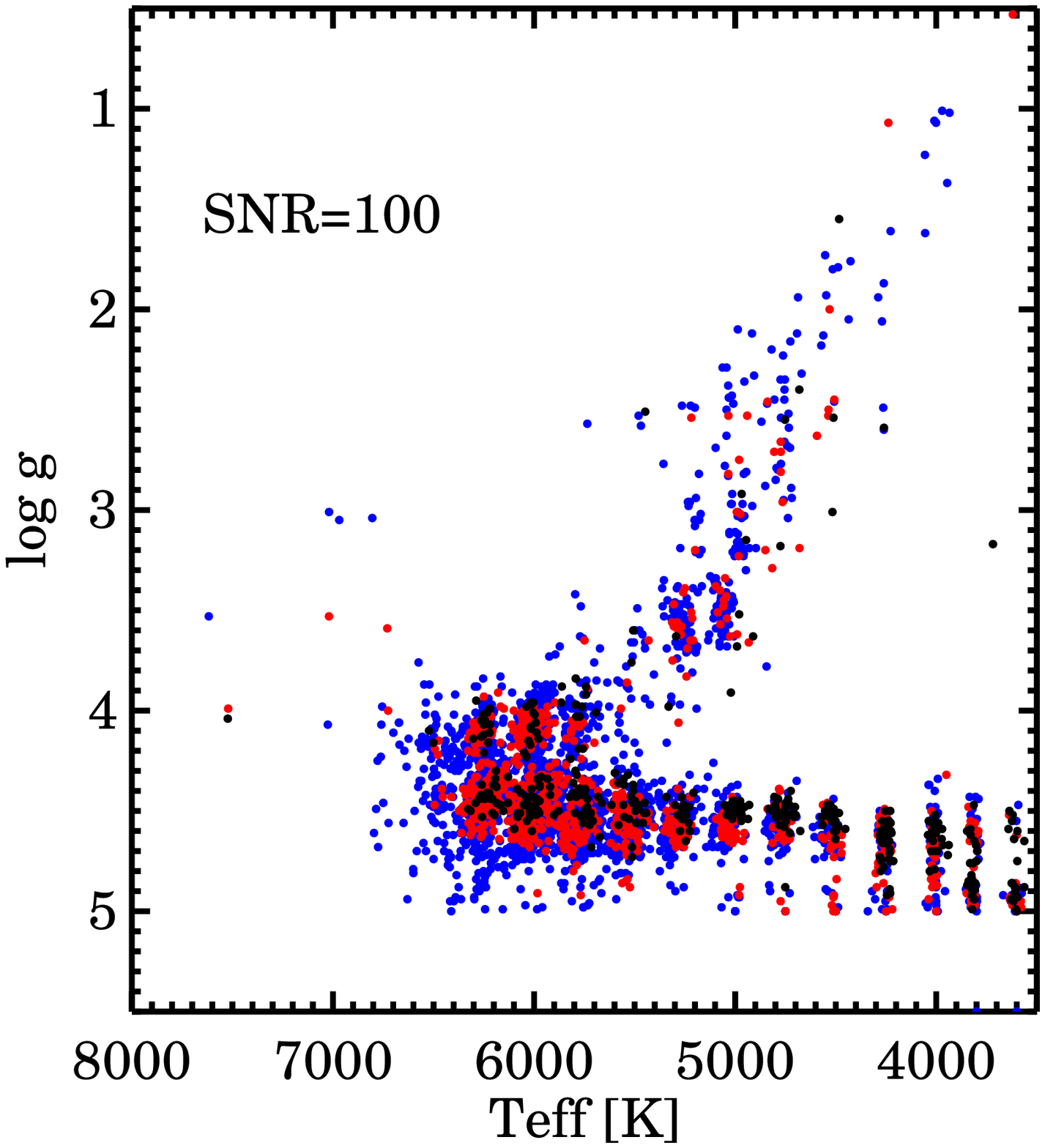} & \includegraphics[width=4.3cm,height=4.3cm]{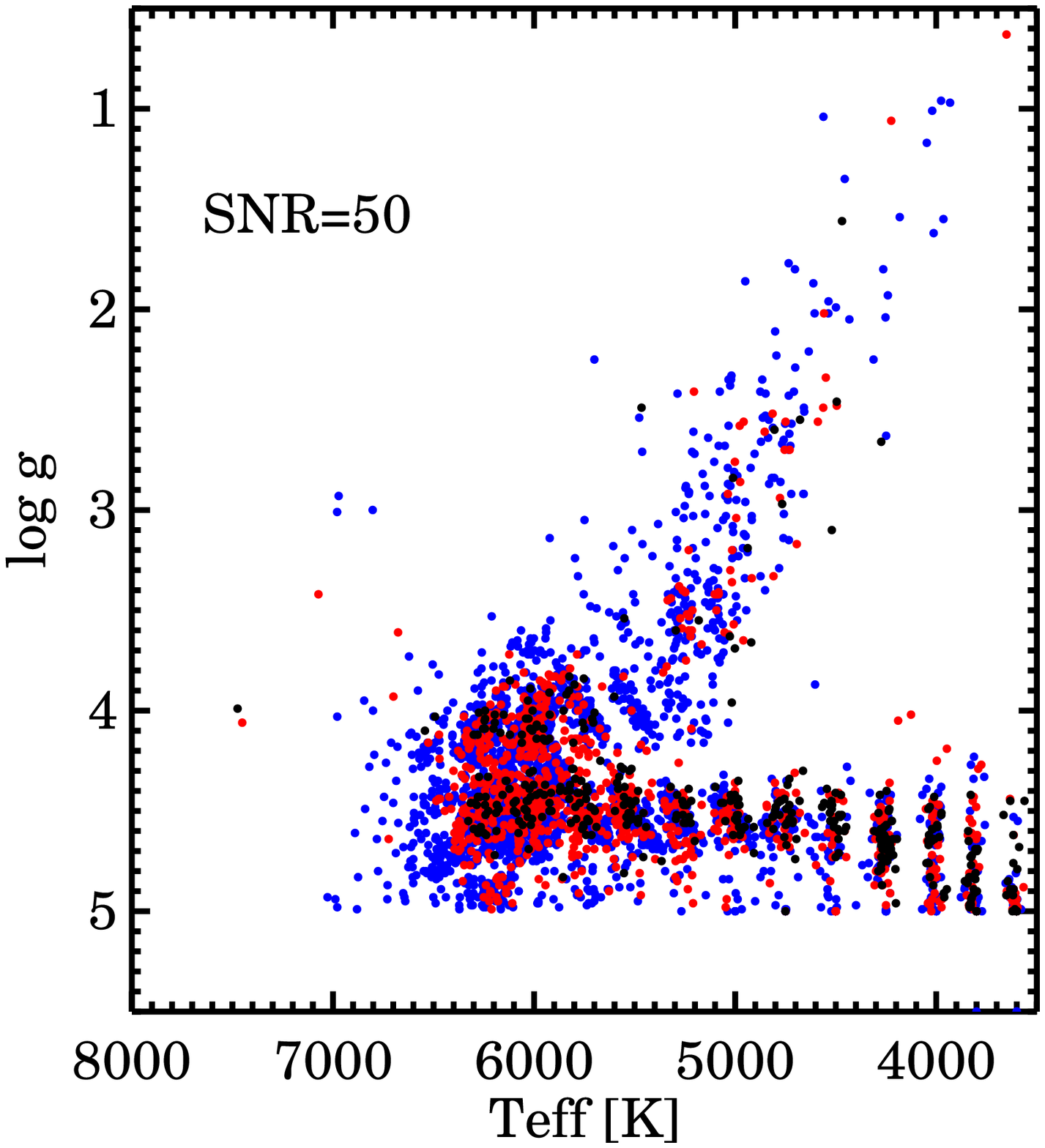}  \\
      \includegraphics[width=4.3cm,height=4.3cm]{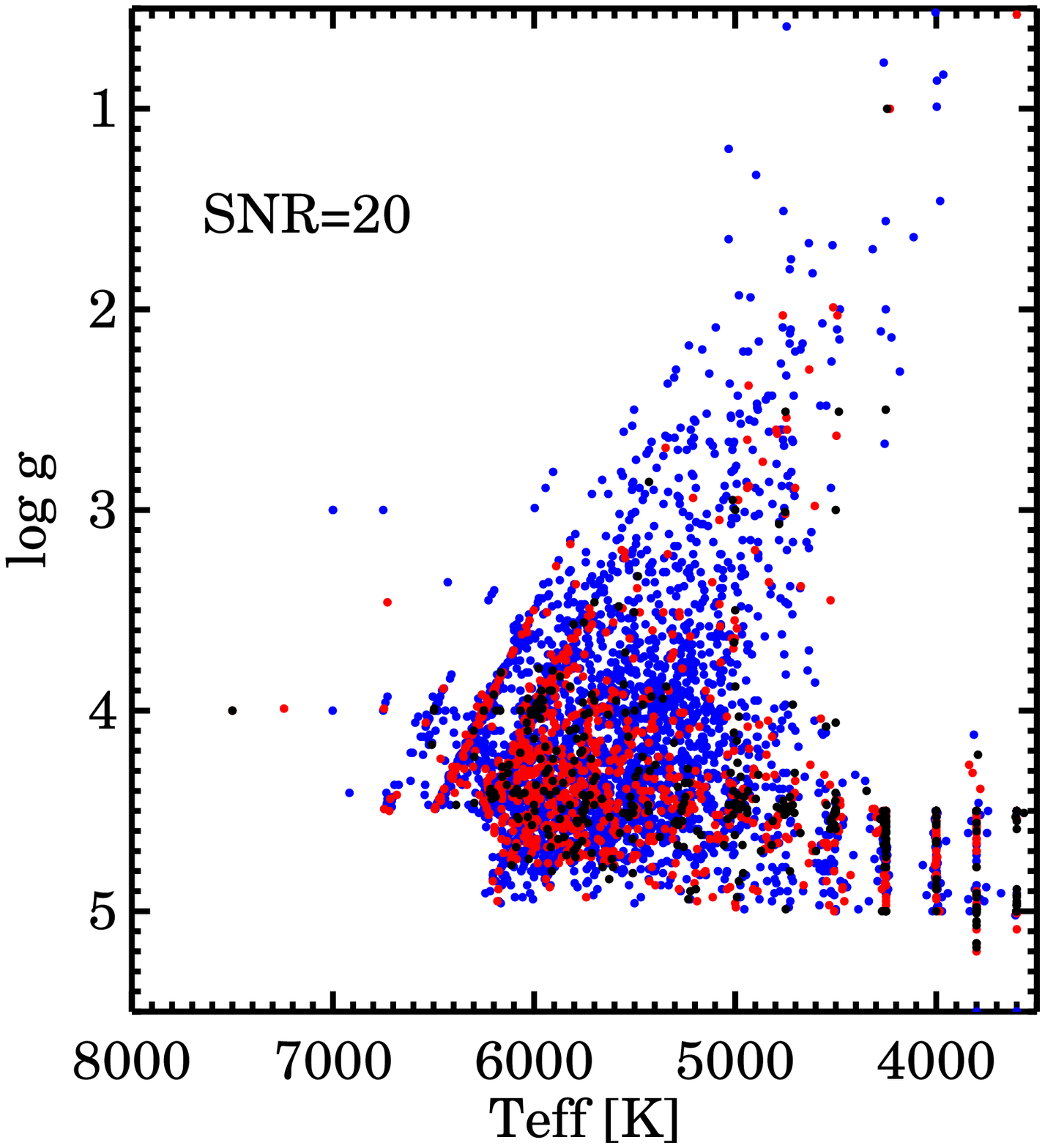} & \includegraphics[width=4.3cm,height=4.3cm]{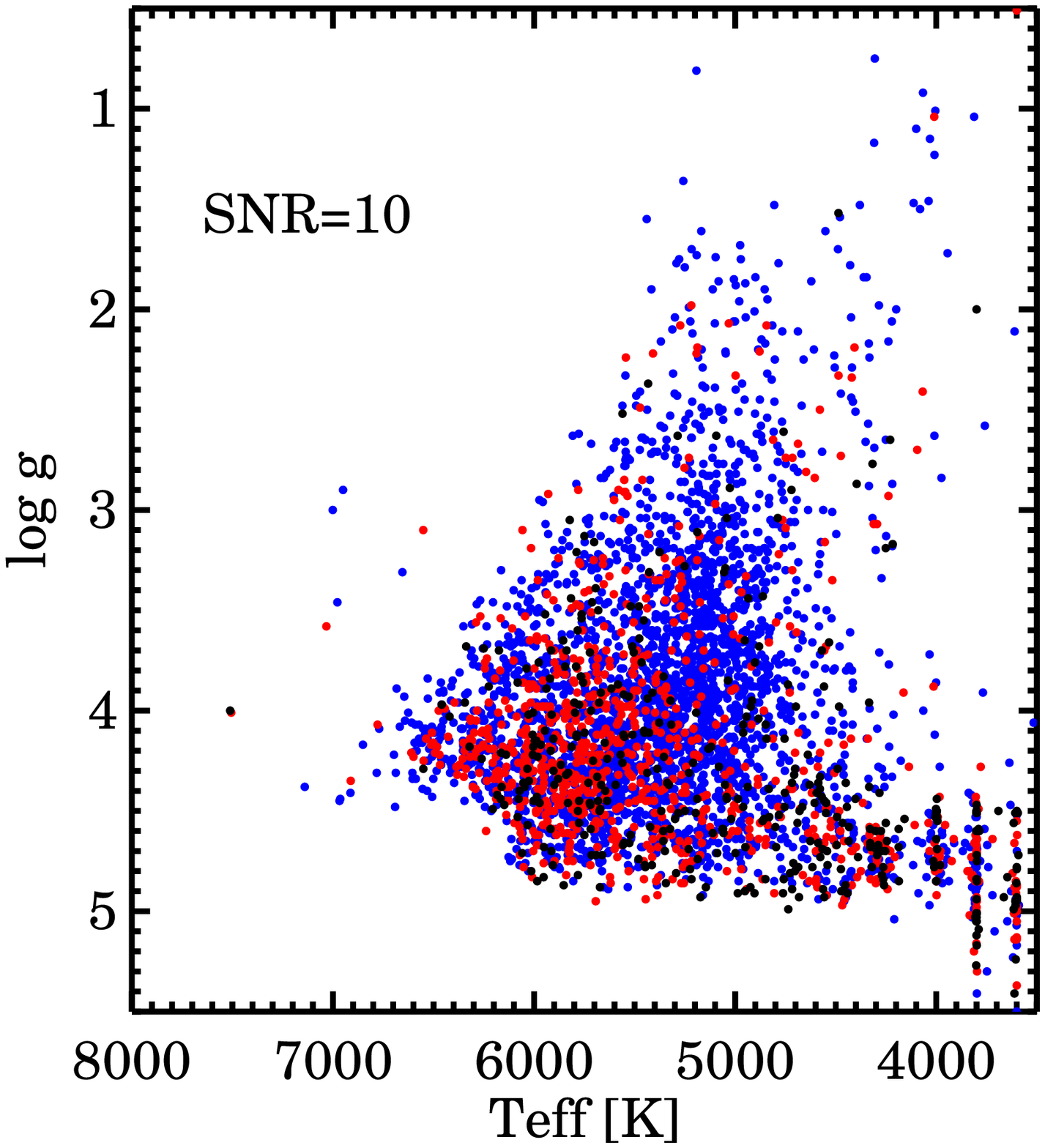}\\
    \end{tabular}
    \caption{Sensitivity of the recovered H--R diagram for simulated
      data to the spectrum signal-to-noise ratio using the final adopted
      pipeline. The colour code is the same as in
      Fig.~\ref{fig:random_grid}. }
\label{fig:HR_Final}
  \end{center}
\end{figure}

Because the parameter derivation is very sensitive to the relative
depths of specific lines, a poorly-determined pseudo-continuum shape can have
sizeable effects. Normalisation is therefore a matter of prime importance
and special care must be taken.  Comprehensive tests were made to minimise 
this effect as much as possible.
 
The final adopted procedure consisted of running DEGAS several times,
renormalising the initially roughly normalised
spectrum after each iteration.  To accomplish this, we use a few consecutive iterations
to fit the shape of the ratio between the interpolated spectrum and the
observed spectrum with a fifth-order polynomial\footnote{A third-degree
  polynomial was also tested, showing either equivalent or
  poorer results than a fifth-order polynomial.} and with symmetric
rejection criteria (clipping away points farther than 0.5 $\sigma$).
The continuum shape of the observed spectrum is then re-adjusted
according to the shape of the fitted function.

\begin{figure*}
  \begin{center}
\includegraphics[width=0.95\textwidth]{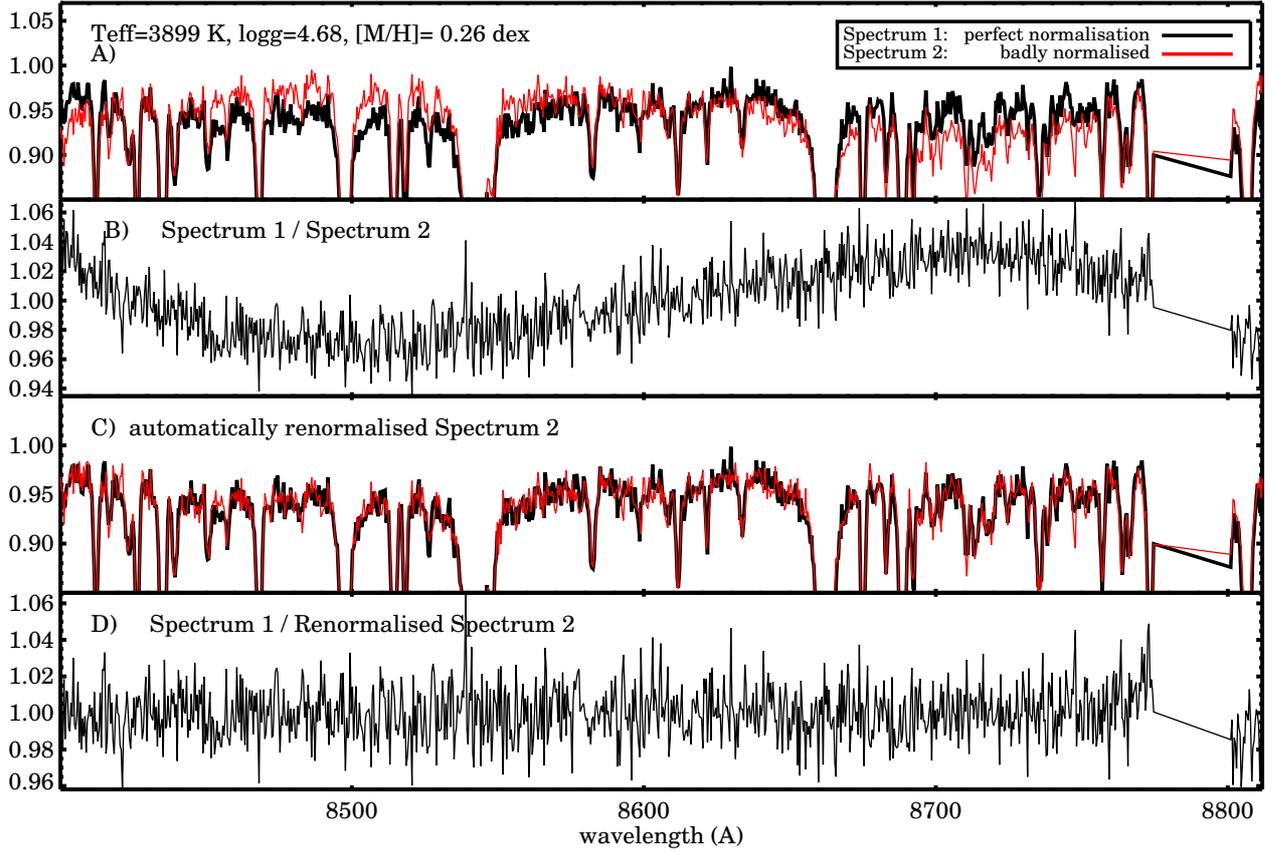} 
      \caption{ Re-normalisation process for a cool metal-rich dwarf star. Plot A) shows in black the theoretical spectrum and in red the spectrum with a modified continuum (see Sect.~ \ref{subsec:normalisation_effects}), which is analysed with our method. The change applied on the input spectrum can be viewed in plot~B) of this figure, where the ratio between the theoretical spectrum and the input is shown. A corrected continuum shape is obtained at the end of our pipeline, as can be seen in the two bottom plots (C and D). }
\label{fig:Renormalisation}
  \end{center}
\end{figure*}

To test the robustness of our procedure, we performed various tests 
with the random spectra grid described in
Sect.~\ref{sec:random_grid}.  We modified the spectral shape (and
hence the continuum) up to 3\% by multiplying the spectrum with either
a slope, a third-degree polynomial, or a fifth-degree
polynomial. The amplitude of 3\% is considered to be a pessimistic
case for roughly normalised spectra corrected for instrumental
effects.  The tests were also run on the perfectly normalised spectra,
passing through the whole pipeline to verify that our
procedure is not degrading the already perfectly normalised input
spectra (see the left column of Fig.~\ref{fig:Q70_Matisse},
Fig.~\ref{fig:HR_Final} and Table~\ref{tab:final_quantiles}).  The
results show that this strategy is quite rapidly converging to a stable solution 
which  minimises the errors caused by normalisation
independent of the deformation of the continuum. 
In addition, as we show in Fig.~\ref{fig:Renormalisation}, 
a good re-normalisation is obtained even for the cool metal-rich stars. Indeed, 
in that case the large wings of the spectral lines and the huge amount of atomic and 
molecular lines can be a challenge to find the continuum.
An unchanged
continuum shape is achieved after a few iterations. The number of
needed iterations depends on the SNR, the departure from the true
continuum and the spectral type.  Practically, this procedure is
performed ten consecutive times and convergence is checked.

It can also be noticed that in some cases, this pipeline derives the
parameters with even better accuracies than the better of the two
individual methods (Tables~\ref{tab:internal_errors_matisse},
\ref{tab:q70_dicho} and \ref{tab:final_quantiles}).  This is because 
the $B^0_\theta(\lambda)$ are not used any more, which better 
constrains the parameter sub-space, as already
discussed in Sect.~\ref{subsec:adopted_strategy}.  Another factor that
explains the better results is that the final adopted pipeline does
not consider grid points in unphysical regions of the H--R diagram.
Nevertheless, the extrapolation properties of MATISSE still derive stellar 
parameters with ``unphysical'' values in some cases, as
seen in the two top plots of Fig.~\ref{fig:HR_Final}.

Finally, the method is robust to deformation of the continuum, with
errors on $T_\mathrm{eff}$, $\log~g$ and [M/H] obtained after the
re-normalisation procedure  almost identical (within 10\%) with the
errors obtained for perfectly normalised spectra.

\section{Application to observed spectra}

\begin{figure*}
  \begin{center}
    \begin{tabular}{ccc}
      \includegraphics[width=5.7cm,height=5.5cm]{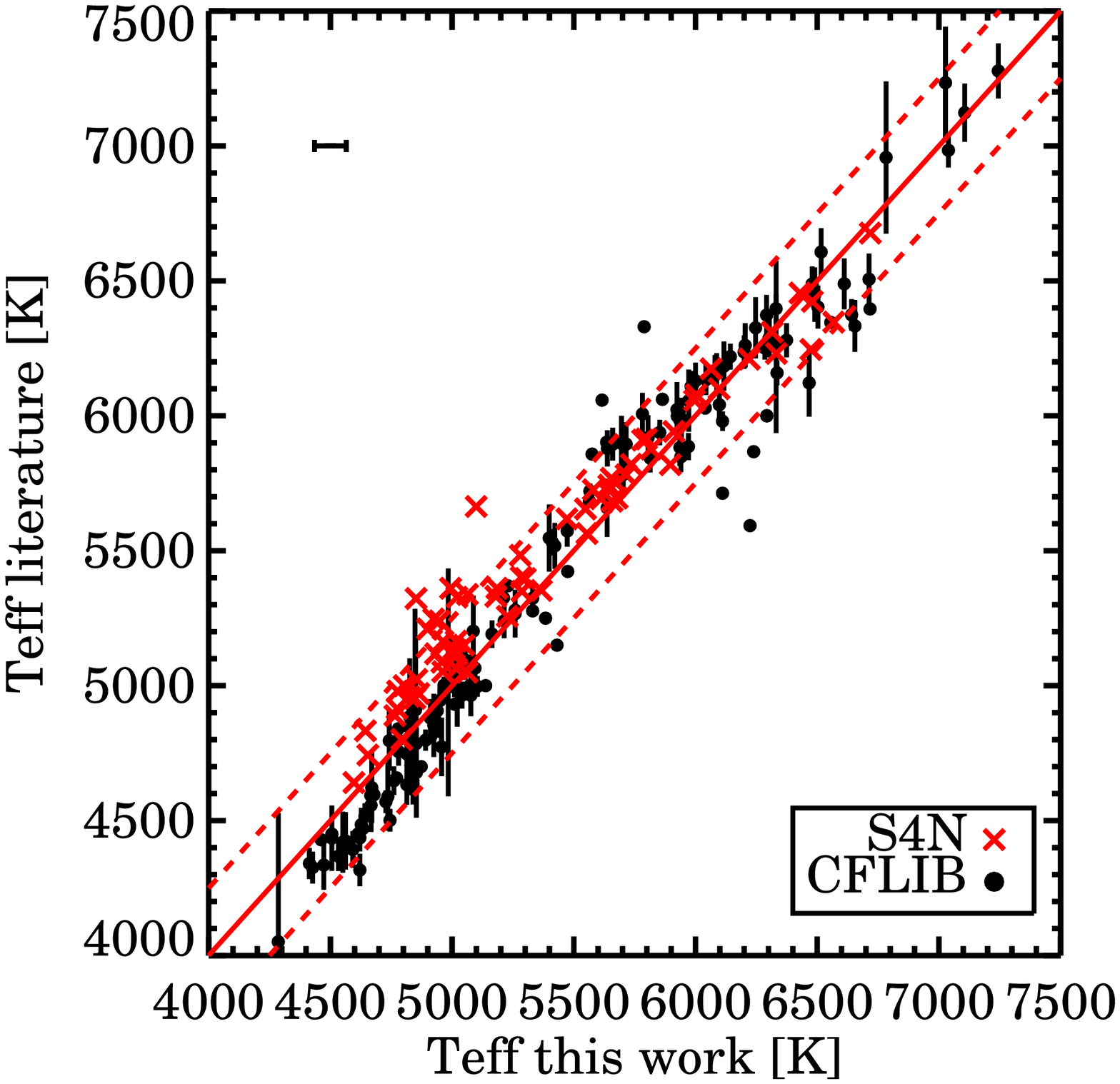} & \includegraphics[width=5.7cm,height=5.5cm]{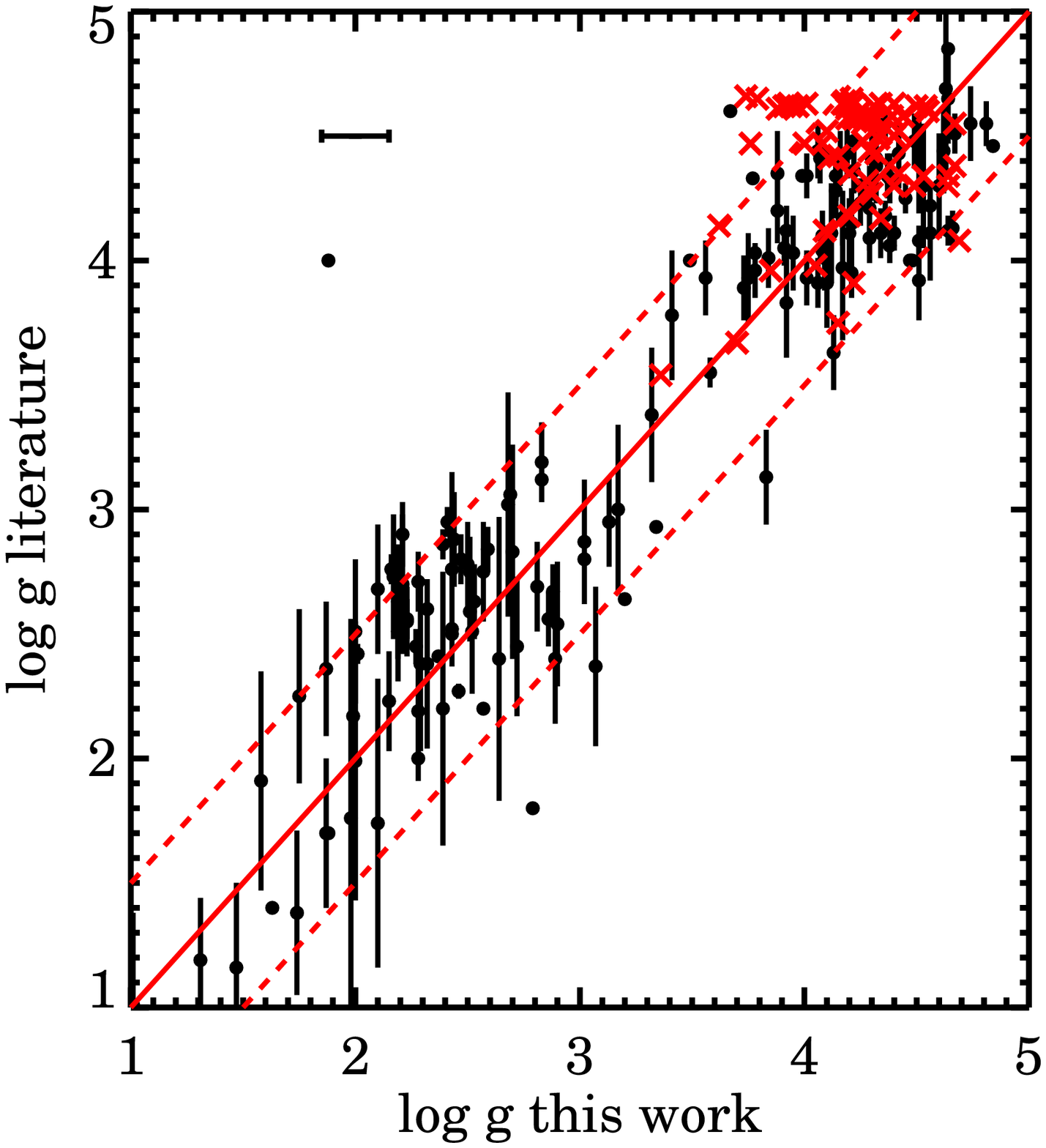}  &
      \includegraphics[width=5.7cm,height=5.5cm]{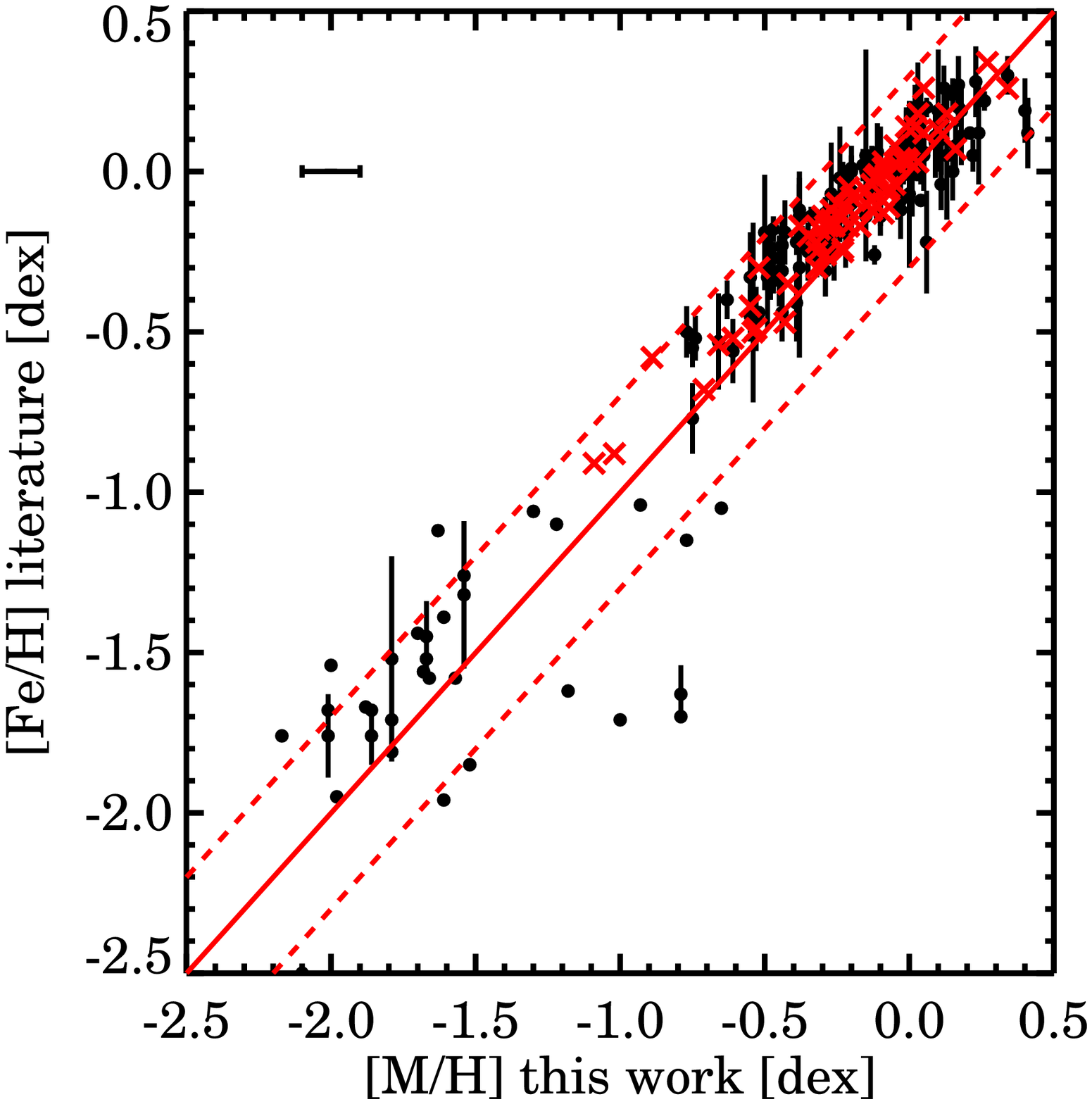} 
    \end{tabular}
    \caption{Comparisons between the results of our pipeline and
      values found in the literature for the $S^4N$ (red crosses) and
      CFLIB (black dots) stellar spectral libraries. The error bars
      shown for the CFLIB spectra represent the dispersion of the
      parameter values found in the literature, when available. The
      red dashed diagonal lines, plotted for an easier interpretation
      of the results, represent errors of $\pm$~250~K, 0.5~dex and
      0.3~dex for $T_\mathrm{eff}$, $\log~g$ and [M/H],
      respectively. Typical errors of our pipeline are represented in
      the upper left corner of each plot.  }
\label{fig:External_errors}
  \end{center}
\end{figure*}

The errors estimated in the previous sections and illustrated in
Table~\ref{tab:final_quantiles} can be viewed as the expected relative
errors of the method when one compares stars within a catalogue of
observed stars. Nevertheless, differences between model spectra and
real stellar spectra can potentially introduce systematic
biases. Therefore we needed to test our pipeline with real data.

We attempted to recover the atmospheric parameters from spectra in 
observed stellar libraries.  It was a challenge to find spectra that
covered the same wavelength region with a sufficiently high resolution and
good parameter estimates. Only two datasets were found: the $S^4N$ from
\cite{Prieto_S4N}, and the CFLIB described in \cite{CFLIB}.  We smoothed the spectra from both
libraries with a Gaussian kernel to match our
working resolving power and re-sampling, and applied the whole
pipeline.

The $S^4N$ catalogue\footnote{See \url{http://hebe.as.utexas.edu/s4n}}
 consists of a survey of 118 stars
in the solar neighbourhood, therefore these stars are mostly metal-rich
dwarfs. The covered spectral range is very broad (3620--9210~\AA),
observed in a single spectral setup at high-resolution (R$\sim$
50~000) and high SNR ($>200$).  Nevertheless, only 68 spectra include
our wavelength range without any inter-order gap.  The effective
temperatures for this catalogue were obtained using 
Str\"{o}mgrem photometry and the (B-V) colour index with the
\citet{Alonso_dwarfs, Alonso_giants} calibrations. Furthermore, these
$T_\mathrm{eff}$ were confirmed by fitting synthetic profiles of
the $H_\alpha$ and $H_\beta$ lines, as described by \citet{Barklem}
and are therefore considered reliable. In addition, the gravities published
for these stars have been derived from the Hipparcos trigonometric
parallaxes and are also very reliable.  Nevertheless, one may expect
differences between physical gravities and gravities obtained
spectroscopically, especially for cool main-sequence stars, for which
the isochrones give roughly the same value of $\log~g$ (see
Fig.~\ref{fig:External_errors}).  In addition, differences are
expected for the determination of the metallicities. The narrow
spectral range and the poor spectral signatures available around the
IR \ion{Ca}{ii} triplet cannot be compared to the plethora of lines
present in the whole wavelength range of the $S^4N$.

Table~\ref{tab:S4N_CFLIB} shows the biases and the dispersions between our derived parameters 
(plotted in Fig.~\ref{fig:External_errors} in red crosses) and the values found in \cite{Prieto_S4N}. 
We find that the biases for $T_\mathrm{eff}$, $\log~g$ and [M/H] are reasonable but not
negligible, of the order of $\sim -$108~K, $-$0.21~dex and
$-$0.08~dex, respectively. The dispersions are 145~K, 0.32~dex and 
0.09~dex, respectively.  Let us note
though that the biases on [M/H] and log~g disappear completely if we
consider the published values derived only from the same wavelength
range as we use here (strictly, the Gaia RVS wavelength range, with
results kindly provided by C.~Allende-Prieto, see \citet{Prieto_RVS}
for a detailed description).

However, the $S^4N$ library alone is not sufficient to test our
method fully. It does not cover the astrophysical parameter space well enough to
fulfil our goals of testing our pipeline over the entire H--R diagram,
and especially on intermediate and low-metallicity stars. 
The CFLIB library\footnote{See \url{http://www.noao.edu/cflib}}
consists of $\sim$~900 high SNR spectra that cover the entire
wavelength range of 3460-9464~\AA~ with a spectral resolution of
1.2~\AA.  This catalogue was conceived to cover the whole H--R
diagram down to metallicities of $\sim$ $-$2.5~dex, with the purpose
to be a testing library for automatic synthesis methods. The
parameters published for these stars, however, are a compilation of
independent studies of several authors, obtained by various
methods. Hence, some dispersion is expected, and the comparison
between the method's results with the published parameters of this
library needs to be performed carefully.

Together with the 900 spectra of the CFLIB catalogue, we used the
on-line database of PASTEL\footnote{\url{http://pastel.obs.u-bordeaux1.fr/}}
 \citep{PASTEL} instead of the published
values in \citet{CFLIB} to obtain updated values for the relevant
stellar parameters.  We chose then to keep only those spectra for which
at least two measurements were available from different authors, and
all stars with [M/H]$<-$~1~dex\footnote{Results of 15 different
  authors were considered to obtain the mean values to which
  we compare the results of our pipeline.} (ignoring the previous
selection criterion for metal poor stars).  In
Fig.~\ref{fig:External_errors} we can see the results obtained for
the 162 CFLIB stars that fulfil our criteria, where the estimates from
our pipeline are plotted against the mean values found in the
literature (black dots). The error bars, if any, represent the
dispersion in the published values. The scatter found is 171~K,
0.42~dex and 0.21~dex for $T_\mathrm{eff}$, $\log~g$ and [M/H],
respectively.  Table~\ref{tab:S4N_CFLIB} shows the error dispersions
 for the whole CFLIB library, and separately for the dwarfs ($\log~g
>$3) and the giants ($\log~g \leq$3).  No significant biases are
found, except for a small effect ($\sim$0.1~dex) in the metallicity of
dwarf stars.  It can be noticed that the stars for which our estimates
are the most distant from the published values are the ones which had only
 one associated measurement. Therefore, it is difficult to know if the
reason for these outliers is an error of this method or an error in
the published values.

We also tested the pipeline on the spectra of the Sun and
Arcturus. The spectra of \citet{Prieto_S4N} and \citet{Sun_Hinkle}
were degraded to SNR $\sim$100~pixel$^{-1}$, with 200 noise realisations. For
each of the considered cases, 400 spectra were processed.  For
the Sun the results are very close to the standard solar
parameter values, except for the metallicity, for which a difference of
--0.1~dex is noticed. Similar parameters are derived for the $T_\mathrm{eff}$ and the [M/H] 
of Arcturus, but $\log~g$ is less well derived, perhaps because of the imperfect
normalisation of the spectrum of \citet{Sun_Hinkle}
(Sect.~\ref{sec:linelist})

We comment briefly on the apparent bias found in metallicity for the
dwarfs. As we can see from Table~\ref{tab:S4N_CFLIB}, a similar bias of
$\sim-0.1$~dex is found for the Sun, the $S^4N$ stars and the dwarfs
of CFLIB. We decided to make no zero-point correction because the
suspected bias is within the error bars found for the metallicity,
while inadequate statistics make it impossible to quantify any bias as
a function of stellar type.  Application of the pipeline to Arcturus
showed no particular bias on the [M/H] parameter. As a consequence,
given our fairly arbitrary separation between metal-rich and metal-poor stars 
and between giants and dwarfs, we decided not to correct
any systematic bias.

Finally, we tested the pipeline by introducing four different values
of white Gaussian noise (SNR$\sim$ 10, 20, 50, 100~pixel$^{-1}$) to the $S^4N$ and
CFLIB spectra. In addition,  to increase the statistics of our
tests, twenty noise realisations for each spectrum were performed
for each value of the SNR.  Table~\ref{tab:S4N_CFLIB_SNR} shows the
error at 70\% of the total distribution for each of the
libraries. Good estimates are found down to SNR$\sim$20, as expected,
validating in this way our method for the case of observed spectra.

\begin{table}[!t]
\centering
\caption{Biases and dispersions for the derived parameters of the $S^4N$ and CFLIB libraries. }
\begin{tabular}{ccccc}
\hline
\hline

                              &    $T_\mathrm{eff}$              &log~$g$                 &$\mathrm{[M/H]}$      \\ 
                               &            (K)                  &    (dex)             &   (dex)             \\  \hline
   $S^4N$ (all)                & -108    $\pm$ 145                &-0.21  $\pm$ 0.32      & -0.08  $\pm$ 0.09  \\
   CFLIB (all)                 & 30      $\pm$ 171                &-0.04  $\pm$ 0.42      & -0.05  $\pm$ 0.21    \\
   CFLIB (dwarfs)              & -27     $\pm$ 156                &0.03   $\pm$ 0.26      & -0.10  $\pm$ 0.10   \\
   CFLIB (giants)              & 91      $\pm$ 118                &-0.05  $\pm$ 0.45      & -0.04  $\pm$ 0.21    \\ \hline 

\end{tabular}
\label{tab:S4N_CFLIB}
\end{table}

\begin{table}[!t]
\centering
\caption{Derived errors for the $S^4N$ and CFLIB libraries, at different signal-to-noise ratios.}
\begin{tabular}{c|ccc|ccc}
\hline
\hline
              & \multicolumn{3}{c|}{$S^4N$}      & \multicolumn{3}{c}{CFLIB} \\ \hline
    SNR       & $T_\mathrm{eff}$   &log~$g$       &$\mathrm{[M/H]}$ & $T_\mathrm{eff}$ &log~$g$    &$\mathrm{[M/H]}$    \\ 
(pixel$^{-1}$)& (K)               &    (dex)   &   (dex)         & (K)             &    (dex) &   (dex)             \\  \hline
   $\sim$100  & 173                 &0.40       & 0.13  & 158                 &0.40       & 0.19 \\
   $\sim$50   & 208                 &0.42       & 0.14  & 175                 &0.40       & 0.21      \\
   $\sim$20   & 270                 &0.53       & 0.15  & 233                 &0.48       & 0.27    \\
   $\sim$10   & 320                 &0.61       & 0.29  & 299                 &0.62       & 0.37      \\ \hline

\end{tabular}
\label{tab:S4N_CFLIB_SNR}
\end{table}


\section{Conclusions}
\label{sect:conclusions}
In the era of large spectroscopic surveys, the automated
parameterisation of stellar spectra in the infra-red ionised calcium
triplet region is a fundamental problem that must be addressed and
solved.  We here presented such a method to derive automatically
values of the effective temperatures, surface gravities and overall
metallicities of observed stars.  Alhough the application discussed in
this paper was to spectra with wavelength coverage around $\sim
8500\AA$, the method we derived is easily adaptable to any other
wavelength range and spectral resolution, provided there exists a grid
of synthetic spectra with the same characteristics as the observed
spectra. Furthermore, the dimensionality of the
parameter space can be increased from that discussed here, if desired,
by adding  the [$\alpha$/Fe] chemical ratio \citep{Gazzano}.

We studied two different approaches to the problem of automated
classification, which led to a new hybrid method that combines the two
approaches.  At high SNR, a projection algorithm, the MATISSE method
is preferred, because of its capability to interpolate between the spectra
of a synthetic grid, its lower errors and the easy interpretation of
its output.  At low SNR, the problem of secondary minima in the
solution for the best-fit synthetic spectrum is better treated with
DEGAS, the algorithm based on pattern recognition.  A quantification of
the internal errors showed that results that are sufficiently accurate for
galactic archaeology surveys which aim to study the full metallicity
distribution ($\sigma_\mathrm{[M/H]}<$0.1~dex), can be obtained down to
SNR$\sim$35 for all metal-rich and intermediate-metallicity stars
([M/H]$<-1$~dex). Furthermore, accurate results
($\sigma_\mathrm{[M/H]}<$0.2~dex) that suffice to separate the
different galactic components can be obtained down to
SNR$\sim$20. Indeed, we obtained $\sigma_\mathrm{[M/H]}<$0.12~dex,
0.18~dex and 0.23~dex for typical old thin disc, thick disc and halo
stars, respectively.

The application of this hybrid method was tested successfully on
stellar libraries of observed targets, validating it for extended surveys
such as Gaia. Our results show for the first time 
the expected accuracies that are possible from the spectra gathered by the RVS
of Gaia at its low sampling.  Furthermore, in addition to deriving
accurate parameters, the method is well suited for big datasets,
because it is not very time-consuming computationally -- processing
$2~10^4$ spectra takes less than an hour on a current laptop.  The
method is already coded in Java and will be delivered to the Gaia Data
Processing and Analysis Consortium (DPAC) as one of the possible nodes
of the whole processing pipeline.

We identified those regions of the H--R diagram where special
care must be taken to avoid systematic biases.  To
illustrate this point, consider a survey that mainly targets giant
stars.  In this case, the derived line-of-sight distances, obtained by
the projection of the estimated stellar atmospheric parameters on
a set of isochrones, can be seriously under-estimated because 
 a giant star can be misclassified as a hot dwarf.  Spurious
effects can therefore be introduced when computing the galactocentric
positions and space velocities of the sample and therefore compromise
any conclusions.  Nevertheless,  to minimise the effect of
this degeneracy, we can consider a synthetic grid composed only of the
spectra of giant stars, and consequently avoid populating regions of the
H--R diagram that should be left empty.  Furthermore, if 
a variety of spectral types are observed, appropriate
photometric ( (B-V) for warm stars, (V-K) for cooler stars)
measurements can also help to constrain the parameter space.

The conclusions presented in this paper will be used in a companion
paper \cite{Kordopatis11b}, where we will present the chemical
and dynamical properties of roughly 700 stars, selected to
probe the galactic thick disc, outside the solar neighbourhood.

\begin{acknowledgements} 
  The authors would like to thank the Mesocentre computer centre of
  the Observatoire de la C\^{o}te d'Azur for computing the grid of
  synthetic spectra and the $B(\lambda)$ functions. We are
  grateful to B.~Edvardsson for providing the grid of MARCS model
  atmospheres, B.~Plez for his molecular line-lists and the improved
  Turbospec code, M.~Irwin for  his very fruitful comments and suggestions as a referee, 
  A.~Robin for her useful advices on the use of the
  Besan\c{c}on model, C.~C.~Worley for the careful reading of this paper, 
  and C.~Allende-Prieto for letting us use
  his re-normalisation routine for IDL. RFGW acknowledges support from
  the US National Science Foundation, through grant AST-0908326, and
  from the Gordon \& Betty Moore Foundation.  Finally, G.K. would like
  to thank the Centre National d'Etudes Spatiales (CNES) and the
  Centre National de Recherche Scientifique (CNRS) for the financial
  support.
\end{acknowledgements}

\bibliographystyle{aa} 
\bibliography{my_article} 

\end{document}